%% file: main.tex
\documentclass[acmsmall]{acmart} 

\newif\ifTR
\TRtrue
\ifTR
\newcommand{\refappendix}[1]{\appref{#1}} %% w/ appendix
\newcommand{\reftr}[1]{\ref{#1}}
\else 
\usepackage{xr}
\newcommand{\refappendix}[1]{the extended version \cite{HistoriaExtendedMeier23}~Appendix~\ref*{tr-app:#1}} %% w/o appendix
\newcommand{\reftr}[1]{\ref*{tr-#1}} 
\fi

\usepackage[nosubfig,nographicx,nonatbib,noxcolor,nohyperref,notheoremenvs]{bec}
\SetSymbolFont{stmry}{bold}{U}{stmry}{m}{n} %supress asmmath font warning
\setboolean{showjedi}{false}
\setboolean{showcomments}{true} % turn off todo comments too
\usepackage{tikz} 
\usetikzlibrary{automata, positioning,arrows,arrows.meta,fit}
\usepackage{subcaption}
\usepackage{setspace} % for "\setstretch" macro used by specbox
\usepackage{empheq}
\newboolean{mhpl}
\newboolean{cbcftl}
\newboolean{mhpl-cbcftl-together}
\newboolean{eval}
\newboolean{relatedwork}
\newboolean{appendix}
\newboolean{overview}

\setboolean{overview}{true}
\setboolean{mhpl}{true}
\setboolean{cbcftl}{true}
\setboolean{mhpl-cbcftl-together}{true}
\setboolean{eval}{true}
\setboolean{relatedwork}{true}
\setboolean{appendix}{true}
\newcommand{\inputsection}[1]{\ifthenelse{\boolean{#1}}{\input{#1}}{\TODO{====== temp exclusion section #1 for build =======}}}

\setcopyright{rightsretained}
\acmPrice{}
\acmDOI{10.1145/3622865}
\acmYear{2023}
\copyrightyear{2023}
\acmSubmissionID{oopslab23main-p556-p}
\acmJournal{PACMPL}
\acmVolume{7}
\acmNumber{OOPSLA2}
\acmArticle{289}
\acmMonth{10}
\received{2023-04-14}
\received[accepted]{2023-08-27}

\renewcommand{\reftxt}[2]{\hyperref[#2]{#1~\ref*{#2}}}
\renewcommand{\lineref}[1]{\reftxt{line}{line:#1}} 
\newcommand{\refprog}[1]{\reftxt{location}{#1}}
\newcommand{\refhistimp}[1]{\reftxt{History Implication}{#1}}

\begin{document}

\ifTR
\title{Historia: Refuting Callback Reachability with Message-History Logics (Extended Version)}
\else
\title{Historia: Refuting Callback Reachability with Message-History Logics}
\fi

\author{Shawn Meier}
\orcid{}
\email{shawn.meier@colorado.edu}
\affiliation{%
  \institution{University of Colorado Boulder}
  \city{Boulder}
  \country{USA}
}

\author{Sergio Mover}
\orcid{0000-0003-1029-9547}
\email{sergio.mover@lix.polytechnique.fr}
\affiliation{%
  % \institution{École Polytechnique}
  \institution{LIX, École Polytechnique, CNRS, Institut Polytechnique de Paris}
  \city{Paris}
  \country{France}
}

\author{Gowtham Kaki}
\orcid{0000-0002-4189-3189}
\email{gowtham.kaki@colorado.edu}
\affiliation{%
  \institution{University of Colorado Boulder}
  \city{Boulder}
  \country{USA}
}

\author{Bor-Yuh Evan Chang}
\authornote{Bor-Yuh Evan Chang holds concurrent appointments at the University of Colorado Boulder and as an Amazon Scholar. This paper describes work performed at the University of Colorado Boulder and is not associated with Amazon.}
\orcid{0000-0002-1954-0774}
\email{evan.chang@colorado.edu}
\affiliation{%
  \institution{University of Colorado Boulder \& Amazon}
  \city{Boulder}
  \country{USA}
}

\input{macros}
\newcommand{\plusmini}{\scalebox{0.8}{$+$}}
\newcommand{\minusmini}{\scalebox{0.8}{$-$}}
\newcommand{\expos}{\ensuremath{\SetName{Ex}^{\plusmini}}}
\newcommand{\exneg}{\ensuremath{\SetName{Ex}^{\minusmini}}}
\newcommand{\newls}{CBCFTL\xspace}
\newcommand{\ignore}[1]{}

\begin{abstract}
This paper considers the callback reachability problem --- determining if a callback can be called by an event-driven framework in an unexpected state.
Event-driven programming frameworks are pervasive for creating user-interactive applications (apps) on just about every modern platform.
Control flow between callbacks is determined by the framework and largely opaque to the programmer.
This opacity of the callback control flow not only causes difficulty for the programmer but is also difficult for those developing static analysis.
Previous static analysis techniques address this opacity either by assuming an arbitrary framework implementation or attempting to eagerly specify all possible callback control flow,
but this is either too coarse to prove properties requiring callback-ordering constraints or too burdensome and tricky to get right.
Instead, we present a middle way where the callback control flow can be gradually refined in a targeted manner to prove assertions of interest.
The key insight to get this middle way is by reasoning about the history of method invocations at the boundary between app and framework code --- enabling a decoupling of the specification of callback control flow from the analysis of app code.
We call the sequence of such boundary-method invocations \emph{message histories} and develop message-history logics to do this reasoning.
In particular, we define the notion of an application-only transition system with boundary transitions, a message-history program logic for programs with such transitions, and a temporal specification logic for capturing callback control flow in a targeted and compositional manner.
Then to utilize the logics in a goal-directed verifier, we define a way to combine after-the-fact an assertion about message histories with a specification of callback control flow.
We implemented a prototype message history-based verifier called \toolname and provide evidence that our approach is uniquely capable of distinguishing between buggy and fixed versions on challenging examples drawn from real-world issues and that our targeted specification approach enables proving the absence of multi-callback bug patterns in real-world open-source Android apps.

\end{abstract}



\maketitle
\input{generated/specMacros}
\section{Introduction}
\label{sec:intro}
The standard approach for creating user-interactive applications (apps) is with event-driven frameworks.
In this programming model, a developer defines \emph{callback} methods that the framework invokes at run time in response to asynchronous events (e.g., starting the application, clicking a button, or a background task finishing).
Since the callbacks modify the state of the application, an unexpected order of callback invocations may lead to a bad application state and a subsequent crash.
To fix such a crash, it is common for the developer to update the application to change or accommodate complex callback interactions.
% In this paper, we reason about callback order and the associated application state for proving fixes applied to the app by the developer. 
To help the developer verify fixes to the crashing app, we present in this paper a technique to reason about callback order and to develop a tool that can automatically prove such fixes correct.

\JEDI{Important: App Dev: How could this callback have happened here? What sequence of callbacks got me here?}
As a specific example of an event-driven framework, we consider Android, a widely-used and complex mobile operating system. \autoref{fig:antennapod-stacktrace-issue-2855} shows a stack trace for a reported crash in AntennaPod, a popular open-source podcast player. But this stack trace is not helpful to the developer because the cause and effect span multiple callback invocations. 
This stack trace shows that there was a \codej{null} dereference in the \codej{call} callback --- but not how the reference became \codej{null}.
In particular, the reference must have been set to \codej{null} in some previous callback invocation before this \codej{call} callback invocation that is not visible in this stack trace.
To make such reasoning even more difficult, the app itself may affect the order of callbacks through the invocation of methods defined by the framework API; we refer to calls from the app to framework API methods as \emph{callins} by analogy to callbacks.
In summary, the developer of the app needs to reason about the order in which the framework could invoke callbacks and how the app invokes callins to understand and fix this crash.

Understanding the order of callbacks in this event-driven programming model is not only challenging for the developer but is also a central challenge for a program verifier attempting to prove an app safe from crashes. The program verifier has access to the app code, but the framework code is unavailable for all intents and purposes.
While this is true for most event-driven frameworks, it is particularly difficult in Android, which consists of thousands of API classes~\cite{https://developer.android.com/reference/classes}, 
evolves quickly,
includes lots of native code, and varies by device with manufacturer customizations.
One approach to the unavailable-framework problem is to analyze the app code assuming an arbitrary framework implementation. This design corresponds to analyzing each top-level callback (i.e., entry point into the app code) as a separate program with an application-only call graph~\cite{DBLP:conf/ecoop/AliL12}.
The advantage of this approach is that it is simple and general (i.e., it can over-approximate any framework implementation by assuming all callbacks can be invoked at any time in the event loop) and thus is the approach generally taken by industrial-scale analyzers for Android apps (e.g., \cite{fuchs:cs-tr-4991,DBLP:conf/ccs/LiangKMLGAH13,mariana-trench-web,DBLP:journals/cacm/DistefanoFLO19}).

\begin{figure}
\begin{lstlisting}[mathescape=false, basicstyle=\footnotesize\ttfamily\fontseries{m}\fontshape{n}]
java.lang.IllegalStateException: #$\ldots$#
Caused by: java.lang.NullPointerException: #$\ldots$#
  at de.danoeh.antennapod.fragment.ExternalPlayerFragment$$Lambda$4.call (Unknown Source:1193)
  #$\ldots$#
\end{lstlisting}
\Description{Null pointer stack trace.}
\caption{\label{fig:antennapod-stacktrace-issue-2855}
A reported stack trace from a confirmed bug~\citep{antennapodbug} that crashed the AntennaPod Android app. 
We have elided multiple lines identifying Android framework methods (using $\ldots$ for 14 elided lines in total).
% There are only two useful pieces of information here: a program variable was \codej{null}, and this crash happened in a callback named ``\codej{call}''.
}
%%% from https://github.com/AntennaPod/AntennaPod/issues/2855
\end{figure}

However, this most-over-approximate framework model is also wildly unrealistic. Without callback-ordering constraints, a verifier cannot possibly prove correct the accepted fix for \autoref{fig:antennapod-stacktrace-issue-2855}. 
Thus, many static analyzers for Android attempt to eagerly encode the callback control flow of core classes of the framework (e.g., the Activity Lifecycle~\cite{activity-lifecycle} modeled by ~\cite{DBLP:conf/pldi/ArztRFBBKTOM14,DBLP:conf/pldi/BlackshearGC15,DBLP:conf/icse/YangYWWR15}).
This approach also has some significant limitations.
For one, it is not feasible to eagerly specify the callback control flow of thousands of framework classes individually --- let alone callback control flow involving relationships between multiple classes.
On top of that, even specifying the behavior for a few framework components results in both soundness and precision issues~\cite{DBLP:conf/pldi/WangZR16,DBLP:conf/ecoop/MeierMC19,DBLP:conf/ndss/CaoFBEKVC15}.
% The problem is that the modeling of the framework is so tightly coupled with the analysis of the app code.

\JEDI{Contribution: Message-history logics: a program logic for the application and a specification logic for the framework.}
The main observation of this paper is that although the most-over-approximate framework model is unrealistic, the application-only approach for analyzing event-driven apps is not completely hopeless either.
In particular, if we decouple the framework modeling from the analysis of the app code, then the approach offers the appealing capability to gradually refine the possible callback control flow as needed and in a targeted manner.
To get a sense, imagine a call graph with a ``framework'' node representing the event loop and outgoing edges to each callback entry (as well as edges from callback nodes to the callin nodes they invoke).
%This graph is essentially a generalization of application-only call graphs~\cite{DBLP:conf/ecoop/AliL12} to apps developed against event-driven frameworks.
We can now consider traces through this graph instantiating callback or callin nodes with object instances;
we refer to such a callback or a callin instantiation generically as a \emph{message}.
Thus, we are interested in reasoning about \emph{message histories} --- sequences of messages obtained by call-return traces through this graph.

For any given framework implementation, not all message histories are realizable at run time. Thus, our key insight is to encode possible framework implementations by abstracting possible message histories. Crucially, this encoding and reasoning about message histories enables decoupling the specification of callback control flow from the abstract interpretation to compute an inductive program invariant. Specifically, in this paper, we make the following contributions:
\begin{itemize}
  \item We define the notion of an application-only transition system that records messages and a message-history program logic (MHPL) to describe and reason about \emph{boundary transitions} between the app and the framework independently of any specification of the framework (\autoref{sec:applog}).
  In MHPL, we consider a backwards-from-error formulation that enables goal-directed reasoning from a state assertion in the app, and we observe that deriving infeasible initial message histories refutes callback reachability.
  To capture consumption and ordering in MHPL, message-history assertions are derived from a fragment of ordered linear logic~\cite{DBLP:conf/tlca/PolakowP99,POLAKOW1999449} --- making use of an ordered linear implication.
  \item We formalize a callback control-flow temporal logic (\newls) to specify \emph{realizable message histories} --- that is fully decoupled from any particular program logic (\autoref{sec:newls}).
  This specification logic enables us to restrict possible callback control flow in a manner that is targeted and compositional.
  To capture possible traces, \newls{} is a specialization of past-time linear temporal logic~\cite{DBLP:conf/lop/LichtensteinPZ85}.
  \item We design an automated reasoning approach for the combination of MHPL assertions and \newls{} specifications (\autoref{sec:decision}).
  To utilize MHPL and \newls{} together in a static verifier, we define an algorithm instantiating \newls{} specifications with MHPL assertions into a single formula describing realizable message histories.
  We then use this encoding to answer queries about
  message-history entailment or
  whether a message history excludes the initial state
  with an off-the-shelf SMT solver.
  \item We empirically evaluate \toolname{}, a prototype goal-directed verifier with MHPL and show that it can refute callback reachability assertions with succinct specifications of callback control flow in \newls{} (\autoref{sec:eval}).
  In particular, we applied it to distinguish between the buggy and fixed versions from 5 real-world multi-callback issues from in-the-wild crashes of Android apps.
  Furthermore, we codified these 5 issues into bug patterns and evaluated the ability to use \toolname to prove the absence of these bug patterns on 47 open-source apps containing over 2 million lines of code: 43\% of the potentially buggy locations could be proven safe using \toolname with no additional modeling of callback control flow and then on a sample of the remaining locations, half of them could be proven safe (or witnessed as buggy) with a small amount of additional modeling.

  %\item We empirically evaluate \toolname{}, a prototype goal-directed verifier with MHPL and show that it can refute callback reachability assertions with succinct specifications of callback control flow in \newls{} (\autoref{sec:eval}).
  %In particular, we provide evidence that our approach is uniquely capable of distinguishing between buggy and fixed versions on challenging examples drawn from real-world issues and that our targeted specification approach enables proving the absence of multi-callback bug patterns in open-source Android apps.
\end{itemize}

\input{overview}

\input{mhpl}
\input{cbcftl}

\input{mhpl-cbcftl-together}
\input{eval}
\input{relatedwork}

\section{Conclusion}

We have described a novel middle way for refuting callback reachability that enables a decoupling of the specification of callback control flow from the abstract interpretation to compute program invariants over an application-only transition system.
This decoupling offers the appealing capability to gradually refine the possible callback control flow as needed and in a targeted manner to prove an assertion of interest, and
it thus moves us past the false dichotomy of either using no modeling or eagerly modeling all callback control-flow constraints.
The key innovation of our approach is an internalization of message histories into the analysis abstraction as a hypothetical (i.e., an ordered linear implication) to capture message histories up to a program location constrained by future messages and parametrized by a separate specification of realizable message histories.
We then define a specification logic for callback control flow (CBCFTL) that carefully specializes past-time linear temporal logic so that we can utilize message-history program logic (MHPL) assertions together with CBCFTL specifications.
Our evaluation provides evidence with a proof-of-concept implementation that our approach can refute callback reachability in challenging examples drawn from real-world issues among open-source apps.

%======================

\begin{acks}
We sincerely thank the anonymous reviewers for their constructive reviews, suggestions, and guidance throughout the revision process. We also cannot thank enough the members of the University of Colorado Programming Languages and Verification Group (CUPLV) for the helpful comments and support through the course of this work.
Specific thanks to Chi Huynh for helping with paper formatting and running experiments, as well as Benno Stein for thoughtful discussions on goal-directed verification.
This research was supported in part by the National Science Foundation under grants
% Evan
CCF-1619282, CCF-2008369, and 
% Sergio
AID/CIEDS project FARO.
\end{acks}

\section*{Data-Availability Statement}

The full implementation and data used for this evaluation are available as an artifact on Zenodo~\cite{conf/oopsla/Meier23Artifact}.
While the full experiments are resource-intensive computationally, a subset of the experiments may be run on an x86 Linux machine with 8GB of memory allocated to a Docker container.

\bibliography{conference.short,bec.short}

\ifTR
\clearpage
\appendix
\input{appendix}
\fi

\end{document}

%% file: generated/specMacros.tex
%%%Generated by bounder
\newcommand{\showSpec}{\specOnly{\ensuremath{\specciwithret{\codej{d}}{\codej{b}}{show}{}}}{ \ensuremath{\notiDir{\specciwithret{\codej{d}}{\codej{_}}{show}{}}}}}

\newcommand{\clickFinishSpec}{
		\ensuremath{\enkwCb~\codej{l.onClick()}}
	\boxright
		\ensuremath{
		\iDir{\enkwCi~\codej{v:=a.findViewById(id)}} 
		\wedge 
		[ \niDir{\enkwCb\enkwRet~\codej{a.onPause()}}{\enkwCb~\codej{a.onResume()}} \vee \iDir{\enkwCi~\codej{a.finish()}} ]
		\wedge 
		\niDir{\enkwCi~\codej{v.setOnClickList(}\enkwNull\codej{)}}{\enkwCi~\codej{v.setOnClickList(l)}} 	
		}
}
\newcommand{\callSpec}{\ensuremath{\speccb{\lsvar{l}}{call}{\lsvar{m}}\boxright \exists \lsvar{s}.~\niDir{\specci{\lsvar{s}}{unsubscribe}{}}{\specciwithret{\lsvar{s}}{\lsvar{_}}{subscribe}{\lsvar{l}} }}}

\newcommand{\getActivityNullSpec}{ \ensuremath{\specciwithret{\enkwNull}{\codej{f}}{getActivity}{}} \boxright \ensuremath{\niDir{\enkwCb~\codej{f.onActivityCreated()}}{\enkwCb\enkwRet~\codej{f.onDestroy()}} \vee \notiDir{\enkwCb~\codej{f.onActivityCreated()}}}}
\newcommand{\onActivityCreatedSpec}{}
\newcommand{\dismissSpec}{\specOnly{\ensuremath{\enkwCi~\codej{d.dismiss()}}}{\ensuremath{\exists \codej{a}.~\iDir{\enkwCi~\codej{d := _.show(a)}} \wedge \notiDir{\enkwCb~\codej{a.onResume()}} \vee \niDir{\enkwCb~\codej{a.onResume()}}{\enkwCb\enkwRet~\codej{a.onPause()}}}}}
\newcommand{\createOnceSpec}{\specOnly{\ensuremath{\enkwCb~\codej{a.onCreate()}}}{\ensuremath{\notiDir{\enkwCb~\codej{a.onCreate()}}}}}
\newcommand{\clickDisableSpec}{\specOnly{\ensuremath{\enkwCb~\codej{l.onClick()}}}{\ensuremath{\iDir{\enkwCi~\codej{v.setOnClickListener(l)}} \wedge \notiDir{\enkwCi~\codej{v.setEnabled(} \enkwFalse \codej{)}} \vee \niDir{\enkwCi~\codej{v.setEnabled(}\enkwFalse\codej{)}}{\enkwCi~\codej{v.setEnabled(}\enkwTrue \codej{)}}}}}
\newcommand{\executeSpec}{\specOnly{\ensuremath{\enkwCi~\codej{t.execute()}}}{\ensuremath{\iDir{\enkwCi\enkwRet ~\codej{t.execute()}}}}}

%% file: overview.tex
\newcommand{\addedhighlight}[1][added]{{\setlength{\fboxsep}{0pt}\colorbox{lightgreenbg}{#1}}}
\newcommand{\removedhighlight}[1][removed]{{\setlength{\fboxsep}{0pt}\colorbox{lightredbg}{#1}}}

\newsavebox{\SBoxUnsubCodeOne}
\begin{lrbox}{\SBoxUnsubCodeOne}
\begin{small}
\lststopn\begin{lstdiffplus}[language=java,alsolanguage=exthighlighting,style=number,linewidth=0.30\linewidth]
this.sub.unsubscribe();
\end{lstdiffplus}\lststartn
\end{small}
\end{lrbox}

\newsavebox{\SBoxUnsubCode}
\begin{lrbox}{\SBoxUnsubCode}
\begin{small}
\usebox{\SBoxUnsubCodeOne} \hlfix{}
\end{small}
\end{lrbox}

\newcommand{\unsubscribecode}{\usebox{\SBoxUnsubCode}}

\lstset{
  language=Java,
  escapechar=|
}
\section{Overview}
\label{sec:overview}

In this section, we illustrate how a developer could go deeper to diagnose and fix the bug causing the crash shown in \autoref{fig:antennapod-stacktrace-issue-2855} (\autoref{sec:distinguish}) and then demonstrate how our approach is able to prove the fixed version correct (\autoref{sec:automated-reasoning-message-histories}).

%We first illustrate the message histories of the buggy and fixed app (\autoref{sec:distinguish}) and motivate our static analysis approach that separates the specification of message histories from the analysis (\autoref{sec:separating}). Then, in \autoref{sec:automated-reasoning-message-histories} we explain the automated reasoning we use for proving the bug-fix correct (i.e., refuting that the assertion failure is reachable).

\subsection{Using Message Histories to Distinguish Bugs from Fixes}
\label{sec:distinguish}

We show a simplified version of an actual pull request submitted to fix the issue in \autoref{fig:playerfragment}.
What distinguishes the buggy and fixed versions --- without and with \lineref{unsubscribe-fix}, respectively --- are the possible \emph{message histories} (i.e., the possible sequences of callbacks and callins).
In \autoref{fig:feasible}, we show a message history witnessing a crash in the buggy version, while in \autoref{fig:infeasible}, we show the analogous message history in the fixed version.
The key to distinguishing the buggy and fixed versions is determining if such message histories are \emph{realizable} at run time.

% First, we explain the bug and fix associated with this example then we illustrate our message history based automated reasoning to prove the fix.

\JEDI{Our motivating example uses a combination of callbacks and callins to load media in the background.}
\autoref{fig:playerfragment} shows part of the \codej{PlayerFragment} class of the AntennaPod app that displays a user interface (UI) and loads some media for playing podcasts. Importantly, the app loads the media in a background task, a thread running asynchronously to the UI thread, to not block the user interface.
The framework invokes the \codej{onCreate} callback of a \codej{PlayerFragment} object when initializing the user interface. At \refprog{line:create}, this callback invokes the \codej{Single.create} callin (from the RxJava library) to start the background task that loads the media.
The subsequent call to \codej{task.subscribe(this)} at \refprog{line:subscribe} registers the \codej{call} callback.
At a later time, the framework will invoke the \codej{call} callback.
The \codej{call} callback uses the \codej{Glide} class to display the media on the \codej{this.act} object.
At any time, the framework can destroy the \codej{PlayerFragment} (e.g., when the user navigates to another part of the app) and invoke the \codej{onDestroy} callback, which at \refprog{line:actnull}, sets the field \codej{this.act} to \codej{null} to prevent memory leaks.

%note: reordered subsection header to be less ambiguous
% \subsection{Distinguishing Fixes from Bugs with Message Histories}
%\subsection{Using Message Histories to Distinguish Bugs From Fixes}
%\label{sec:distinguish}

% Importantly, multiple background tasks may be running at once for a variety of purposes. 
% Values shared between messages such as the first argument of \codej{subscribe} and the reciever of \codej{call} and the value returned by \codej{subscribe} are used by the application to interact with specific instance of a task.  Such values may indicate which instance is finishing or be used by the application to select a task to unsubscribe or cancel.

\newcommand{\histwidth}{0.25\linewidth}
\newcommand{\patchxleftmargin}{10em}
\newcommand{\minipageWidthRight}{1.5\linewidth}
\begin{figure}
\begin{subfigure}[b]{0.60\linewidth}\small %\hspace{0.2cm}
\begin{subfigure}[b]{0.94\linewidth}\small  
\lststopn\begin{lstlisting}[language=Java,alsolanguage=exthighlighting,style=number,xleftmargin=\patchxleftmargin,numbersep=-5pt]
  class PlayerFragment extends Fragment
  implements Action1<Playable> {
    Activity act = null;
    Subscription sub = null;
    void onCreate() { |\lstbeginn|
      this.act = $\ldots\text{\sl get non-}$null Activity$\ldots$;|\label{line:precreate}|
      var task = Single.create($\ldots$); |\label{line:create}|
      this.sub = task.subscribe(this); |\label{line:subscribe}|
    }|\lststopn|
    void call(Playable media){|\lststartn|
      assert(this.act != null); |\label{line:actnull}|
      Glide.with(this.act).load(media);|\label{line:npe}\lststopn|
    }
    void onDestroy() {|\lststartn|
      this.act = null;|\label{line:set-null}|\end{lstlisting}
  \begin{lstdiffplus}[language=Java,alsolanguage=exthighlighting,style=number,xleftmargin=\patchxleftmargin,numbersep=-5pt]
      this.sub.unsubscribe();|\label{line:unsubscribe-fix}|\end{lstdiffplus}
  \begin{lstcont}[style=number,xleftmargin=\patchxleftmargin,numbersep=-5pt]
    }|\label{line:destRet}\lststopn|
  }|\lststartn|\end{lstcont}
\end{subfigure}\hspace{-7.6cm}%
\begin{subfigure}[b]{0.8\linewidth}\small
  \input{figures/appOnlySide}%  
\end{subfigure}
% \captionsetup{margin={0.3cm,0cm}}
\caption{A patch adding line \ref{line:unsubscribe-fix} fixing the crash~\cite{antennapodfix}. On the left side, we illustrate an application-only transition system with a location $\enkwFwk$ representing the framework location inside the event loop with call and return edges $\rightharpoonup$ for callbacks. Callins are captured by self-loops as they affect framework state but are not on the event loop.
%, our starting point for static analysis (a single framework node with call edges $\rightharpoonup$). 
%We take the verification approach of starting with an over-approximate model where the framework is a single node in the control flow graph refining as needed.
% The graph augments the app with transitions to and from a single framework node, \enkwFwk.
% Call and return edges are indicated with $\rightharpoonup$ and connect the framework node to the corresponding line numbers.
%
We further indicate that the framework location $\enkwFwk$ is annotated with an inductive invariant to prove the assertion safe (expanded later in \autoref{fig:refgraph}).
}
\label{fig:playerfragment}
\end{subfigure}%
\begin{minipage}[b]{\minipageWidthRight}
\hspace{0.3cm}\begin{subfigure}[t]{\histwidth}\small\footnotesize
\begin{lstlisting}[frame=single,language=Java,alsolanguage=exthighlighting,style=number]
$\enkwCb$ f.onCreate();|\label{line:onCreate-cbmsg_}|
  $\enkwCi$ t = Single.create($\ldots$);
  $\enkwCi$ s = t.subscribe(f);
$\enkwCb\enkwRet$ f.onCreate();|\label{line:onCreate-cbretmsg_}
\end{lstlisting}
\begin{lstlisting}[frame=tlr,language=Java,alsolanguage=exthighlighting,style=number,firstnumber=last]
$\enkwCb$ f.onDestroy();|\label{line:onDestroy-cbmsg_}|
\end{lstlisting}
% \begin{lstdiffplus}[frame=lr,language=Java,alsolanguage=exthighlighting,style=number]
%   $\enkwCi$ s.unsubscribe();|\label{line:unsubscribe-cimsg}|
% \end{lstdiffplus}
\begin{lstcont}[frame=lrb,language=Java,alsolanguage=exthighlighting,style=number]
$\enkwCb\enkwRet$ f.onDestroy();|\label{line:onDestroy-cbretmsg_}|
\end{lstcont}
\begin{lstlisting}[frame=tlr,language=Java,alsolanguage=exthighlighting,style=number,firstnumber=last]
$\enkwCb$ f.call(m);|\label{line:call-cbmsg_}\lststopn|
  $\enkwErr$ a == null;
\end{lstlisting}
    \caption{%
      A realizable message history witnessing an exception ($\enkwErr$) in the buggy version.
      %and an unrealizable message history refuting the reachability of callback \codej{call} with \codej{this.act == null} for the buggy and fixed versions of (\subref{fig:playerfragment}), respectively.
    }
    \label{fig:feasible}
\end{subfigure}

\hspace{0.3cm}\begin{subfigure}[b]{\histwidth}\small\footnotesize
\lststartn
\begin{lstlisting}[frame=single,language=Java,alsolanguage=exthighlighting,style=number]
$\enkwCb$ f.onCreate();|\label{line:onCreate-cbmsg}|
  $\enkwCi$ t = Single.create($\ldots$);
  $\enkwCi$ s = t.subscribe(f);
$\enkwCb\enkwRet$ f.onCreate();|\label{line:onCreate-cbretmsg}
\end{lstlisting}
\begin{lstlisting}[frame=tlr,language=Java,alsolanguage=exthighlighting,style=number,firstnumber=last]
$\enkwCb$ f.onDestroy();|\label{line:onDestroy-cbmsg}|
\end{lstlisting}
\begin{lstdiffplus}[frame=lr,language=Java,alsolanguage=exthighlighting,style=number]
  $\enkwCi$ s.unsubscribe();|\label{line:unsubscribe-cimsg}|
\end{lstdiffplus}
\begin{lstcont}[frame=lrb,language=Java,alsolanguage=exthighlighting,style=number]
$\enkwCb\enkwRet$ f.onDestroy();|\label{line:onDestroy-cbretmsg}|
\end{lstcont}
\begin{lstlisting}[frame=tlr,language=Java,alsolanguage=exthighlighting,style=number,firstnumber=last,backgroundcolor=\color{lightredbg}]
$\enkwCb$ f.call(m);|\label{line:call-cbmsg}\lststopn|
  $\enkwErr$ a == null;
\end{lstlisting}
    \caption{%
      An unrealizable message history witnessing an exception ($\enkwErr$) in the fixed app. This cannot happen at run time.
      %and an unrealizable message history refuting the reachability of callback \codej{call} with \codej{this.act == null} for the buggy and fixed versions of (\subref{fig:playerfragment}), respectively.
    }
    \label{fig:infeasible}
\end{subfigure}
\end{minipage}
\Description{Motivating example.}
\caption{%
The crash from \autoref{fig:antennapod-stacktrace-issue-2855} arises in the buggy version of the code (without \reftxt{line}{line:unsubscribe-fix}) when a \codej{null} value is passed to \codej{Glide.with} from the \codej{this.act} field at \reftxt{line}{line:npe} in the \codej{call} callback.
To prevent the crash, the developer creates the fixed version by adding \codej{unsubscribe} at \reftxt{line}{line:unsubscribe-fix}.
Message histories shown by sub-figures~(\subref{fig:feasible}) and~(\subref{fig:infeasible}) show how a crash might be reached in both the buggy and fixed apps (without any reasoning about their realizability).
What we prove is that the message history for the fixed version shown in (\subref{fig:infeasible}) is unrealizable.
%The key to understanding the fix and our method of automated reasoning is identifying that Message The message history in sub-figure \subref{fig:infeasible} is not realizable at runtime.
In the message histories, single-letter identifiers represent run-time instances (e.g., \codej{f} is an instance of \codej{PlayerFragment}) and boxes are drawn around callback invocations.
% \codej{Glide.with} expects its argument to be non-\codej{null}.
% Whether or not this assertion fails depends on the ordering of callback invocations.
% The \codej{f.call(m)} callback invocation must happen after \codej{f.onDestroy()} (where \codej{f.act} is set to \codej{null}) to witness the crash in the buggy version (like we show in (\subref{fig:infeasible})) --- making this bug particularly pernicious.
% The fix is to prevent the \codej{f.call(m)} callback from happening after \codej{f.onDestroy()} by unsubscribing it there (with the \addedhighlight{} \codej{this.sub.unsubscribe()} call in \codej{onDestroy}).
%The fix only works if a \codej{call} cannot occur after an \codej{unsubscribe} where the receiver of \codej{call} is connected to the receiver of \codej{unsubscribe} through the \codej{subscribe} callin.
}
\label{fig:bug}
\end{figure}

\JEDI{message history witnessing defect explanation}

%Developers explain crashes such as this in terms of the sequence of callbacks and other methods.
%We refer to such methods as \emph{messages} when they cross the app-framework boundary.
%Similar to the developer, we reason about the crash as a sequence of messages, which we call a \emph{message history}.
% [sentence], which [...] - ... refers to the last substantative before the comma. substantative can be a clause with or without a verb. Look up "dependent clauses" parts of the sentence that cannot stand alone
% We generically term any transition across the app-framework boundary a \emph{message} and name such a sequence of messages a \emph{message history}.
As noted above, \autoref{fig:feasible} shows a crashing message history, that is, a sequence callback entries ($\enkwCb$), callin calls ($\enkwCi$), and callback returns ($\enkwCb\enkwRet$).
The arguments and return values of each message (e.g., \codej{f}, \codej{t}, \codej{s}, and \codej{a}) represent run-time addresses of objects.
%Here, we use such message histories to explain the crash and later formalize the notion of a message history for our message history logic.
%
The app crashes if the \codej{call} callback is invoked after the \codej{PlayerFragment} is paused.
%This is shown by the message history of \autoref{fig:feasible}.
After the $\codej{f.onCreate()}$ callback invocation (i.e., \reftxt{messages}{line:onCreate-cbmsg_}--\ref{line:onCreate-cbretmsg_}), 
the framework disposes the user interface and invokes the $\codej{f.onDestroy()}$ callback, setting the \codej{this.act} field to \codej{null} (i.e., app transitions between \reftxt{messages}{line:onDestroy-cbmsg_}--\ref{line:onDestroy-cbretmsg_}).
Then, the background process completes triggering the \codej{call} callback (i.e., \reftxt{message}{line:call-cbmsg_}).
However, the \codej{this.act} field is now \codej{null} causing the call to \codej{Glide.with(this.act)} to crash (represented by the \codej{assert} on \reftxt{line}{line:actnull} of \autoref{fig:playerfragment}).

\JEDI{fix is a fix because msg history witnessing problem is infeas}
The fixed version adds a call to \codej{this.sub.unsubscribe()} at \lineref{unsubscribe-fix}, which unsubscribes the \codej{call} callback preventing its invocation after \codej{onDestroy}.
In \autoref{fig:infeasible}, we show a message history that, while similar looking to the crashing message history of \autoref{fig:feasible}, is \emph{not realizable} with respect to the framework implementation. There is no execution that can generate the message history of \autoref{fig:infeasible} because the callback at \reftxt{message}{line:call-cbmsg} is \removedhighlight{} by the \addedhighlight{} \reftxt{message}{line:unsubscribe-cimsg} in the fixed version.
Such a minimal difference between the ``realizable crashing'' message history and the ``unrealizable safe'' message history highlights the automated reasoning challenge in distinguishing between the buggy and fixed versions of the app.
%
%That is, an abstract interpretation that raises an alarm on the buggy version while proving the fixed version safe (i.e., refuting the assertion failure at line \lineref{actnull} to be reachable) must determine if the message histories that can reach the exception are \emph{realizable} at runtime.

%The goal of this paper is to automatically reason about such minimal differences to prove safety in event-driven apps.

\newcommand*\numcircled[1]{\raisebox{.5pt}{\textcircled{\raisebox{-.9pt} {#1}}}}

\newcommand{\factdef}[2]{\hypertarget{Fact#1}{\begin{center}Informal Spec #1: ``#2''} \end{center}}
\newcommand{\factref}[1]{\hyperlink{Fact#1}{Informal Spec #1}}

\subsection{\toolname: Refuting Callback Reachability with Message-History Logics}
\label{sec:automated-reasoning-message-histories}

% We first generate the application-only control-flow graph shown on the left of \autoref{fig:playerfragment} representing an exploded supergraph of application methods from the application-only call graph mentioned earlier.
% All possible callbacks and callins are connected to the singleton framework node via \emph{boundary transitions} (e.g. $\enkwCb~\codej{this.onCreate()}$ from \autoref{fig:playerfragment}).
% The boundary transitions represent places where the framework makes non-deterministic choices of callbacks as well as their arguments and return values for callins (formalized in \autoref{sec:apponly-transition-system}).
% While clearly sound, this approach cannot prove the assertion in the fixed app on its own as it contains the message history \autoref{fig:infeasible}.

Here, we provide an overview of our static analysis abstraction and our framework specification logic that enables refining paths through the \emph{application-only transition system} by reasoning about the realizability of message histories. 
In \autoref{fig:playerfragment}, we illustrate an application-only transition system, which consists of app transitions from the app code augmented with \emph{boundary transitions} to a single, distinguished framework location $\refstate{}$.
Boundary transitions represent places where the framework makes non-deterministic choices of callback invocations as well as return values of the callin invocations (formalized in \autoref{sec:apponly-transition-system}).
The application-only transition system has the benefit of being a sound framework model by default (i.e., without further specification, is the most-over-approximate framework model) but clearly admits many unrealizable paths.
%%% Evan: This exponential point doesn't below here, I think. This paragraph is about the modeling. This challenge comes later.
%And worse, even assuming each callback is just invoked once, enumerating all possible callback sequences is clearly exponential.
Our key insight is to internalize the concept of realizable message histories into the static analysis abstraction.
This internalization of realizable message histories into the abstract domain enables decoupling the specification of possible callback control flow from the abstract interpretation to compute an inductive program invariant.
Such an approach is in contrast with the ones that \emph{eagerly} augment the interprocedural control flow graph with framework-specific control flow. At a high level, our approach shares some conceptual similarity with context-free language (CFL) reachability-based analysis~\cite{reps1998:program-analysis} in reasoning about realizability in the analysis but for imposing callback control flow instead of call-return semantics.

To describe our static analysis, we define Message-History Program Logic (MHPL), a program logic with an ordered linear implication for capturing assumptions about future messages. Then, to enable specifying callback-control flow, we introduce Callback-Control Flow Temporal Logic (CBCFTL), a past-time temporal logic for specifying constraints on the past message history given the present message. Finally, to automate reasoning about the realizability of message histories, we define an algorithm for instantiating CBCFTL specifications with MHPL assertions --- combining the inferred assumptions from a backwards-from-error static analysis and specified callback-control--flow constraints. In the rest of this section, we demonstrate our technique by walking through the verification of the bug-fix from \autoref{fig:playerfragment}.

\subsubsection{Message-History Program Logic (MHPL)}
\label{sec:overview-mhpl}

% note: analyses is plural of anlysis, plural used in this sentence
Program analyses often compute invariants for each location in the program. 
%For the locations inside a loop, the invariant needs to be inductive for it to capture all possible states in all iterations of the loop. 
%In a reactive application, such as AntennaPod, every location inside a callback is technically a loop location as every callback can be called infinitely often from the event loop. Consequently, proving program properties requires computing an inductive invariant for \emph{each location} in the application. 
Our approach relies on computing such an invariant that abstracts the message histories and application states that may reach the assertion failure using a novel message-history program logic (MHPL).
% novel message-history program logic (MHPL) for this purpose. The inductive invariant at a location over-approximates all possible message histories and states that may reach the assertion failure from that location. 
If the invariant excludes the initial state of the program (e.g., the empty message history), then there is no way the assertion failure can be reached from the initial state, letting \toolname prove that the assertion failure is impossible. 
Here, we demonstrate such a proof on our running example.
% We present a novel message-history program logic (MHPL) to abstract these message histories.

% Conversely, if the invariant (not necessarily inductive) includes the initial state, it means that the message histories and states that may reach the assertion failure via the location include the initial state of the program, which in turn means that there exists a possible path to assertion violation. \toolname raises an alarm in that case. We shall now demonstrate this approach on our running example.

\begin{figure}\centering\footnotesize%[H]%\begin{wrapfigure}{r}{0.6\textwidth}
\input{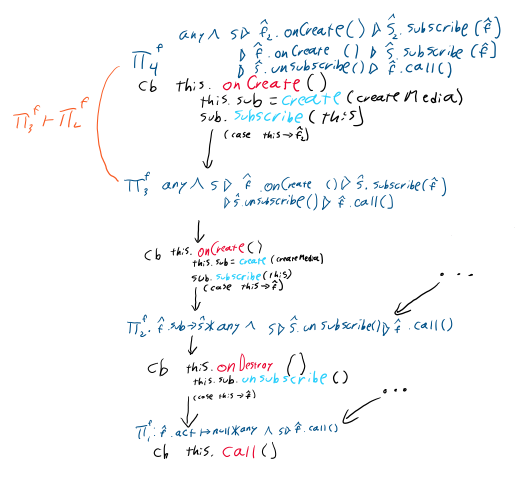}
\caption{An inductive invariant for the fixed app consisting of abstract message histories and abstract app states at each location in the application-only transition system (\autoref{fig:playerfragment}) that may reach the assertion failure. 
For brevity, this figure excludes less interesting transitions such as the case where the failing \codej{call} invocation is preceded by another invocation of \codej{call}. 
However, all transitions are considered by the verifier.
The abstract message histories only show messages from the user provided specification for clarity.
Note, there is only one framework location and that we use the subscripts (e.g., $\refstate{1}$, $\refstate{2}$) to indicate disjunctive elements of the abstract state at the framework location.
With the specification of realizable message histories (\autoref{sec:overview-cbcftl}) combined with the abstract message histories (\autoref{sec:overview-combined}), we can show that this invariant has reached a fixed point and that it excludes the initial state proving the assertion safe.
%   Refuting reachability of an assertion failure.
%   We illustrate a derivation in our message history program logic (MHPL) to prove that \codej{this.act} cannot be \codej{null} at \refprog{line:actnull} in the fixed version of \autoref{fig:playerfragment}.
%   The analysis is goal-directed by starting with the assertion failure condition shown as the abstract state $\mkshortabsmemnostore{\ref{line:actnull}}{\ldots}$ and proceeds backwards (upwards in the figure) to attempt to derive a contradiction on all backwards execution paths.
%   The nodes in round boxes show the derivation within a callback showing boundary-transition messages, relevant app-code statements, and program invariants at key locations.
%   Edges represent the event-driven control flow from the framework $\quies$ location to a callback entry.
%   Abstract states along the edges between callbacks labeled with $\refstate{n}$ are each individual disjuncts of the full abstract state at the $\quies$ location.
%   Single-letter identifiers in this figure represent existentially-quantified, symbolic variables (e.g., \codej{f} represents some instance of \codej{PlayerFragment} that may or may not be different from $\codej{f}_2$).
}
\label{fig:refgraph}
\end{figure}

While analyzing the application, \toolname maintains an invariant map, mapping each program location to its current invariant (represented as $\invpstate$ in \autoref{sec:refuting-with-mhpl}). 
Individual locations in the application are labeled with line numbers, and the framework location representing the event loop is labeled with $\refstate{}$.
We show a representation of this invariant map for our example in \autoref{fig:refgraph}. 
% with locations for $\refstate{}$ \TODO{usedef?} and individual line numbers.
% goal-directed in that we are computing the invariant map specifically for a single target assertion (as opposed to abstracting all application behavior).
% First we explain the meaning of individual abstract states and how the invariant map is initialized, and then, we explain how we iteratively compute the rest of the invariant map.
%In order for this invariant map to converge and be used to prove safety, \toolname additionally need a specification of realizable message histories capturing the facts mentioned earlier (\autoref{sec:overview-cbcftl}) and a way to reason about the combination of specifications and abstract message histories (\autoref{sec:overview-combined}).
In a goal-directed, backwards-from-error formulation, the invariant map is initialized with the state $\mkshortabsmem{\ref{line:actnull}}{\abstracebase}{\codej{f.act} \ptsto \enkwNull \lsep \codej{this} \ptsto \codej{f}}$ positioned right before the assertion failure.
There are two parts of the abstract state (which we separate with a centered dot $\cdot$).
On the right, there is an abstraction of the heap or store of the app for the assertion failure;
in particular, it says an object $\code{f}$ has a field $\code{act}$ that points to $\codej{null}$ and $\codej{this}$ points to $\code{f}$.
We use intuitionistic separation logic~\cite{ishtiaq+2001:assertion-language,reynolds2002:separation-logic:} to describe the relevant heap or store, but our approach is parameterized by essentially whatever logic one wishes to use to reason about the app state.
The left component is more interesting --- it is our abstraction of the possible message histories to reach this location (\abstrace~defined in \autoref{sec:refuting-with-mhpl}).
In particular, any execution that witnesses this assertion failure must have done so with a realizable message history, written $\abstracebase$.
Intuitively, $\abstracebase$ denotes the set of all realizable message histories as dictated by the framework. Note that $\abstracebase$ is not ``top'' or ``true'', which would concretize to all message histories --- realizable or unrealizable.

Next, we compute the abstract message history at the location just
before the \codej{call} entry.  This location is the $\quies$ location
(i.e., the framework event loop).  We write the message from this
transition as $\enkwCb~\codej{f.call(m)}$ where \codej{f} and
\codej{m} are symbolic variables corresponding to the values bound to
the formal parameters \codej{this} and \codej{media} of \codej{call},
respectively.
% Applying the pre-transformer (algorithmic applications of the Hoare
% triples in \secref{refuting-with-mhpl}) on this transition yields
% $\mkshortabsmem{\refstate{1}}{\ronepre
% \abstracebase}{\codej{f}.\codej{act} \ptsto \enkwNull} \;$.  The
% subscripted $\refstate{1}$ indicates that this is the first disjunct
% of the abstract state at the $\quies$ location.
To capture the effect of invoking \codej{call}, the abstract message
history is updated to $\ronepre\abstracebase$ at the abstract message
history labeled by $\refstate{1}$ in \autoref{fig:refgraph} (this
abstract state also removes the \codej{this} variable due to popping
the stack).  Intuitively, $\ronepre\abstracebase$ denotes any message
history to which \codej{f.call(m)} can be appended to obtain a
realizable message history. Operationally, $\ronepre\abstracebase$ can
be thought of as the set obtained by starting with the set of realizable message histories
$\abstracebase$, removing those that do not end with
\codej{f.call(m)}, and truncating the remaining ones to remove
\codej{f.call(m)} from the end.
%with values chosen for \codej{f} and \codej{m} to the current message history implies the whole message history is realizable''.
%Message histories corresponding to unrealizable choices for \codej{f} and \codej{m} are excluded from this abstraction.

% In other words, if \codej{f.call(m)} cannot be the next message is excluded.
% Importantly, the syntax ``$\abstracecons{}{\enkwCb~\codej{f.call(m)}}$'' does not correspond to any part of the abstracted message histories.
% We use the ordered linear implication operator \abstracecons{}{} because a message matching \codej{f.call(m)} appended to the message history implies the whole message history is realizable (there always exists a message matching the premise, therefore the premise cannot be false).

\JEDI{excludes-initial}
Having computed an abstract state at a framework location ($\quies_1$ in
Fig.~\ref{fig:refgraph}), we can now check if the abstract state excludes 
the initial state; if it does not, then we have not yet found a proof that the assertion failure is unreachable.
%\footnote{Note that the invariant need not be inductive to alarm as one realizable path 
%is sufficient to demonstrate the violation. On contrary, refuting the 
%assertion violation requires an inductive invariant as the proof needs 
%to show that the assertion is valid on all the paths. }
The message history abstraction of the initial state is simply a
singleton set containing an empty message history.  If the framework
could invoke \codej{f.call(m)} as the first callback, then
\codej{f.call(m)} alone is a realizable message history (i.e.,
$\codej{f.call(m)} \in\abstracebase$). Consequently,
$\ronepre\abstracebase$ is a set that contains an empty message
history, hence it includes the initial state, prompting an alarm.

However, in reality the framework cannot invoke \codej{f.call(m)} as the first callback (i.e., $\codej{f.call(m)}$ is not a realizable message history).
Crucially, since the set of realizable message history $\abstracebase$ of the framework is not available (i.e., it is defined in the framework implementation), \toolname can use separately-provided specifications of realizable message histories.
The following informal specification requires $\codej{f.call(m)}$ to be preceded by the invocation of $\enkwCb~\codej{s=t.subscribe(f)}$ for the message history to be realizable:

\factdef{1}{The framework may only invoke \codej{call} if \codej{subscribe} has been invoked in the past, and \codej{unsubscribe} has not been invoked since \codej{subscribe}.}

\noindent With this targeted specification,
$\ronepre\abstracebase$ no longer contains the empty message history and excludes the initial state. \toolname therefore
(correctly) does not raise an alarm at $\quies_1$.
%
% While our example uses more specifications, here we introduced only \factref{1} for clarity (all the specifications we used can be found in \refappendix{speclist}).

\JEDI{Pre-transformer repeated application}

This transformation of the abstract state from location \numcircled{5}
to $\quies_1$ is done by an abstract pre-transformer, which is applied
repeatedly to all predecessor transitions of an updated state until reaching a
fixed point.  Applying the
pre-transformer on the \codej{onDestroy} callback results in the
abstract state $\{\refstate{2}:\rtwopre\ronepre\abstracebase \cdot
...\}$ denoting the message histories that end in
$\enkwCi~\codej{s.unsubscribe()}$ followed by
$\enkwCb~\codej{f.call(m)}$.
%For clarity, we omit messages not mentioned by the facts shown in \autoref{sec:devint}.
Each new
abstract state is added to the invariant at a location as a
disjunctive clause (i.e., the $\enkwFwk$ location has an invariant of the form
$\{\refstate{1} \vee \refstate{2} \vee ...\}$).
\JEDI{We cannot exclude initial on buggy apps}
Continuing so on the buggy version from \autoref{fig:playerfragment} (and using the necessary specifications)
would yield a state whose abstract message history is
$\abstracecons{\abstracebase}{\abstracecons{\enkwCb~\codej{l.call(m)}}{\abstracecons{\enkwCi~\codej{s=t.subscribe(l)}}{\enkwCb~\codej{f.onCreate()}}}}$.
Note that this state includes the initial state raising an alarm, as it satisfies the
constraint imposed by \factref{1}.

\JEDI{Entailment}
As messages are parametrized by symbolic variables, we consider an unbounded number of possible message instances at each predecessor step.
Even if one restricts to one possible message instance for each callback method, considering all possible predecessor callbacks makes the proof
search exponential.
Fortunately, we are able to join abstract states by merging disjunctions, and
merging is required to find a fixed point in most cases.
Given a disjunction of abstract states, one disjunct may be merged with another 
if the first disjunction implies, or \emph{entails}, the one it is being merged with. 
% The entailment is defined point-wise on abstract message histories and abstract heaps. 
By merging new abstract states from backward transitions when possible,
we can reach a fixed point on some paths being explored. In
\autoref{fig:refgraph}, for example, $\enkwFwk_4$ is merged with
$\enkwFwk_3$. No further backward transitions need
to be explored from $\refstate{4}$ because they have been explored
from $\refstate{3}$.
%We refer to the check allowing such a merging as
%\emph{entailment} (noted by the judgment $\entail{}{}$ defined in
%\autoref{sec:applog}).

% Clauses in the state are merged when one clause represents a subset of the message histories of an existing clause.\TODO{}
% We refer to the merging of abstract states as entailment and use the judgment $\entail{}{}$ to indicate one abstract state entails another.
% When a clause is merged, no more backwards transitions from that state need to be considered, the state it is merged into has already caused the transitions to be explored. 

For presentation, Fig.~\ref{fig:refgraph} shows only a few of the
transitions and abstract states computed for the running example.
In practice, \toolname computes the fixed point at the framework location
after considering all backward transitions and shows that the
resultant inductive invariant excludes the initial state (for the
running example with the fix). In other words, it proves that no
realizable message history reaches the assertion violation in the
fixed version of AntennaPod. 

\subsubsection{Callback Control-Flow Temporal Logic (\newls)}
\label{sec:overview-cbcftl}

% The explanation thus far has treated the framework specification informally (e.g. \factref{1}). 
\newls is a novel language for formally writing specifications of realizable message histories.
A specification written in \newls consists of a conjunction of \emph{history implications}.  
With no history implications, the \newls specification places no restrictions on the realizable message history.
% The \newls specification $\enkwTrue$ with no history implications places no restrictions on the callback control flow (i.e., all the message histories are realizable). 
Each additional history implication targets one message that the framework can control, such as the invocation of
the callback \codej{call} or \codej{onCreate}. 
A history implication $\absmsg
\boxright \directive$
says that whenever a message satisfying the target abstract message
$\absmsg$ occurs, then the preceding message history must satisfy the
temporal formula $\directive$.\footnote{The $\absmsg
  \boxright \directive$ history implication can be seen as the
  first-order, past-time linear temporal logic formula
  $\square(\absmsg \rightarrow \mathop{\mathbf{Y}} \directive)$ that
  combines always (or globally) $\square$, %from linear temporal logic (LTL),
  implication $\rightarrow$, and yesterday (or previous)
  $\mathbf{Y}$ from past-time linear temporal logic (ptLTL).} 
% The history implication then
% implies a \emph{temporal formula} about the message history up until
% the target abstract message occurs. 
Temporal formulas are drawn from a
syntactically restricted fragment of past-time linear temporal logic over
finite traces. 
Such a structure for \newls is natural for three reasons:
\begin{inparaenum} \item it allows the developer of the framework model to target callbacks or callins with history implications as needed, \item history implications are compositional, and \item
    specifications may be cleanly combined with abstract message
    histories to automatically check excludes-initial and entailment
    (as described in \autoref{sec:overview-combined}).
\end{inparaenum} 
% Every additional history implication places a new restriction, and is added
% as a conjunction to the existing specification. 

% To illustrate $\newls$, we shall consider how the History Implication
% for \factref{1} may be written. 

% (full syntax, restrictions, and
% semantics are described in \autoref{sec:newls}).  
% For \factref{1}, the abstract message, $\absmsg$, is $\enkwCb~\codej{l.call(m)}$.

For \factref{1}, the history implication captures what must be
true of the message history when $\enkwCb~\codej{l.call(m)}$ is next.  
First, there must exist a subscription object \codej{s} that was returned from invoking \codej{subscribe}.
This object \codej{s} must be returned from the same invocation of
\codej{subscribe} that \codej{l} was passed to registering the
\codej{call} callback.  The subscription object \codej{s} is the
only parameter to the method \codej{unsubscribe} which must not have
come since the invocation of \codej{subscribe}.  Invoking
\codej{unsubscribe} on other objects will have no effect
on the target \codej{call}.  All of this is captured by History Implication \ref{spec:call}.

\begin{specDef}\label{spec:call}
For all objects \codej{l} and \code{m}, if the framework invokes \codej{l.call(m)}, then for some (subscription) object \code{s}, the message $\enkwCi~\codej{s.unsubscribe()}$ has \textbf{N}ot happened \textbf{S}ince $\enkwCi~\codej{s = _.subscribe(l)}$.
\specbox{\callSpec}
\end{specDef}

Note that \emph{since}, \textbf{S}, is the past-time dual of
\emph{until}, \textbf{U}, in LTL.  The \textbf{NS} operator has a
built-in ``not'' and restricts nesting to maintain decidability
(discussed in \autoref{sec:newls}).  Underscore \codej{_} in the above
specification is simply a shorthand for a locally
existentially-quantified variable (i.e., ``don't care'').

A key feature of \newls is that it
handles quantified values such as the listener object \codej{l} and
the subscription object \codej{s}.  Since the history implication
applies any time a \codej{call} is invoked, the listener object
\codej{l} is universally quantified.  Reasoning about such
quantifier alternation is often undecidable.  
However, the restrictions we have chosen for \newls allow them to be combined with abstract message histories such that automated reasoning is feasible.

% Both judgments
% are described formally in \secref{decision}. These decisions involve
% combined reasoning with MHPL and CBCFTL, which is described in the
% next subsection.

\subsubsection{Combining Abstract Message Histories with Callback Control-Flow}
\label{sec:overview-combined}

% Abstract message histories may be combined with \newls specifications
% allowing us to automatically check two properties needed by the
% message history program logic, excludes-initial and entailment
% (detailed in \secref{decision}).  For both properties, the first step
% is to combine the specification with the abstract state producing a
% single temporal formula.  We carefully designed temporal formula such
% that they may be compiled to a first order logic formula restricted to
% a decidable fragment of logic commonly supported by SMT solvers,
% Extended
% EPR~\cite{DBLP:conf/frocos/Korovin13,DBLP:journals/pacmpl/PadonLSS17}.

% Excludes-initial may be decided for a temporal formula by compiling to
% first order logic and checking if the formula is unsatisfiable when
% the message history is required to be empty.  The excludes-initial
% check is always decidable.  For entailment of two states, both states
% are combined with the specification into two temporal formulae.  If no
% message history exists in the first temporal formula that is not also
% satisfied by the second, then the first formula entails the second.
% Entailment is not in this decidable fragment of logic. However, Z3
% tends to be able to solve such formula quickly in practice.

Next, we consider how to interpret and automatically reason about the meaning of abstract message histories.
Consider the abstract state just before the \codej{call} callback with the abstract message history transition, $\ronepre\abstracebase$, which says that the next message must be $\enkwCb~\codej{f.call(m)}$.
First, excludes-initial needs to be proven (i.e., this abstract state does not contain the initial state), and then, entailment checks if it should be merged with any equivalent or weaker abstract states.
Both of these steps rely on a first-order logic encoding of abstract message histories that we explain here.
We prove the resulting first-order logic encoding to be decidable for the \emph{excludes-initial} judgment and computable in
practice for the \emph{entailment} judgment. 

This encoding starts by combining the abstract state $\ronepre\abstracebase$ with \reftxt{History Implication}{spec:call}. 
% The combination requires that at some point in the past, \codej{subscribe} must have been called on \codej{f} returning some \codej{s'}, and since then, \codej{unsubscribe} has not been invoked \codej{s'}.
The first step of combining abstract message histories with temporal formula is \emph{instantiation} (i.e., the \TirName{instantiate-yes} judgment in \autoref{sec:decision}). 
Intuitively, instantiation turns the "next" message from the abstract state into requirements on the message history so far.
For convenience, the output of instantiation is represented by the same language of temporal formula as is used in the history implications.
Temporal formula \eqref{eq:inst} shown below results from instantiating \reftxt{History Implication}{spec:call} on the abstract message history $\ronepre\abstracebase$.
Note how fresh variables are introduced for the existentially quantified values, but the values from the \codej{call} are retained.
\begin{equation}\label{eq:inst} \exists \lsvar{s'}, \lsvar{t'}.\;
  \niDir{\specci{\lsvar{s'}}{unsubscribe}{}}{\specciwithret{\lsvar{s'}}{\lsvar{t'}}{subscribe}{\lsvar{f}}
  } \end{equation} 
\noindent
Such a temporal formula may be converted to first-order logic and checked for excludes-initial and entailment via SMT solver as explained in \secref{overview-cbcftl}.
  
This temporal formula excludes the initial state because there must exist a \codej{subscribe} call in the message history (i.e., the judgment $\notinit{\abstrace}$ from section \autoref{sec:decision}).
% Intuitively, the first order logic interpretation of this formula can prove excludes-initial since the formula is not satisfiable with an empty message history.
Here, we also see why targeted specification is desirable for performance reasons.  
Each history implication can add constraints that prevent states from being merged via entailment.
The result of combining two sound history implications is always sound.
However, such combinations may impact performance by increasing the abstract state disjunctions at a location.

The next abstract message history shown by the invariant map at
$\refstate{2}$ adds the \codej{unsubscribe} call,
$\rtwopre\ronepre\abstracebase$.  The previously instantiated formula
needs to be updated for the
$\enkwCi~\codej{s.unsubscribe()}$ message.
That is, we must consider two cases: \begin{inparaenum}
  \item this \codej{unsubscribe} matches the \codej{unsubscribe} from the previous step (deriving a contradiction), and 
  \item this \codej{unsubscribe} is irrelevant to the previous step.
\end{inparaenum}
Combining these cases into the temporal formula is referred to as \emph{quotienting} (i.e., the \TirName{Quotient-not-since} judgment from \autoref{sec:decision}).
Additionally, if there was a history implication targeting \codej{unsubscribe}, it would also need to be instantiated (since there is not, the \TirName{instantiate-no} rule applies instead).
Combining these steps results in temporal formula \eqref{eq:quot}.
\begin{equation}\label{eq:quot}
\exists \lsvar{s'}. \left( \specci{\lsvar{s'}}{unsubscribe}{} \neq \specci{\lsvar{s}}{unsubscribe}{} \right) \wedge 
\niDir{\specci{\lsvar{s'}}{unsubscribe}{}}{\specciwithret{\lsvar{s'}}{\lsvar{_}}{subscribe}{\lsvar{f}}
}
\end{equation}

We attempt to eagerly merge abstract states using \emph{entailment}.
%%% Evan: Unnecessary to use the judgment form here, I think. And the judgment form is for message histories not the full state. Would need to reference the later section, at least.
% (i.e., the $\entail{\abstrace}{\abstrace'}$ judgment).
While as noted above $\refstate{4}$ merges with $\refstate{3}$, the abstract states at $\refstate{1}$  and $\refstate{2}$ cannot be merged since they restrict the app heap differently. 
%%% Evan: This reference to fwk_4 seems to come out of nowhere, so I think it must be stale. I added the dependent clause to the above sentence.
%As a result, no further steps back from $\refstate{4}$ need to be considered.
However for presentation, we illustrate entailment on the abstract message histories from $\refstate{1}$  and $\refstate{2}$, ignoring the app heap.
To determine entailment, we algorithmically search for a message history represented by temporal formula \eqref{eq:quot} and not by temporal formula \eqref{eq:inst}.  
If no such message history exists, then this disjunction has not progressed toward the initial state and can be dropped.
This entailment holds: the added constraint $\left( \specci{\lsvar{s'}}{unsubscribe}{} \neq \specci{\lsvar{s}}{unsubscribe}{} \right)$ simplifies to $\lsvar{s'} \neq \lsvar{s}$ and does not add any message histories to the abstraction over those represented by temporal formula \eqref{eq:inst}.  Note with the abstract app heap, a concrete app heap where \codej{f.act} points to a non-null value is represented by $\refstate{2}$ but not $\refstate{1}$, so these states may not be merged overall.

%% file: figures/appOnlySide.tex
\newcommand{\abovenode}{0.15cm}
\newcommand{\smallfont}[1]{\scalebox{.75}{#1}}
\newcommand{\sidexcoord}{-3.00}
\newcommand{\sidexcoordcode}{-2.60}
\newcommand{\fwkrightx}{-4.61}
\newcommand{\onedotstwo}{1 ~$...$~ 2}
\newcommand{\sixdots}{6 ~$...$~}
\newcommand{\aotrans}[1]{$#1$}
\newcommand{\vnodesep}{0.35cm}
\newcommand{\gxshiftarrows}{-0.95cm}
\newcommand{\cishiftarrows}{0.30cm}
\newcommand{\morexshiftarrows}{-0.19cm}
\newcommand{\ciin}{25}
\newcommand{\ciout}{30}
\newcommand{\cixshiftdesc}{0.0cm}
\newcommand{\ciyshiftdesc}{0.05cm}
\newcommand{\outincilooseness}{3}
\tikzset{location node/.style={draw=black, rounded corners, inner sep=0.2em, outer sep=0}}
\tikzset{msg node/.style={draw=none, inner sep=0, outer sep=-0.1em}}

\begin{tikzpicture}[every node/.style={rectangle, draw=none, rounded corners, every edge/.style={pos=0.4,draw=none}}, 
    descr/.style={above=\abovenode,anchor=west,xshift=\gxshiftarrows}, cbalign/.style={above=\abovenode,anchor=west,xshift=\gxshiftarrows+\morexshiftarrows}, 
    cidescr/.style={anchor=east, xshift=\cixshiftdesc, yshift=\ciyshiftdesc}, bendleftci/.style={bend left=-75,looseness=\outincilooseness}]

%%%%%%%%%%%%%%%%%%%%%%%%%%%%%%%
\node (fwk) [location node, node distance=4.2cm, text width=0.2cm, minimum height=3.0cm, minimum width = 0.8cm, anchor=north, draw]  at (-5,0){
    \small \enkwFwk:
    \begin{tabular}{c}
        \rotatebox{90}{     
            \footnotesize        
            $\{\refstate{1}:...\} \vee$
            $\{\refstate{2}:...\} \vee$
            $\{\refstate{3}:...\} \vee$
            $\{\refstate{4}:...\} $}
      \end{tabular}};

% % [Arrows on Side only]
%% onCreate
\draw[arrows = {-Stealth[harpoon]}] (\fwkrightx, -1.79) -- node[cbalign]{\hspace{0.4em}\smallfont{\aotrans{\enkwCb~\codej{this.onCreate()}}}} (\sidexcoordcode,-1.79); %,xshift=-0.19cm
\draw[arrows = {-Stealth[harpoon]}] (\sidexcoord, -2.47) to [bendleftci] node[cidescr]{\smallfont{\aotrans{\enkwCi~\codej{$...$}}}} (\sidexcoord,-2.75); %,xshift=-0.00cm
\draw[arrows = {-Stealth[harpoon]}] (\sidexcoord, -2.82) to [bendleftci] node[cidescr]{\smallfont{\aotrans{\enkwCi~\codej{$...$}}}} (\sidexcoord,-3.10); % ,xshift=-0.90cm
\draw[arrows = {Stealth[harpoon]-}] (\fwkrightx, -3.21) -- node[descr]{\smallfont{\hspace{0.55em}\aotrans{\enkwCb\enkwRet~\codej{$...$}}}} (\sidexcoord,-3.21); 
% \node[draw,dotted, xshift=0.15cm,  fit=(\fwkrightx, -1.74) (\fwkrightx, -3.25)] {};

%% call
\draw[arrows = {-Stealth[harpoon]}] (\fwkrightx, -3.55) -- node[cbalign]{\hspace{0.37em}\smallfont{\aotrans{\enkwCb~\codej{this.call(m)}}}} (\sidexcoordcode,-3.55); 
\draw[arrows = {-Stealth[harpoon]}] (\sidexcoord,-4.25) to [bendleftci] node[cidescr]{\smallfont{\aotrans{\enkwCi~\codej{$...$}}}} (\sidexcoord,-4.55); 
\draw[arrows = {Stealth[harpoon]-}] (\fwkrightx, -4.62) -- node[descr]{\hspace{0.45em}\smallfont{\aotrans{\enkwCb\enkwRet~\codej{$...$}}}} (\sidexcoord,-4.62); 

%% onDestroy
\draw[arrows = {-Stealth[harpoon]}] (\fwkrightx, -4.95) -- node[cbalign]{\hspace{0.39em}\smallfont{\aotrans{\enkwCb~\codej{this.onDestroy()}}}} (\sidexcoordcode,-4.95); 
\draw[arrows = {-Stealth[harpoon]}] (\sidexcoord,-5.60) to [bendleftci] node[cidescr]{\smallfont{\aotrans{\enkwCi~\codej{$...$}}}} (\sidexcoord,-5.89); 
\draw[arrows = {Stealth[harpoon]-}] (\fwkrightx, -5.95) -- node[descr]{\hspace{0.46em}\smallfont{\aotrans{\enkwCb\enkwRet~\codej{$...$}}}} (\sidexcoord,-5.95);

\end{tikzpicture}

%% file: figures/proofTree.tex
\newcommand{\calShort}{
	$\enkwCb~\codej{this.call(media);} \mkshortabsmem{\ref{line:actnull}}{\abstracebase}{\codej{f.act} \ptsto \enkwNull \lsep \codej{this} \ptsto \codej{f}}\codej{;assert(this.act != null)}$
}%
\newcommand{\ondShort}{
	$\enkwCb~\codej{this.onDestroy();}
	\mkshortabsmem{\ref{line:set-null}}{\rtwopre\ldots}{\codej{f.sub} \ptsto \codej{s} \lsep \codej{this} \ptsto \codej{f}}
	\codej{;this.act = null;}
	\enkwCi~\codej{this.sub.unsubscribe()}$
}%
\newcommand{\oncShort}[1]{
		$\speccb{\codej{this}}{onCreate}{}\codej{;}#1\codej{;}
		\specstaticciwithret{\codej{tsk}}{create}{$\ldots$}\codej{;}
		\specciwithret{\ensuremath{\codej{this.sub}}}{\codej{tsk}}{subscribe}{\codej{this}}$
}%
\newcommand{\subs}{
$\entail{\refstate{4}}{\refstate{3}}$
}%
\begin{tikzpicture}[node distance=0.70cm, every node/.style={rectangle, draw=black, rounded corners, every edge/.style={pos=0.4,draw=none}}]
\node (cal)[] {\calShort};
\node (R1)[above of=cal, draw=none] {
$\mkshortabsmem{\refstate{1}}{\ronepre \abstracebase}{\codej{f}.\codej{act} \ptsto \enkwNull}$ %
};
\node (ond)[above of=R1]{\ondShort};
% \node (ondr)[right of=ond, node distance = 3cm]{...};
\node (R2)[above of=ond, draw=none] {
	$\mkshortabsmem{\refstate{2}}{\rtwopre \ronepre \abstracebase}{\codej{f}.\codej{sub} \ptsto \codej{s}}$ %
};
\node (onc)[above of=R2]{\oncShort{\mkshorterabsmem{\ref{line:precreate}}{... \lsep \codej{this} \ptsto \codej{f}}}};
\node (R3)[above of=onc, draw=none]{$\mkshortabsmemnostore{\refstate{3}}{\rthreepre{\codej{s}}{\codej{f}}{\codej{t}} \rtwopre \ronepre \abstracebase}{}$};
% \node (ondr2)[right of=onc, node distance = 3cm]{...};
\node (onctwo)[above of=R3]{\oncShort{\mkshorterabsmem{\ref{line:precreate}}{... \lsep \codej{this} \ptsto \codej{f}_2}}};
\node (R4)[above of=onctwo, draw=none, text width=13.25cm,node distance=1cm]{$\mkshortabsmem{\refstate{4}}{\begin{array}[t]{@{}l@{}}\rthreepre{\codej{s}_2}{\codej{f}_2}{\codej{t}_2} \\ \rthreepre{\codej{s}}{\codej{f}}{\codej{t}} \rtwopre \ronepre \abstracebase}{\codej{f}_2 \neq \codej{f}}\end{array}$};
% \node (ondr3)[right of=onctwo, node distance = 3.5cm]{...};

\path[->] (R1) edge (cal);
\draw (ond) edge[-] node[draw=none,right]{
%\begin{minipage}{3.5cm}
%(case: \codej{f} $\mapsto \codej{f}$)
%\end{minipage}
} (R1);
% \draw (ondr) edge[-] node[draw=none,right,text width=0.75cm]{(case: other)} (R1.east);
% \draw (ondr2) edge[-] node[draw=none,right,text width=0.75cm]{(case: other)} (R2.east);
% \draw (ondr3) edge[-] node[draw=none,right,text width=0.75cm]{(case: other)} (R3.east);
\draw (onc) edge[-] node[draw=none,right]{
	% (case: \codej{f} $\mapsto \absof{f}$)
} (R2);
\draw (onctwo) edge[-] node[draw=none,right]{
%\begin{minipage}{4cm}
% (case: \codej{f} $\mapsto \absof{f}_2$)
%\end{minipage}
} (R3);

\path[->] (R2) edge (ond);
\path[->] (R3) edge (onc);
\path[->] (R4) edge (onctwo);
\draw (R4.west) edge[dashed, bend right, left,->, looseness=1] node[draw=none]{$\entail{}{}$} (R3.west);
\end{tikzpicture}

%% file: mhpl.tex
\section{Message-History Program Logic (MHPL)}
\label{sec:applog}

In this section, we explain the process of proving an application safe by showing no realizable message history can reach the assertion failure.
First, we define the notion of an application-only transition system that records a \emph{message history} during execution (\autoref{sec:apponly-transition-system}). 
Executions in this transition system, such as reaching the assertion failure, may be restricted based on whether they are realizable (e.g., with a user provided specification).
The application-only transition system provides a concrete semantics for a message-history program logic (MHPL) to reason about realizable message histories by adding the message history $\hist$ to the concrete state as ghost state.
Using MHPL, we can abstract message histories backwards from an assertion proving that no failing message history is realizable (\autoref{sec:refuting-with-mhpl}).

\subsection{An Application-Only Transition System with Message Histories}
\label{sec:apponly-transition-system}

\autoref{fig:apponly-transition-system} defines the syntax and semantics of a program that uses \emph{boundary transitions} to provide semantics to an app absent of the hidden framework implementation.
Conceptually, all framework code is merged within a single framework location $\enkwFwk$ (as illustrated in \autoref{fig:playerfragment} from \secref{overview}).
Our semantics are non-deterministic when the framework chooses the arguments for a callback invocation or the return value for a callin.
Execution simply ``gets stuck'' on an unrealizable boundary transition from the framework.

Boundary transitions $\btrans$ append a message to the message history $\hist$ capturing all interaction between the app and framework.
A boundary transition is a crossing of the app-framework boundary via
a callback invocation $\cfedge{\quies}{ \msgcb{\msgid}{\varseq} }{\apploc}$ from the framework back to the app,
a callback return $\cfedge{\apploc}{ \msgcbret{\var'}{\msgid}{\varseq} }{\quies}$ from the app into the framework,
or a callin invocation $\cfedge{\apploc}{ \msgci{\var'}{\msgid}{\varseq} }{\apploc'}$ from the app into the framework and back.
App transitions $\trans$ represent app code, for example, consisting of standard operations like reading and writing to the application heap.
%% Evan (Sep 10): Dropping this half of the sentence b/c our view now is creating objects is a callin to the "allocation framework".
% and creating new objects.
A message history $\hist \bnfdef \histemp \bnfalt \histcons{\hist}{\msg}$ is a sequence of messages with $\histemp$ being the empty sequence.
The application-only transition system is parametrized by a set of realizable message histories $\wfhist$ representing actions possible under the real framework.

% Method names, $\msgid$, are unique identifiers for the method procedure (e.g., a fully qualified and disambiguated name).
Method names $\msgid$ are a fully qualified and disambiguated name for method procedures.
We assume we can identify a method as being an app (i.e., a callback) method or a framework (i.e., a callin) method based on the method identifier $\msgid$ (e.g., app methods that override a framework type are callbacks in the case of Android).
The key part of the program state is recording a message history where messages $\msg$ are instances of boundary transitions; that is, a callback invocation with bound values $\msgcb{\msgid}{\valseq}$, a callback return $\msgcbret{\val'}{\msgid}{\valseq}$, or a callin invocation $\msgci{\val'}{\msgid}{\valseq}$.
Values, $\val$, may be compared with equality, created by the app, or created by the framework.

Callbacks and callins use a sequence of parameters as program variables $\var$ and return a value; we write a sequence with an overline (e.g., $\varseq$ for a sequence of variables). 
For simplicity, we assume that variable scoping and shadowing is handled by translation to this language (e.g., via alpha-renaming).
A callback return $\msgcbret{\var'}{\msgid}{\varseq}$ says that it returns the value in variable $\var'$ --- for the corresponding callback $\msgcb{\msgid}{\varseq}$; for simplicity, we assume an A-normal form where program expressions are evaluated in internal app transitions and bound to variable $\var'$ here.
For a callin invocation $\msgci{\var'}{\msgid}{\varseq}$, variable $\var'$ is the variable to bind the return value of the invocation.

We see transitions as control-flow edges between two program locations $\loc$.
A program location can be the framework location $\quies$ or an app location $\apploc$.
The framework location $\quies$ represents all control locations inside the framework.
A program $\prog$ is then a set of boundary $\btrans$ or app transitions $\trans$ for some unspecified syntax of app transitions. % that captures internal app transitions (where we write $\progemp$ for empty).
Conceptually, a program $\prog$ is the control-flow graph for each app callback augmented with boundary control-flow edges into and back from the framework location $\quies$.

A program state $\pstate$ is a memory $\mem$, at program location $\loc$, which consists of a message history $\hist$ with a boundary stack $\msgstack$ and an app store $\appstore$. 
A boundary stack $\msgstack$ is a stack ensuring that the message history consists of matching calls and returns.
%%% Evan 9/10: I think this reference to the appendix will be weird without it being there, so rewriting the sentence.
%Since we assume that callbacks may not be nested inside of callbacks (discussed in \refappendix{assume}), this stack may have at most one activation $\msgstackframe$.
If we assume that callbacks may not be nested inside of callbacks, this stack may have at most one activation $\msgstackframe$.
Like app transitions, the specific form of app stores $\appstore$ is unspecified, except we assume it supports looking up the value for an app variable $\maplookup{\appstore}{\var}$ and initializing variables $\appstore\extmap{\var}{\val}$. An app state $\appstate$ is then a pair of an app location $\apploc$ and an app store $\appstore$.

\begin{figure}\small{ 
  \begin{mathpar}

    \begin{array}{Rr@{\;}r@{\;}l}
      boundary transitions & \btrans & \bnfdef &
        \cfedge{\quies}{ \msgcb{\msgid}{\varseq} }{\apploc}
        \bnfalt \cfedge{\apploc}{ \msgcbret{\var'}{\msgid}{\varseq} }{\quies}
        \bnfalt \cfedge{\apploc}{ \msgci{\var'}{\msgid}{\varseq} }{\apploc'}
    \end{array}\\

    \text{app transitions}\quad \trans
 
    \text{message histories}\quad\hist\bnfdef \histemp \bnfalt \histcons{\hist}{\msg}

    \text{realizable message histories}\quad\wfhist

    \\

    \text{method names}\quad\msgid

    \text{messages}\quad\msg\bnfdef
        \msgcb{\msgid}{\valseq}
        \bnfalt \msgcbret{\val'}{\msgid}{\valseq}
        \bnfalt \msgci{\val'}{\msgid}{\valseq}

    \text{values}\quad\val

    \\
  
    \text{variables}\quad\var
  
    \text{program locations}\quad \loc \bnfdef \quies\bnfalt\apploc

    \text{app locations}\quad \apploc

    \\
  
    \text{programs}\quad\prog \bnfdef \progemp\bnfalt\progcons{\prog}{\btrans}\bnfalt\progcons{\prog}{\trans}

    \text{program states}\quad\pstate\bnfdef\mkpstatemem{\mem}{\loc}

    \\
    
    \text{memories}\quad\mem\bnfdef\mkmem{\hist}{\appstore}{\msgstack}

    \text{app stores}\quad\appstore

    \\

    \text{boundary stacks}\quad\msgstack\bnfdef\msgstackemp \bnfalt \msgstackcons{\msgstack}{\msgstackframe}

    \text{boundary activations}\quad\msgstackframe\bnfdef\msgcb{\msgid}{\valseq}
  
    \text{app states}\quad\appstate\bnfdef\mkappstate{\appstore}{\apploc}
  \end{mathpar}}
  \begin{mathpar}
    \fbox{%
    $\jeval{\pstate}{\btrans}{\pstate'}$
    \\
    $\jstep[\prog]{\pstate}{\pstate'}$
    % \\
    % $\jstep[\prog]{\pstate}{\pstate'}$
    % \\
    % $\jeval{\appstate}{\trans}{\appstate'}$
    }
    \\ 
  
    \infer[c-callback-invoke]{
      %\msgci{\val'}{\msgid'}{...\appval...} \in \hist\;\text{for all $\appval \in \valseq$} \\
      \wfhistjudge{\histcons{\hist}{\msgcb{\msgid}{\valseq}}}
      % \jrealhist{ \histcons{\hist}{\msgcb{\msgid}{\valseq}} }
    }{
      \jeval
        { \mkpstate{\hist}{\appstore}{ \msgstack }{\quies} }
        { \cfedge{\quies}{ \msgcb{\msgid}{\varseq} }{\apploc} }
        { \mkpstate{ \histcons{\hist}{\msgcb{\msgid}{\valseq}} }{ \appstore\fmtseq{\extmap{\var}{\val}} }{ \msgstackcons{\msgstack}{\msgcb{\msgid}{\valseq}} }{\apploc}  }
    }
    \\ %\rule{0pt}{2ex}\\ % extra space hack: no idea why the line spacing in mathpar is being weird
    
    \infer[c-callback-return]{
      \msg = \msgcbret{ \maplookup{\appstore}{\var'} }{\msgid}{\valseq} \\
      \valseq = \fmtseq{ \maplookup{\appstore}{\var} } \\
      \wfhistjudge{\histcons{\hist}{\msg}}
      % \jrealhist{ \histcons{\hist}{\msg} }
    }{
      \jeval
        { \mkpstate{\hist}{\appstore}{ \msgstackcons{\msgstack}{\msgcb{\msgid}{\valseq}} }{ \apploc } }
        { \cfedge{\apploc}{ \msgcbret{\var'}{\msgid}{\varseq} }{\quies} }
        { \mkpstate{ \histcons{\hist}{\msg} }{\appstore}{ \msgstack }{\quies}  }
    }
    \\

    \infer[c-callin-invoke]{
      \valseq = \fmtseq{ \maplookup{\appstore}{\var} } \\
      \msg = \msgci{ \val' }{\msgid}{\valseq} \\
      %\msgci{\val}{\msgid'}{...\appval...} \in \hist\;\text{if $\val' = \appval$} \\
      %\jseen{\hist}{\val'} \\
      \wfhistjudge{ \histcons{\hist}{\msg} }
      % \jrealhist{ \histcons{\hist}{\msg} }
    }{
      \jeval
        { \mkpstate{\hist}{\appstore}{ \msgstack }{ \apploc } }
        { \cfedge{\apploc}{ \msgci{\var'}{\msgid}{\varseq} }{\apploc'} }
        { \mkpstate{ \histcons{\hist}{\msg} }{\appstore\extmap{\var'}{\val'} }{ \msgstack }{\apploc'}  }
    }
    \\

    \infer[c-app-step]{
      \jappeval{ \mkappstate{\appstore}{\apploc} }{ \trans }{ \mkappstate{\appstore'}{\apploc'} } \\
      \trans \in \prog \\
      \apploc = \preof{\trans} \\
      \apploc' = \postof{\trans}
    }{
      \jstep[\prog]%
        { \mkpstate{\hist}{\appstore}{\msgstack}{\apploc} }
        { \mkpstate{\hist}{\appstore'}{\msgstack}{\apploc'} }
    }
      
    \infer[c-boundary-step]{
      \jeval{ \pstate }{ \btrans }{ \pstate' } \\
      \btrans \in \prog
    }{
      \jstep[\prog]%
        { \pstate }
        { \pstate' }
    }
    \\ %\rule{0pt}{2ex}\\ % extra space hack: no idea why the line spacing in mathpar is being weird

    \text{initial program state}\quad\initpstate = \mkpstate{\histemp}{\initappstore}{\msgstackemp}{\quies}

    \text{initial app store}\quad\initappstore
    \\

  \end{mathpar}
  \caption{\label{fig:apponly-transition-system}%
    An application-only transition system with boundary transitions and message histories.
    The message history~$\hist$ component of the program state records the execution of boundary transitions $\btrans$ between the app and framework.
    We use the judgment $\jstep[\prog]{\pstate}{\pstate'}$ to represent a single step over either an app or boundary transition in the application.
    The transition system is parametrized by a set of realizable message histories~$\wfhistjudge{\hist}$.
  }
\end{figure}

Boundary transitions $\btrans$ are particularly interesting as they capture the non-deterministic or unobserved behavior of the framework and record the action in the ``ghost state'' $\hist$ for the message history.
The boundary transition judgment form $\jeval{\pstate}{\btrans}{\pstate'}$ says, ``In program state $\pstate$, executing the boundary transition $\btrans$ results in an updated program state $\pstate'$ and is realizable under realizable message histories \wfhist.''
This judgment form captures the realizable executions of boundary transitions.
To execute a callback invocation transition $\cfedge{\quies}{ \msgcb{\msgid}{\varseq} }{\apploc}$ via rule \TirName{c-callback-invoke}, the program state is at the framework location $\quies$, values $\valseq$ are chosen for the callback parameters $\varseq$ non-deterministically (conceptually by the framework) and initialized in the app store \smash{$\appstore\fmtseq{\extmap{\var}{\val}}$}, and the callback activation $\msgcb{\msgid}{\valseq}$ is pushed on the boundary stack $\msgstack$.
Then to record this boundary transition execution, this callback message $\msgcb{\msgid}{\valseq}$ is appended onto the current message history $\hist$.
We want to capture that depending on the current message history $\hist$, this callback invocation transition may not be realizable.
This realizability of message histories is captured by checking if the new message history is realizable with $\wfhistjudge{\histcons{\hist}{\msgcb{\msgid}{\valseq}}}$.
%% Evan 9/10: The following seem to be duplicated with more detailed description of the rules further below.
%%Both of these rules perform the same check that the message history is well-formed.
%Mirroring the callback invocation, the callback return rule assumes that there is a callback on the boundary stack, $\msgstackcons{\msgstack}{\msgcb{\msgid}{\valseq}}$, reads the value for the return parameter, $\appstore\extmap{\var'}{\val'}$, and finally appends the message history with a return message, $\msgcbret{\val'}{\msgid}{\valseq}$.
%The callin invoke rule reads arguments from the app store, $\appstore\fmtseq{\extmap{\var}{\val}}$, then the framework chooses a return value, $\val'$, which is assigned to the return variable, $\var'$ in the app store, and finally, the callin message, $\msgci{\val'}{\msgid}{\valseq}$ is added to the history.
%As a consequence of these semantics, every prefix of a realizable message history must also be realizable.

Executing a callback return transition $\cfedge{\apploc}{ \msgcbret{\var'}{\msgid}{\varseq} }{\quies}$ via rule \TirName{c-callback-return} is then the expected symmetric operation. The return value is read out of the app store $\maplookup{\appstore}{\var'}$, the callback activation $\msgcb{\msgid}{\valseq}$ is popped off the boundary stack, and control goes into the framework location $\quies$.
The premise $\valseq = \fmtseq{ \maplookup{\appstore}{\var} }$ enforces that argument variables are not modified by the callback which simplifies the formalism.
The callback return message $\msgcbret{\maplookup{\appstore}{\var'}}{\msgid}{\valseq}$ is similarly appended onto the current message history $\hist$ to record the execution of the callback return transition --- and checked for realizability.
Note that to connect the callback invocation with its return, the callback return message includes the method name $\msgid$ and actual arguments $\valseq$ from the callback activation.

The callin invocation transition $\cfedge{\apploc}{ \msgci{\var'}{\msgid}{\varseq} }{\apploc'}$ via rule \TirName{c-callin-invoke} is symmetric to callback invocation and return together, that is, the arguments for the callin $\valseq$ are read from the app store, and the callin return value from the framework $\val'$ is chosen non-deterministically (conceptually by the framework) and then bound to variable $\var'$. Then, the callin message $\msgci{\val'}{\msgid}{\valseq}$ is appended onto the current message history $\hist$ and checked for realizability.
It should be noted that \TirName{c-callback-invoke}, \TirName{c-callback-return}, and \TirName{c-callin-invoke} together capture that the framework cannot modify the app store $\appstore$ except through invoking callback methods.
%There is no arbitrary modification of the app store after returning from a callback,
%while at the framework location $\quies$, and before invoking another.
This formalizes one aspect of the so-called separate compilation assumption of \citet{DBLP:conf/ecoop/AliL12}, which considers the consequences of the assumption that framework is developed separately and compiled in the absence of the app.
As a consequence of these semantics checking realizability at each boundary transition, every prefix of a realizable message history must also be realizable.

The application-only transition system is then given by the transition relation judgment form $\jstep[\prog]{\pstate}{\pstate'}$ that says, ``Program state $\pstate$ steps to $\pstate'$ in program $\prog$ by either a boundary transition $\btrans$ or app transition $\trans$ under realizable message histories $\wfhist$.''
Straightforwardly, the \TirName{c-app-step} and \TirName{c-boundary-step} rules simply state that we can either take a step with an app transition $\trans$ or a boundary transition $\btrans$ in the program $\prog$ (depending on the program location).
The transition semantics of app transitions are left unspecified $\jappeval{\appstate}{\trans}{\appstate'}$.
We assume that app transitions themselves do not read or write framework state directly, as the app store $\appstore$ is separate from the framework state.
%% Evan 9/10: This is already stated above
%App state $\appstate$ may persist between callbacks.
%
And finally, concrete executions are given by the reflexive-transitive closure of this single-step transition relation from an initial program state $\initpstate$.
We write $\jstepstar[p]{\pstate}{\pstate'}$ for the reflexive-transitive closure of $\jstep[\prog]{\pstate}{\pstate'}$.

\subsection{Refuting Callback Reachability with Message-History Program Logic}
\label{sec:refuting-with-mhpl}

The ultimate aim of MHPL is to prove statically that a program assertion cannot fail.
We start with the \emph{error condition}, $\abspstate$, an abstract state just before the assertion such that the assertion may fail (e.g., $\codej{f.act} \ptsto \enkwNull$ from the running example in \secref{overview} representing app memories where there exists a framework object with a null \codej{act} field).
We refute the reachability of the error condition with the judgment form $\jrefute[\prog]{\loc}{\abspstate}$.
This judgment form is read as, ``No concrete program state satisfying the abstract state $\abspstate$ is reachable in program $\prog$ with realizable message history specification $\SpecSet$.''
We use \newls, defined in \secref{newls}, to abstract the set of reachable message histories $\wfhist$ in the concrete semantics.
% However, note that MHPL is defined independently of any particular specification language; other specification languages could also be used with this logic.
We use the judgment $\semincluded{\hist}{\SpecSet}$ to say that a message history $\hist$ is captured by a specification $\SpecSet$ and note the set of message histories in a specification as $\wfhistspec$ (i.e., the concretization of a specification $\SpecSet$ is the set of message histories $\wfhistspec$).
Since the specification is an input to our algorithm, we assume that $\SpecSet$ is a sound abstraction of realizable message histories
(i.e., $\wfhistspec \subseteq \wfhist$ for the set of realizable message histories $\wfhist$ in the concrete semantics).
%(i.e., if $\wfhistjudge{\hist}$, then $\semincluded{\hist}{\SpecSet}$, where we write $\semincluded{\hist}{\SpecSet}$ for a concrete message history $\hist$ modeled by a specification $\SpecSet$).

Our proof technique works in a goal-directed manner: 
we over-approximate the set of the states that may reach the given error condition $\abspstate$ with a program-state invariant, $\invpstate$.
If the initial program state $\initpstate\colon \mkpstate{\histemp}{\initappstore}{\msgstackemp}{\quies}$ is excluded from the program state-invariant, $\invpstate$, then the location of $\abspstate$ cannot be reached with any concrete state satisfying abstract state error condition $\abspstate$.
For some abstract state $\abspstate$, the \emph{excludes-initial} judgment, $\notinit{\abspstate}$, holds only if the concretization of $\abspstate$ must not contain the initial state $\initpstate$.
% We write $\notinit{\abspstate}$ for the \emph{excludes-initial} judgment that states when the abstract program state excludes --- i.e., must not contain in its concretization --- the initial program state $\initpstate$.
While an abstract state $\abspstate$ may be excludes-initial in any of its components (e.g., the abstract app store), the particularly interesting component here is its abstract message history $\abstrace$.
%that is excludes-initial if it does not include the empty message-history sequence $\histemp$.
Thus, in subsequent sections, we focus in on the excludes-initial judgment on abstract message histories.
Excludes-initial for message histories, $\notinit{\abstrace}$, holds if $\histemp$ is not in the concretization of $\abstrace$.

\begin{figure}\small
%{ %\setstretch{1.2} % extra space hack: no idea why the line spacing in mathpar is being weird
\begin{mathpar}
  \text{abstract messages}\quad\absmsg\bnfdef
      \msgcb{\msgid}{\symvalseq}
      \bnfalt \msgci{\symval'}{\msgid}{\symvalseq}
      \bnfalt \msgcbret{\symval'}{\msgid}{\symvalseq}

  \text{symbolic variables}\quad\symval

  \text{assignments}\quad\valua\bnfdef \valuaemp \bnfalt \valuaext{\valua}{\symval}{\val}

  % \text{pure constraints}\quad\puredom
  %\bnfdef\puretrue\bnfalt\pureand{\puredom_1}{\puredom_2}\bnfalt\cdots\bnfalt\pureseen{\symval}

  \text{abstract message histories}\quad\abstrace\bnfdef
    \abstracebase
    \bnfalt\abstracecons{\abstrace}{\absmsg}

  \text{realizable message histories specs}\quad\SpecSet

  \text{abstract app stores}\quad\absappstore\bnfdef\absappstoreTrue\bnfalt\absappstoreAnd{\absappstore_1}{\absappstore_2}
  \bnfalt\absappstorePtsto{\var}{\symval}\bnfalt\cdots
  \bnfalt\absmemFalse
  \bnfalt\absmemOr{\absappstore_1}{\absappstore_2}

  \text{abstract app states}\quad\absappstate\bnfdef\mkabsappstate{\absappstore}{\apploc}{\puredom}

  \text{abstract memories}\quad
  \absmem\bnfdef\mkabsmem{\abstrace}{\absappstore}{\absmsgstack}{\puredom}\bnfalt\absmemFalse\bnfalt\absmemOr{\absmem_1}{\absmem_2}

  \text{abstract program states}\quad\abspstate\bnfdef\mklocstate{\absmem}{\loc}
  %\mkabspstate{\abstrace}{\absappstore}{\absmsgstack}{\loc}{\puredom}\bnfalt\mklocstate{\bot}{\loc}

  \text{abstract boundary stacks}\quad\absmsgstack\bnfdef\absmsgstackemp
  \bnfalt\absmsgstackcons{\absmsgstack_1}{\msgcb{\msgid}{\symvalseq}}
  %\bnfalt\absmsgstackframe
  %\text{abstract boundary activations}\quad\absmsgstackframe\bnfdef
  % \bnfalt\msgcb{\msgid}{\symvalseq}

  \text{abstract program-state invariants}\quad\invpstate\bnfdef\invemp\bnfalt\invcons{\invpstate}{\loc}{\abspstate}

  \\
%\end{mathpar}}
%\begin{mathpar}

  \fbox{%
  $\withvalua{\hist} \models_{\SpecSet} \abstrace$
  }
  \\
  \inconc{\withvalua{\hist}}{\abstracebase} \quad\text{iff}\quad \hist \models \SpecSet

  \inconc{\withvalua{\hist}}{\abstracecons{\abstrace}{\absmsg}} \quad\text{iff}\quad \text{$\semincluded{\withvalua{\msg}}{\absmsg}$ implies $\inconc{\withvalua{\histcons{\hist}{\msg}}}{\abstrace}$} \text{ and } \histcons{\hist}{\msg}\models\SpecSet
  %
  % \concof{\mkabspstate{\abstrace}{\absappstore}{\absmsgstack}{\loc}{\puredom}} 
  %   \defeq \SetST{\mkpstate{\hist}{\appstore}{\msgstack}{\loc}}{ \text{exists}~\valua~\text{s.t.}~\withvalua{\hist} \in \concof{\abstrace}~\text{and}~
  %   \withvalua{\msgstack} \in \concof{\absmsgstack}~\text{and}~\withvalua{\appstore} \in \concof{\absappstore}
  %   }
  % 
  % \concof{\mklocstate{\bot}{\loc}} \defeq \set{} 
  %
  %\concof{\abstracebase}\defeq \SetST{ \withvalua{\hist} }{ \text{$\wfhistjudge{\hist}$ for any $\valua$}  }
  %
  %\concof{\abstracecons{\abstrace}{\absmsg}}\defeq \SetST{ \withvalua{\hist} }{ 
  %  \text{if $\withvalua{\msg} \in \concof{\absmsg}$, then $\withvalua{\histcons{\hist}{\msg}} \in \concof{\abstrace}$} }
  \\ %\rule{0pt}{2ex}\\ % extra space hack: no idea why the line spacing in mathpar is being weird

  \fbox{%
  $\jhoare{\abspstate'}{\btrans}{\abspstate}$
  % \\
  % $\jtransok[\SpecSet]{\invpstate}{\btrans}$
  % \\
  % $\entail[\SpecSet]{\abspstate}{\abspstate'}$
  % \\
  % $\jhoare{\absappstate'}{\trans}{\absappstate}$
  % \\
  % $\jtransok{\invpstate}{\trans}$
  % \\
  % $\entail[]{\absappstate}{\absappstate'}$
  }
  % \\ %\rule{0pt}{2ex}\\ % extra space hack: no idea why the line spacing in mathpar is being weird

  \infer[a-callback-invoke]{
  }{
    \jhoare
      { \mkabspstate{ \abstracecons{\abstrace}{\msgcb{\msgid}{\symvalseq}} }{ \absappstore }{ \absmsgstack }{\quies}{ \pureandseq{\puredom}{ \pureseen{\symval} }} }
      { \cfedge{\quies}{ \msgcb{\msgid}{\varseq} }{\apploc} }
      { \mkabspstate
          { \abstrace }
    { \absappstoreAndSeq{\absappstore}{ \absappstorePtsto{\var}{\symval} } }
    { \absmsgstackcons{\absmsgstack}{\msgcb{\msgid}{\symvalseq}} }
    {\apploc}{\puredom}
      }
  }
  \\ %\rule{0pt}{2ex}\\ % extra space hack: no idea why the line spacing in mathpar is being weird
  \infer[a-callback-return]{
    %\symval',\symvalseq \notin \mkabsmem{ \abstrace }{\absappstore}{ \absmsgstack }{\puredom} 
    \absappstore = \absappstoreAndSeq{\absappstoreAnd{\absappstore'}{\absappstorePtsto{\var'}{\symval'}}}{ \absappstorePtsto{\var}{\symval} }
  }{
    \jhoare
      { \mkabspstate
          { \abstracecons{\abstrace}{ \msgcbret{\symval'}{\msgid}{\symvalseq} } }
    %{ \absappstoreAndSeq{ \absappstoreAnd{\absappstore}{ \absappstorePtsto{\var'}{\symval'} } }{ \absappstorePtsto{\var}{-}} }
    { \absappstore }
          { \absmsgstackcons{\absmsgstack}{\msgcb{\msgid}{\symvalseq}} }
          { \apploc }{\puredom}
      }
      { \cfedge{\apploc}{ \msgcbret{\var'}{\msgid}{\varseq} }{\quies} }
      { \mkabspstate{ \abstrace }{\absappstore}{ \absmsgstack }{\quies}{\puredom}  }
  }
  \\ %\rule{0pt}{2ex}\\ % extra space hack: no idea why the line spacing in mathpar is being weird

  \infer[a-callin-invoke]{
    \absappstore = \absappstoreAndSeq{ \absappstore' }{ \absappstorePtsto{\var}{\symval} }
  }{
    \jhoare
      { \mkabspstate
          { \abstracecons{\abstrace}{ \msgci{ \symval' }{\msgid}{\symvalseq} } }
    %{ \absappstoreAnd{\absappstore}{ \absappstorePtsto{\var'}{-} } }
    { \absappstore }
    { \absmsgstack }{ \apploc }{\pureand{\puredom}{\pureseen{\symval'}}}
      }
      { \cfedge{\apploc}{ \msgci{\var'}{\msgid}{\varseq} }{\apploc'} }
      { \mkabspstate
          {\abstrace}
          { \absappstoreAnd{\absappstore}{ \absappstorePtsto{\var'}{\symval'} } }
          {\absmsgstack}{\apploc'}{\puredom}
      }
  }

  \\
  \fbox{
    $\jtransok[\SpecSet]{\invpstate}{\btrans}$
  }

  \infer[a-boundary-step]{
    \entail{ \maplookup{\invpstate}{\postof{\btrans}} }{ \abspstate } \quad
    \jhoare{\abspstate'}{\btrans}{\abspstate} \quad
    \entail{ \abspstate' }{ \maplookup{\invpstate}{\preof{\btrans}} }
  }{
    \jtransok[\SpecSet]{\invpstate}{\btrans}
  }

  \\
  \fbox{
    $\jtransok{\invpstate}{\trans}$
  }

  \infer[a-app-step]{
    \entail[]{ \appof{\maplookup{\invpstate}{\postof{\trans}}} }{ \absappstate } \quad
    \jhoare{\absappstate'}{\trans}{\absappstate} \quad
    \entail[]{ \absappstate' }{ \appof{\maplookup{\invpstate}{\preof{\trans}}} }
  }{
    \jtransok{\invpstate}{\trans}
  }
  \\
  \fbox{%
  $\jmaywitness[\prog]{\invpstate}{\loc}{\abspstate}$
  \\
  $\jrefute[\prog]{\loc}{\abspstate}$
  % \\
  % $\notinit{ \abspstate }$
  }
  \\ %\rule{0pt}{2ex}\\ % extra space hack: no idea why the line spacing in mathpar is being weird

  \infer[a-inductive]{
    %\entail{ \mklocstate{\absmem}{\loc} }{ \maplookup{\invpstate}{\loc} } \\
    \entail{ \abspstate }{ \maplookup{\invpstate}{\locof{\abspstate}} } \\
    \jtransok[\SpecSet]{\invpstate}{\btrans}\;\;\text{for all $\btrans \in \prog$} \\
    \jtransok{\invpstate}{\trans}\;\;\text{for all $\trans \in \prog$}
  }{
    %\jmaywitness[\prog]{\invpstate}{\loc}{\mklocstate{\absmem}{\loc}}
    \jmaywitness[\prog]{\invpstate}{\loc}{\abspstate}
  }

  \infer[a-refute]{
    \jmaywitness[\prog]{\invpstate}{\loc}{\abspstate} \\
    \notinit{ \maplookup{\invpstate}{\quies} }
  }{
    \jrefute[\prog]{\loc}{\abspstate}
  }
\end{mathpar}
\caption{\label{fig:refute-with-mhpl}%
Refuting callback reachability with a MHPL. 
We abstract the application-only transition system with Hoare triples over app and boundary transitions and an abstract program state invariant $\invpstate$.
The location of an abstract state is noted with $\locof{\abspstate}$ and looking up the state at a location in the invariant is noted with $\invpstate(\loc)$.
Executing backwards, an abstract message history $\abstrace$ becomes conditional in messages observed in the future execution. An abstract realizable message history $\SpecSet$ is parameter and is the abstract analogue of the concrete set of realizable message histories $\wfhist$. 
% The $\SpecSet$ is what is specified by a \newls specification (as defined in \secref{newls}).
% We use the judgment $\jtransok[\SpecSet]{\invpstate}{\btrans}$ to indicate that the invariant map is consistent with a given transition via the  $\jhoare{\abspstate'}{\btrans}{\abspstate}$ judgment.
% Mirroring the boundary step we assume an app step exists, $\jtransok{\invpstate}{\trans}$.
% These semantics depend on excludes initial, $\notinit{ \abspstate }$, and entailment, $\entail[\SpecSet]{\abspstate}{\abspstate'}$, (defined in \secref{decision}).
}
\end{figure}

% \TODO{
% - We assume a data-relevance relation that defines when an app transition is relevant to the current heap state. That is, an app transition is data-relevant iff taking its pre on the current state can strengthen the heap state.
% - An interface transition is historically-relevant iff taking its pre on the current state can strengthen the history state.
% - A f: cbret m is jump-relevant iff (a) an data-relevant app transition is backwards-reachable from f in the app control-flow or (b) an interface transition is backwards-reachable from f in the app control-flow.
% }
% \TODO{
% - Separate compilation assumption via well-formed traces
% - Need to state refutation condition with non-initial message history.
% }

\subsubsection{An Abstract Semantics with Message Histories}

In \autoref{fig:refute-with-mhpl}, we define MHPL,
which abstracts the application-only transition system from \secref{apponly-transition-system},
to derive refutations with respect to message histories.
To abstract messages $\absmsg$, we replace concrete values with symbolic variables $\symval$.
Symbolic variables are existentially quantified across each part of the abstract program state with an assignment, $\valua$.
That is, a concrete state $\pstate$ satisfies the concretization relation of an abstract state, $\abspstate$ (i.e., $\pstate \models \abspstate$) if a $\valua$ exists such that each part of $\pstate$ satisfies the concretization relation with each part of $\abspstate$ (e.g., $\inconc{\withvalua{\hist}}{\abstrace}$).

%which we essentially leave unspecified, except showing top/true $\puretrue$, conjunction $\pureand{\puredom_1}{\puredom_2}$, and the $\pureseen{\symval}$ constraint that captures when a value $\val$ abstracted by $\symval$ has been seen by the framework --- and should satisfy $\jseen{\hist}{\val}$ for all possible message histories $\hist$ abstracted by the current state.
%Note that we include the $\pureseen{\symval}$ predicate simply as an example of the kind of realizability constraints we can capture by reifying message histories.
%%% Evan: [Commenting the last bit of the above b/c the judgment is only in the appendix.
%
% We write $\valua$ for an assignment from symbolic variables $\symval$ (i.e., existential variables) to concrete values $\val$, and thus pure constraints $\puredom$ concretize to assignments, $\valua$.
%

%Intuitively, an abstract message history $\abstrace$ captures the set of concrete message histories reaching a given program location $\loc$.
%An abstract message history can be $\abstracebase$, which concretizes to all \emph{realizable} message histories.
%Note that since we only care about realizable message histories, we do not include a $\top$ abstract message history, which would concretize to all message histories.
An abstract message history $\abstrace$ captures the set of message histories reaching a given program location $\loc$ under the realizable message history specification $\SpecSet$.
An abstract message history can be $\abstracebase$, which corresponds to all realizable message histories under $\SpecSet$.
Note that since we only care about \emph{realizable} message histories, we do not include a $\top$ abstract message history corresponding to \emph{all} message histories.
Since our logic explores backwards, it adds constraints on \emph{future} boundary transitions to the abstract message history $\abstrace$ as they are encountered.
% Now we want to go backwards in execution in our analysis, so we need to be able to constrain an abstract message history $\abstrace$ with a message from a boundary transition that we know must be executed next in the future.
The key is to see this constraint as an ordered linear implication on the right $\abstracecons{\abstrace}{\absmsg}$, which informally says, ``For all messages satisfying $\absmsg$, appending that message to the current message history implies that the new message history satisfies $\abstrace$.'' 
In the middle part of \autoref{fig:refute-with-mhpl}, we give a precise concretization relation between a message history with an assignment $\withvalua{\hist}$ and an abstract message history: $\withvalua{\hist} \models_{\SpecSet} \abstrace$.
%the concretization of abstract message histories $\concof{\abstrace}$ that concretizes to message history-assignment pairs $\withvalua{\hist}$.
%Returning to the top part of \autoref{fig:refute-with-mhpl}, we remark that the program logic is parametrized by a specification of realizable message histories $\SpecSet$ (as previously noted).

The rest of an abstract program state is straightforward.
We do not care specifically about the form of abstract app stores, except that like concrete app stores, we need a way to look up a (symbolic) value for a variable and to initialize variables.
To do that, we use intuitionistic separation logic~\cite{ishtiaq+2001:assertion-language,reynolds2002:separation-logic:} to indicate arbitrary store $\absappstoreTrue$, separating-conjunction of two stores $\absappstoreAnd{\absappstore_1}{\absappstore_2}$, a singleton points-to or cell for program variables $\absappstorePtsto{\var}{\symval}$, 
% a constraining a store with pure constraints $\pureand{\absappstore}{\puredom}$, 
an infeasible store $\absmemFalse$, or a disjunction of stores $\absmemOr{\absappstore_1}{\absappstore_2}$, which we usually consider in disjunctive normal form.
An abstract app state $\absappstate\bnfdef\mkabsappstate{\absappstore}{\apploc}{\puredom}\bnfalt\cdots$
is then an abstract app store and location $\apploc$.
An abstract memory $\absmem\bnfdef\mkabsmem{\abstrace}{\absappstore}{\absmsgstack}{\puredom}\bnfalt\cdots$ is then a product of an abstract message history $\abstrace$, an abstract boundary stack $\absmsgstack$, and an abstract store $\absappstore$ --- or a disjunction of such products.
% and pure constraints $\puredom$ (or a disjuction of such products).
An abstract program state $\abspstate\bnfdef\mklocstate{\absmem}{\loc}$ is simply an abstract memory $\absmem$ at a program location $\loc$.
%While we do not formalize it here, we consider a disjuctive program state in the implementation (i.e., with a set of such products).
% For abstract boundary stacks $\absmsgstack$, we consider arbitrary boundary stacks $\top$, appending abstract boundary stacks $\absmsgstackcons{\absmsgstack_1}{\absmsgstack_2}$ (i.e., with ``fuse'', an ordered conjunction), or an abstract boundary activation $\msgcb{\msgid}{\symvalseq}$.
For abstract boundary stacks $\absmsgstack$, we consider arbitrary boundary stacks  $\top$, or appending a boundary activation $\absmsgstackcons{\absmsgstack}{\msgcb{\msgid}{\symvalseq}}$.
Finally, we consider abstract program-state invariants $\invpstate$ to be a set of abstract program states $\abspstate$, which we also treat as a map from locations $\loc$ to the abstract state $\abspstate$ at that location 
(i.e., $\maplookup{\invpstate}{\loc} = \abspstate$ iff $\abspstate = \mklocstate{\absmem}{\loc}$ and $\abspstate \in \invpstate$).

We describe the abstract semantics of boundary transitions $\btrans$ as Hoare triples $\jhoare{\abspstate'}{\btrans}{\abspstate}$, except that we are interested in backwards over-approximating triples instead of forwards. That is, we read the judgment as, ``If there is an execution of the boundary transition $\btrans$ to a post-state satisfying $\abspstate$, the pre-state of that execution satisfies $\abspstate'$'' (\reftxt{Lemma}{lem:stepsound}).

\begin{lemma}[hoare triple soundness] 
If~$\jhoare{\abspstate'}{\btrans}{\abspstate}$ and $\jeval{\pstate'}{\btrans}{\pstate}$ such that $\pstate \models_{\SpecSet} \abspstate$ and $\wfhist \subseteq \wfhistspec$, then $\pstate' \models_{\SpecSet} \abspstate'$. 
%As a corollary, if $\jrefute[]{\prog}{\SpecSet}{\apploc}$ then for all $\pstate$ such that $ \vdash \jstepinsstar{\pstate}{\prog}{\pstate'}$ and $\locof{\pstate'} = \apploc$ implies $\pstate \neq \initpstate$.
\label{lem:stepsound}
\end{lemma}

%(i.e., the judgment should satisfy the following soundness condition: if $\jhoare{\abspstate'}{\btrans}{\abspstate}$ and $\jeval{\pstate'}{\btrans}{\pstate}$ such that $\inconc{\pstate}{\abspstate}$, then $\inconc{\pstate'}{\abspstate'}$, where we write $\inconc{\pstate}{\abspstate}$ for a program state $\pstate$ modeled by an abstract program state $\abspstate$ under $\SpecSet$).
%$\pstate \in \concof{\abspstate}$, then $\pstate' \in \concof{\abspstate'}$).

The abstract semantics rules for boundary transitions $\btrans$ follow closely their concrete counterparts (assuming a structural rule for disjunction of memories $\absmem$).
The \TirName{a-callback-invoke} rule captures computing the pre-condition of the callback invocation transition $\cfedge{\quies}{ \msgcb{\msgid}{\varseq} }{\apploc}$ and shows moving from an assertion on the abstract boundary stack $\absmsgstackcons{\absmsgstack}{\msgcb{\msgid}{\symvalseq}}$ to a hypothetical next message in the abstract message history $\abstracecons{\abstrace}{\msgcb{\msgid}{\symvalseq}}$.
In detail, it first asserts that the post-app memory has bindings for the callback parameters $\absappstoreAndSeq{\absappstore}{ \absappstorePtsto{\var}{\symval} }$ and has the corresponding callback activation on top of the boundary stack $\absmsgstackcons{\absmsgstack}{\msgcb{\msgid}{\symvalseq}}$.
Then, we drop the parameter bindings and pop the callback activation.
Finally, we update the abstract message history with the abstract message corresponding to the callback invocation, $\abstracecons{\abstrace}{\msgcb{\msgid}{\symvalseq}}$.
As an example from \secref{overview}, the abstract state $\mkshortabsmem{\ref{line:actnull}}{\abstracebase}{\codej{f.act} \ptsto \enkwNull \lsep \codej{this} \ptsto \codej{f}}$ just after the entry of \codej{call} produces the pre-state $\mkshortabsmem{\refstate{1}}{\ronepre \abstracebase}{\codej{f}.\codej{act} \ptsto \enkwNull}$ at the framework location which proceeds \codej{call}.

Continuing to mirror the abstract semantics, the \TirName{a-callback-return} rule pushes a hypothetical callback message on the boundary stack corresponding to the callback that would have just returned in the concrete execution.
Similar to abstract callback invoke (\TirName{a-callback-invoke}), abstract callback return (\TirName{a-callback-return}) and abstract callin invoke (\TirName{a-callin-invoke}) add hypotheticals to the abstract message history.
The main difference is how each updates the abstract app store.
For \TirName{a-callback-return}, the return value and its relationship to other symbolic variables is unknown, therefore %, a fresh symbolic variable is introduced that is pointed to by the return variable. For example, the post state
%$\mkshortabsmem{\refstate{1}}{\ronepre \abstracebase}{\codej{f}.\codej{act} \ptsto \enkwNull}$
% $\mkshortabsmem{\refstate{1}}{\ronepre \abstracebase}{\ldots}$
we ensure that the separation logic domain has materialized return values and arguments for the callback via $\absappstoreAndSeq{\absappstoreAnd{\absappstore'}{\absappstorePtsto{\var'}{\symval'}}}{ \absappstorePtsto{\var}{\symval} }$.
For example, the post state
%$\mkshortabsmem{\refstate{1}}{\ronepre \abstracebase}{\codej{f}.\codej{act} \ptsto \enkwNull}$
% $\mkshortabsmem{\refstate{1}}{\ronepre \abstracebase}{\ldots}$
from the running example is transferred over the return of the \codej{onDestroy} callback to produce the pre-state $\mkshortabsmem{\ref{line:destRet}}{\abstracecons{\ronepre\abstracebase}{\enkwCb\enkwRet~\codej{f}_2\codej{.onDestroy()}}}{\ldots}$.
Note that we elide the return value here, as \codej{onDestroy} is \codej{void}.
%Note 2: the abstract message $\enkwCb\enkwRet~\codej{f}_2\codej{.onDestroy()}$ can be soundly dropped as it does not appear in a history implication.
The value $\codej{f}_2$ may or may not alias \codej{f}, as there is a case split from the separation logic materialization (\autoref{fig:refgraph} from \secref{overview} shows just the aliased case).
%(Note 1: that we ignore the return value as \codej{onDestroy} is \codej{void}. Note 2: the abstract message $\enkwCb\enkwRet~\codej{f}_2\codej{.onDestroy()}$ can be soundly dropped as it does not appear in a history implication. Note 3: the value $\codej{f}_2$ may or may not be equivalen to \codej{f} but there is a case split caused by the materialized heap cells due to maintaining the separation logic constraints, the overview shows the aliased case).

Finally, \TirName{a-callin-invoke} removes a program variable from the post app store corresponding to the return value and introduces fresh symbolic variables to the pre-store bound to the arguments of the callin invoke.
For example, the post state $\mkshortabsmem{\ref{line:subscribe}}{\abstracecons{\abstracebase}{...}}{\codej{task} \ptsto \codej{t} \ast \codej{this}\ptsto\codej{f}}$ transferred over the \codej{create} call creates the pre-state $\mkshortabsmem{\ref{line:create}}{\abstracecons{\abstracecons{\abstracebase}{...}}{\enkwCi~\codej{t}=\codej{create(...)}}}{\codej{this}\ptsto\codej{f}}$.

\iffalse
Finally, for the callin invocation transition $\cfedge{\apploc}{ \msgci{\var'}{\msgid}{\varseq} }{\apploc'}$, the calling semantics in rule \TirName{a-callin-invoke} are as follows: the post-condition has the return value $\symval'$ of the call bound to $\var'$ in the app memory $\absappstoreAnd{\absappstore}{ \absappstorePtsto{\var'}{\symval'} }$,
%and then the pre-condition has an unknown, fresh symbolic value bound to $\var'$ (which we write as $\absappstorePtsto{\var'}{-}$),
while the app memory $\absappstore$ must have bindings for the callin call arguments $\absappstoreSeq{\absappstorePtsto{\var}{\symval}}$.
%For the separate compilation assumption, we assert in the pre-condition that the callin return value has been seen by the framework $\pureand{\puredom}{\pureseen{\symval'}}$. \TODO{$\pureseen{\symval}$ from the post state may be discharged in the pre state if it aliases an argument}
And again, to abstractly record the execution of this callin invocation transition, we assume the corresponding message on the right \smash{$\abstracecons{\abstrace}{ \msgci{ \symval' }{\msgid}{\symvalseq} }$}.
\fi

We write $\appof{\abspstate} = \absappstate$ for the projection of an abstract program state $\abspstate$ to an abstract app state $\absappstate$ that drops the abstract message history $\abstrace$ and abstract boundary stack $\absmsgstack$ components.
The app transitions are checked for being inductive in the analogous way with the \TirName{a-app-step} rule defining the judgment form $\jtransok{\invpstate}{\trans}$, which similarly depends on an abstract semantics for app transitions $\jhoare{\absappstate'}{\trans}{\absappstate}$ and an entailment judgment for abstract app states $\entail[]{\absappstate}{\absappstate'}$.
%
% We write $\feasible{ \mkabsmem{\abstrace}{\absappstore}{\absmsgstack}{\puredom} }$ for the judgment that states which such a program-state tuple is unsatisfiable (i.e., is bottom).

% Full definitions of these judgments may be found in \refappendix{extsemantics}.

%%% TODO: We should do this. Not hard, I think.
\iffalse
These abstract semantics are sound backwards over-approximations:
\begin{lemma}[Sound Abstract Semantics of Boundary Transitions]
  If $\jhoare{\abspstate'}{\btrans}{\abspstate}$ and $\jeval{\pstate'}{\btrans}{\pstate}$ such that $\inconc{\pstate}{\abspstate}$, then $\inconc{\pstate'}{\abspstate'}$.
\end{lemma}
\fi

We then check that the abstract program-state invariants $\invpstate$ are inductive for executing backwards a given boundary transition $\btrans$ with the judgment form $\jtransok[\SpecSet]{\invpstate}{\btrans}$. This judgment says, ``In abstract program-state invariants $\invpstate$, executing boundary transition $\btrans$ backwards is inductive when constrained by realizable message histories defined by specification $\SpecSet$.''
The \TirName{a-boundary-step} defines this judgment and captures the backwards over-approximation.
Specifically, it depends on an entailment judgment $\entail{ \abspstate }{ \abspstate' }$ that is parametrized by the realizable message histories specification~$\SpecSet$ (i.e., it should satisfy the following soundness condition: if $\entail{ \abspstate }{ \abspstate' }$ and $\inconc{\pstate}{\abspstate}$, then $\inconc{\pstate}{\abspstate'}$).
%
% Similar to the excludes-initial judgment, in later sections we focus on the entailment of message histories, $\entail{\abstrace}{\abstrace'}$ (Definition \ref{def:entail})
% \begin{definition}[Sound Entailment]
% 	For all $\hist$, if $\entail{\abstrace}{\abstrace'}$ and $\inconc{\withvalua{\hist}}{\abstrace}$ then $\inconc{\withvalua{\hist}'}{\abstrace'}$ for some $\valua'$.
% 	\label{def:entail}
% \end{definition}
%
To get the post- and pre-locations of a boundary transition $\btrans$, we write $\postof{\btrans}$ and $\preof{\btrans}$, respectively.
Then, the rule chooses some $\abspstate$ that over-approximates $\maplookup{\invpstate}{\postof{\btrans}}$ --- i.e., $\entail{ \maplookup{\invpstate}{\postof{\btrans}} }{ \abspstate }$, applies the abstract semantics for $\btrans$ --- i.e., $\jhoare{\abspstate'}{\btrans}{\abspstate}$, and checks that $\maplookup{\invpstate}{\preof{\btrans}}$ over-approximates $\abspstate'$ --- i.e., $\entail{ \abspstate' }{ \maplookup{\invpstate}{\preof{\btrans}} }$.
This rule is the backwards over-approximating version of the usual Hoare rule of consequence.
A similar judgment may be written for an app transition $\jtransok{\invpstate}{\trans}$ (e.g., as in \citet{DBLP:conf/pldi/BlackshearCS13}).
% The app transitions are checked for being inductive in the analogous way with the \TirName{a-app-step} rule defining the judgment form $\jtransok{\invpstate}{\trans}$, which similarly depends on an abstract semantics for app transitions $\jhoare{\absappstate'}{\trans}{\absappstate}$ and an entailment judgment for abstract app states $\entail[]{\absappstate}{\absappstate'}$.
%
% Note that for clarity, we write the abstract-semantic rules above with explicitly materialized points-to for all message arguments (e.g., \smash{$\absappstoreSeq{\absappstorePtsto{\var}{\symval}}$}) and return values.
For clarity, the semantics is written with explicitly materialized points-to for message arguments and return values (e.g., \smash{$\absappstoreSeq{\absappstorePtsto{\var}{\symval}}$}).
If such values do not otherwise constrain the abstract state, they may be summarized into the top store $\absappstoreTrue$ without precision loss.

\begin{lemma}[boundary-step soundness]
If $\jtransok[\SpecSet]{\invpstate}{\btrans}$ and $\jeval{\pstate'}{\btrans}{\pstate}$ such that $\pstate \models_{\SpecSet} \maplookup{\invpstate}{\postof{\btrans}}$ and $\wfhist \subseteq \wfhistspec$, then $\pstate' \models_{\SpecSet} \maplookup{\invpstate}{\preof{\btrans}}$.
\label{lem:stepsound2}
\end{lemma}

To describe an inductive program invariant, we define the \emph{may-witness} judgment form $\jmaywitness[\prog]{\invpstate}{\loc}{\abspstate}$ that says, ``Abstract program-state invariants \smash{$\invpstate$} is inductive executing backwards from abstract program state $\abspstate$ --- at location $\locof{\abspstate}$ --- in program $\prog$.''
The \TirName{a-inductive} rule that defines this judgment form simply checks that each boundary transition $\btrans$ and each app transition $\trans$ in program $\prog$ are inductive.

\begin{lemma}[inductive soundness]
If~$\jmaywitness[\prog]{\invpstate}{\loc}{\abspstate}$
and $\jstepstar[p]{\pstate'}{\pstate}$
such that $\pstate \models_{\SpecSet} \abspstate$
and $\wfhist \subseteq \wfhistspec$,
then  $\pstate' \models_{\SpecSet} \maplookup{\invpstate}{\locof{\pstate'}}$.
% If~$\jmaywitness[\prog]{\invpstate}{\loc}{\abspstate}$ and $\pstate \models_{\SpecSet} \abspstate$ and $\jstepstar[p]{\pstate'}{\pstate}$ then there exists a $\abspstate'$ such that $\pstate' \models \maplookup{\invpstate}{\locof{\abspstate'}}$.
\label{lem:inductivesound}
\end{lemma}

Finally, to derive a refutation of reachability $\jrefute[\prog]{\loc}{\abspstate}$, the \TirName{a-refute} rule says that we derive an inductive program invariant \smash{$\invpstate$} from \smash{$\abspstate$} --- i.e., \smash{$\jmaywitness[\prog]{\invpstate}{\loc}{\abspstate}$}, and we derive that the program invariant at the entry location $\quies$ excludes the initial (concrete) program state --- i.e., \smash{$\notinit{ \maplookup{\invpstate}{\quies} }$}.

\newsavebox{\InitStateBox}  %% hack to make "init" not italics.
\savebox{\InitStateBox}{
$\initpstate$
}
\begin{theorem}[refute soundness]
If~$\jrefute[\prog]{\loc}{\abspstate}$ and $\jstepstar[p]{\pstate'}{\pstate}$ such that $\pstate \models_{\SpecSet} \abspstate$ and $\wfhist \subseteq \wfhistspec$, then $\pstate' \not = \usebox{\InitStateBox}$.
\label{thm:refutesound}
\end{theorem}

\subsubsection{Abstract Interpretation with Message Histories}

While we have described a checking system with the may-witness judgment form \smash{$\jmaywitness[\prog]{\invpstate}{\loc}{\abspstate}$}, we can consider a direct approach to computing an inductive program invariant \smash{$\invpstate$} from an error condition \smash{$\abspstate$} via a backwards abstract interpretation.
The invariant map $\invpstate$ is initialized with the error condition just before the assertion ($\mkabsmem{\abstracebase}{\absappstore}{\top}{\puredom}$ where $\absappstore$ negates the assertion condition) and $\bot$ at other locations.
We then proceed with a standard worklist algorithm. 
When the invariant map is updated at a location, all transitions to that location are added to the worklist.
Each transition in the worklist and abstract state at the post-location are processed by a transfer function (based on the Hoare triples defined above) producing a pre-condition that is joined into the invariant map.
Pre-conditions at a location are eagerly merged with existing disjuncts both to avoid an updated state at a location and for efficiency.
Merging is done automatically via the entailment check, $\entail{ \mkabsmem{\abstrace}{\absappstore}{\absmsgstack}{\puredom} }{ \mkabsmem{\abstrace'}{\absappstore'}{\absmsgstack'}{\puredom'} }$.
If a new pre-condition cannot be merged with an existing disjunct, it is added to the existing disjunctions.
At the framework location $\enkwFwk$, all callback return boundary transitions, $\cfedge{\apploc}{ \msgcbret{\var'}{\msgid}{\varseq} }{\quies}$, are added to the worklist.
We alarm if we cannot prove that the invariant at $\enkwFwk$ excludes initial.
%at any point.
% We eagerly check excludes initial for each update to the framework location and alarm if it can't be proven.
If a fixed point is reached that excludes the initial state, $\notinit{ \mkabsmem{\abstrace}{\absappstore}{\absmsgstack}{\puredom} }$, then we have refuted the reachability of the assertion failure.
Intuitively, we have now captured the abstract state at all locations that may step to the assertion failure, and excludes initial is proving that no message history can go from the initial state of the program to the assertion and fail.
% until reaching the framework location, $\quies$.

Excludes-initial, $\notinit{ \abstrace}$, and entailment of abstract message histories, $\entail{\abstrace}{\abstrace}$ are automated via SMT (and described in \autoref{sec:decision}). Existing techniques can combine the message history SMT encoding with other parts of the abstract state (e.g., using \citet{DBLP:conf/cav/PiskacWZ13} for separation logic).

%% file: cbcftl.tex
\section{Callback Control-Flow Temporal Logic (\newls)}
\label{sec:newls}

% In section \ref{sec:overview-cbcftl}, we showed how \newls can be used to write a specification, $\SpecSet$, that targets specific callbacks critical to proving an assertion safe.
% As we see in \autoref{sec:overview}, distinguishing between a buggy and fixed app involve reasoning about when a callback cannot happen, that is, the callback control flow that the framework imposes on the app.
% Without further restriction, the application-only transition system from \autoref{sec:apponly-transition-system} essentially says that every callback can happen at any time.
% The application-only transition system and the message history program logic from \autoref{sec:applog} are, however, parametrized by a notion of realizable message histories that restricts the callback control flow. 

% In \autoref{sec:overview}, we showed how \newls can target messages (such as the \codej{call} callback) restricting the realizable message histories and \autoref{sec:applog} showed a program logic that can utilize such restrictions to prove an assertion.
%Here, we focus on how \newls is designed to balance the precision needed for expressing facts about the framework in a specification and the need to automate  

In this section, we describe the Callback Control-Flow Temporal Logic (\newls) that we use to express the specification $\SpecSet$ of the realizable message histories.
We design \newls as a compromise between the expressiveness required to specify callback control flow and the need of the abstract interpretation to automate
judging
excludes-initial ($\notinit{\abstrace}$)
and
entailment ($\entail{\abstrace}{\abstrace'}$), which are parametric in the specification language used to express $\SpecSet$.
% The design of \newls and is a compromise between the expressiveness of the specification logic and the need to automate
%Precision refers to whether facts like History Implication \ref{spec:call} can be expressed. \TODO{(skm) Check the language in the intro after Sergio rewrite and mirror it here.}
% The judgments excludes-initial and entailment need to be automated for abstract interpretation.
% and performant

\JEDI{Why not FO-LTL?}
As we observe in \autoref{sec:overview}, a specification of realizable message histories must be able to express:
\begin{inparaenum}
  \item quantification over message values (e.g., the subscription object \codej{s} from History Implication \ref{spec:call}); and
  \item constraints on what messages must have or have not happened in the past (e.g., \codej{subscribe} or \codej{unsubscribe}).
\end{inparaenum}
These requirements suggest \newls should be a linear temporal logic (LTL)~\cite{DBLP:books/daglib/0077033} interpreted over finite sequences (i.e., message histories)~\cite{DBLP:conf/ijcai/GiacomoV13}, with \emph{first-order} quantification of message arguments, and \emph{past-time} temporal operators~\cite{DBLP:conf/lop/LichtensteinPZ85}.
In principle, the excludes-inital and entailment judgment could be reduced to checking the satisfiability of first-order LTL (FO-LTL) formulas, but it is undecidable~\cite{DBLP:journals/iandc/SongW16} and is limited in ready-to-use implementations.

% Instead, we restrict specifications so that given a target message, reasoning about the histories leading to that message with temporal formula is decidable.
Instead, we restrict the \newls{} syntax such that reasoning about message histories leading to a target message is decidable (i.e., with history implications consisting of a target message and temporal formula). In \autoref{sec:decision}, we show that such a problem can be in turn reduced to the satisfiability of the fragment of \emph{temporal formulas} of \newls{}.
%($\directive$ in \autoref{fig:cbcfsemantics}).
%
We show how to use such a subproblem to decide the excludes-initial ($\notinit{\abstrace}$) judgment and to obtain a semi-algorithm for judging entailment ($\entail{\abstrace}{\abstrace'}$).
The syntactic restriction of \newls carefully controls the use of features of the logic, such as negation and quantifier alternation, that complicate automated reasoning.
In particular, the restrictions are such that we can encode a temporal formula $\directive$ of \newls{} in an equisatisfiable formula in the Extended Effectively Propositional (Extended EPR) logic~\cite{DBLP:conf/frocos/Korovin13,DBLP:journals/pacmpl/PadonLSS17}.
This section first gives the syntax and semantics of \newls and then explains the encoding of the temporal formula fragment in Extended EPR.

%%%%%%%%%%%%%%%%%%%%%%%%%%%%%%%%%%%%%%%%%%%%%%%%%%%%%%%%%%%%%%%%%%%%%%%%%%%%%%%%%%%%%%%%%%%%%%%%%%%%%%%%%%%%%%%%
%\paragraph{A Temporal Logic for Expressing Realizable Message Histories}
\subsection{A Temporal Logic for Expressing Realizable Message Histories}

\input{cbcftlsyntax}

\autoref{fig:cbcfsemantics} describes the \newls syntax and semantics. 
A \newls specification $\SpecSet$ is a conjunction of \emph{history implications} $\ltrule \bnfdef \specOnly{\absmsg}{\directive}$.
Each history implication targets an abstract message, $\absmsg$, controlled by the framework (e.g., invocation of the \codej{call} callback) and a temporal formula, $\directive$, that must hold before the framework outputs that message.
%Each \emph{History Implication}, $\ltrule$, consists of a temporal operator, $\absmsg \boxright \directive$, involving a target abstract message $\absmsg$ and a \emph{temporal formula}, $\directive$.
%
%A history implication implicitly quantifies free variables in the target message, $\absmsg$, universally.
While not captured in the syntax, history implications $\ltrule$ are closed formulas where the variables of the abstract message in the antecedent are implicitly universally quantified, and
we assume that the temporal formula $\directive$ in the consequent is quantified so that
its free variables are a subset of the variables of $\absmsg$ (i.e.,
$\fv{\directive} \subseteq \fv{\absmsg}$ where $\fv{\cdot}$ yields the set of free variables of a formula).
\Secref{decision} will explain how abstract message histories combine with history implications leaving only temporal formula, which motivates this design.

% One of our contributions is to algorithmically reason about the combination of MHPL assertions and CBCFTL specifications. \TODO{redundant}
% History implications enable iteratively hypothesizing the appending of the next message $\absmsg$ to ``instantiate,'' which is made more evident by the form of abstract message histories $\abstracecons{\abstracebase}{\abstrace}$ and history implications $\specOnly{\absmsg}{\directive}$ sharing $\absmsg$ in both antecedents.

Temporal formulas $\directive$ include restricted versions of
standard past-time temporal operators ($\iDir{\ltmsg}$ for \textbf{O}nce, $\notiDir{\ltmsg}$ for \textbf{H}istorically \textbf{N}ot, and $\niDir{\ltmsg_2}{\ltmsg_1}$ for \textbf{N}ot \textbf{S}ince),
equality and disequality between variables, and
positive Boolean combinations.
In particular, the temporal operators apply only to individual symbolic messages $\ltmsg$
and do not allow for explicit negations
(although some negations are implicit in the history implication $\boxright$ and in the temporal operators $\mathbf{HN}$ and $\mathbf{NS}$).
Symbolic messages $\ltmsg$ are essentially abstract messages $\absmsg$, except we allow for a local existential quantification of variables $\exists\symval.\ltmsg$, which is convenient for ``don't care'' arguments (e.g., the \codej{_} in \refhistimp{spec:call}).
%is either a \emph{past} temporal operator ($\iDir{}$ - \emph{Once}, $\notiDir{}$ - \emph{Historically Not}, and $\niDir{}{}$ - \emph{Not Since}), a constant ($\text{true}$ or $\text{false}$), an equality between variables, or a positive Boolean combination of formula. The logic does not allow for explicit negations (although some negations are implicit in the $\boxright$ operator and in the temporal operators $\notiDir{}$ and $\niDir{}{}$).
%
The structure of \newls specifications $\SpecSet$ also limits the nesting of the temporal operators: past temporal operators are always nested in a future temporal operator ($\boxright$), and the only future temporal operator is history implication $\specOnly{\absmsg}{\directive}$.
While not explicitly shown in the syntax, we also restrict quantifiers such that $\forall\symval.\directive$ may only contain conjunctions and disjunctions of $\notiDir{\ltmsg}$ and equality/disequality of symbolic variables.
% While not explicitly shown in the syntax, we also tacitly assume a syntactic well-formedness constraint on the use of quantifiers: the subformula $\directive$ of universal quantification $\forall\symval.\directive$ is restricted to $\notiDir{\ltmsg}$ or the propositional forms to limit quantifier alternation.

The most interesting part of temporal formulas $\directive$ are the temporal operators:
$\notiDir{\ltmsg}$ states that $\ltmsg$ has historically not (i.e., has never) occurred in the past,
$\iDir{\ltmsg}$ that $\ltmsg$ occurred at least once in the past,
and $\niDir{\ltmsg_2}{\ltmsg_1}$ that $\ltmsg_2$ has not occurred since $\ltmsg_1$ occurred.
The operators restrict
 standard past-time temporal operators so that for a given message $\ltmsg$, we either positively look back for the time $\ltmsg$ occurs or negatively rule out $\ltmsg$ at each time in the past.
Thus, the $\iDir{\ltmsg}$ operator is directly the $\mathbf{O}$nce operator from past-time LTL, while
$\notiDir{\ltmsg}$ and $\niDir{\ltmsg_2}{\ltmsg_1}$ are syntactic restrictions for the appropriate negations within
$\mathbf{H}$istorically and $\mathbf{S}$ince
(i.e., $\notiDir{\ltmsg} \defeq \mathbf{H}~\neg\ltmsg$ and 
$\niDir{\ltmsg_2}{\ltmsg_1} \defeq \neg\ltmsg_2~\mathbf{S}~\ltmsg_1$ in past-time LTL).
Additionally, we allow standard boolean combinations of operators as well as equality of variables.

A model of a specification $\SpecSet$ is a (concrete) message history $\hist$, which is a finite sequence of messages $\msg$.
Message histories $\hist$ are zero-indexed by positions $i \in [0,\histlen{\hist})$, and we write $\hist[i]$ for the message at position $i$ in $\hist$ and $\histlen{\hist}$ for the length of $\hist$.
A message history satisfies a history implication $\semincluded{\hist}{\specOnly{\absmsg}{\directive}}$ iff for all positions $i$ in the message history (i.e., $i \in [0,\histlen{\hist}]$), if a concrete message $\msg$ with an assignment for its variables $\valua$ models $\absmsg$ and is $\hist[i]$, then the prefix of $\hist$ up to $i - 1$ must satisfy the temporal formula $\directive$ (under the assignment $\valua$).
A model for a temporal formula $\directive$ is a tuple $\modeltriple{\omega}{\assignment}{i}$ of a message history, an assignment, and a position in $\hist$.
The past-time temporal operators apply to the prefix of the message history $\hist$ up to (and including) position $i$, and the relation is undefined for any position outside the valid range of indices of the message history $\hist$ (e.g., $-1$).
%
% Note needed.
%As a \newls specification $\SpecSet$ is a conjunction of history implications, a message history satisfies a \newls specification $\semincluded{\hist}{\SpecSet}$ iff $\hist$ satisfies each history implication.

The temporal operators captures the looking back ``positively'' for a message (e.g., for $\iDir{\ltmsg}$, there must exist a $j \in [0,i]$ where the message at $j$ models $\ltmsg$ --- i.e., $\modeltriple{\hist}{\assignment}{j} \models \ltmsg$) or ``negatively'' when ruling out one (i.e., for $\notiDir{\ltmsg}$, all the messages at $k \in [0,i]$ must not model $\ltmsg$).
The temporal operator $\niDir{\ltmsg_2}{\ltmsg_1}$ is a more general version combining ``positively'' looking for $\ltmsg_1$ (i.e., $\exists j \in [0,i].\;\modeltriple{\hist}{\assignment}{j} \models \ltmsg_1$) while ``negatively'' ruling out $\ltmsg_2$ (i.e., $\forall k \in (j,i].\;\modeltriple{\hist}{\assignment}{k} \not \models \ltmsg_2$).
All three temporal formula reference the judgment $\modeltriple{\hist}{i}{\msg}\models \ltmsg$ to match a symbolic message to a position $i$ in a message history $\hist$ under a variable assignment $\valua$.

%%%%%%%%%%%%%%%%%%%%%%%%%%%%%%%%%%%%%%%%%%%%%%%%%%%%%%%%%%%%%%%%%%%%%%%%%%%%%%%%%%%%%%%%%%%%%%%%%%%%%%%%%%%%%
%\paragraph{Encoding Temporal Formula Into Extended EPR}
\subsection{Encoding Temporal Formula Into Extended EPR}
\label{sec:decidable}
Here, we describe how we encode temporal formula into Extended EPR~\cite{DBLP:conf/frocos/Korovin13,DBLP:journals/pacmpl/PadonLSS17}, a decidable fragment of first-order logic.
In brief, effectively propositional (EPR) is a first-order logic fragment where closed formulas converted into prenex normal form have the quantifier prefix $\exists\forall$ without any function symbols.
Extended EPR adds function symbols as long as the quantifier alternation graph does not contain cycles.
The quantifier alternation graph is a directed graph where the nodes are sorts and the edges are defined by functions (or $\forall x.\exists y.\;\ldots$) from the sort of the argument to the sort of the value.

\JEDI{challenge here is representing message histories in uninterpreted domain}
To encode a temporal formula $\directive$, we model message histories $\hist$ with uninterpreted functions over uninterpreted sorts.
We use an uninterpreted function $\tracefn\colon \unatset \rightarrow \msgset$ from history indices $\unatset$ to message instances $\msgset$.
To capture message instances, we use a function $\namefn\colon \msgset \rightarrow \nameset$ from message instances to message names $\nameset$ (i.e., representing the message kind, like $\enkwCb$, and the method name)
and a function $\argfn\colon \msgset \rightarrow \argset \rightarrow \valset$ from messages instances to arguments indices $\argset$ to values $\addrset$ (i.e., representing the arguments of the message instance).
%The challenge of modeling message histories in the uninterpreted domain is the utilization of values from interpreted domains such as natural numbers for the indices of a message history.
%%
%%
% We show axioms over 5 uninterpreted sorts and 4 uninterpreted functions to constrain the first order logic satisfiability.
% The sorts we use are:
% \begin{inparaenum}
% 	\item \unatset~for indices,
% 	\item \msgset~for messages,
% 	\item \addrset~for values, 
%   \item \argset~for argument indices, and
%   \item \nameset~for combinations of message types and interface names.
% \end{inparaenum}
% %
% Additionally, we use three uninterpreted functions to represent
% concrete traces.
% \begin{inparaenum}
% \item Less than or equals: $\leq: \unatset \rightarrow \unatset \rightarrow \boolset$.
% \item Histories: $\tracefn : \unatset \rightarrow \msgset$.
% \item Names: $\namefn : \msgset \rightarrow \nameset$.
% \item Arguments: $\argfn : \msgset \rightarrow \argset \rightarrow \valset$.
% \end{inparaenum}

\JEDI{indices and arguments may be axiomatized}
% We introduce axioms that make our encoding precise in the representation of message histories.
Then to describe ordering constraints on messages in a message history, we use a set of ordering axioms (referred to as $\histAxiomsFol$). We use an uninterpreted function $\leq: \unatset \rightarrow \unatset \rightarrow \boolset$ and axiomatize a total ordering on $\unatset$ (like \citet{DBLP:journals/pacmpl/PadonLSS17}), as well as an axiom for zero (i.e., $\forall \foli \in \unatset.\; 0 \leq \foli$ where $0$ is a variable).
Argument indices \argset{} are finite and bounded to the largest arity found in the framework methods.
% detail, removed for clarity and space
%(plus a variable for the return value position as an argument index)
Message names $\nameset$ are also finite and bounded by the framework interface definition.
As such, we precisely represent the needed ordering constraints on messages in message histories.

%\JEDI{unique message per index, unique name for msg, value for each parameter}
%The rest of the axioms to construct a message history are enforced by the definition of functions.  For each index, we have a message.  All messages have a type and name.  Each parameter of a message has an argument (note that the arity of each message associated with a name is defined by the interface).

\iffalse
We use \msgAtFolRel{\foli}{\absmsg}{\fol} to represent the constraint that a given abstract message is at an index in the message history. 
Indices ( by $\foli$, $\folj$, and $\folk$) in our first-order logic encoding are encoded as a total order over an uninterpreted sort, $\unatset$ (shown to be in extended EPR by \citet{DBLP:journals/pacmpl/PadonLSS17}).
This total order gives us a less than or equal $\foli \leq \folj$, less than $\foli < \folj$, and equality, $\foli = \folj$, relations over indices.
Additionally, we add a constraint giving us a zero: $\forall \foli \in \unatset.~ 0 \leq \foli$.
The following constraints can then be applied to model the indices without introducing edges in the quantifier alternation graph where $\leq$ is an uninterpreted function $\unatset \rightarrow \unatset \rightarrow \boolset$.
\begin{enumerate}
\item $\forall \foli \in \unatset.~ \foli \leq \foli$
\item $\forall i,j,k \in \unatset.~ i \leq j \wedge j \leq k \implies i \leq k$
\item $\forall i,j \in \unatset.~ i \leq j \wedge j \leq i \implies i = j$
\item $\forall i,j \in \unatset.~ i \leq j \vee j \leq i$
\item $\forall i \in \unatset.~ 0 \leq i$
\end{enumerate}
\fi

Given the above, the encoding of temporal formula $\directive$ is now direct.
We can encode an abstract message $\absmsg$ (i.e., an unquantified symbolic message $\ltmsg$) at an index $\foli \in \unatset$ using the $\tracefn$, $\namefn$, and $\argfn$ functions.
To be able to encode the length of a message history, we introduce a distinguished variable $\folLen$.
Then, we can encode the past-time temporal operators ($\iDir{\ltmsg}$, $\notiDir{\ltmsg}$, and $\niDir{\ltmsg_2}{\ltmsg_1}$) using the encoding of an abstract message at an index, $0$, $\leq$, and $\folLen$.
Here, we leverage the restriction that the temporal operators apply only to individual symbolic messages $\ltmsg$.
With respect to Extended EPR, it is clear that the quantifier alternation graph from the function symbols is acyclic.
And then, the encoding of temporal formula $\directive$ described above stays in Extended EPR because of the careful control of negation to prevent introducing any $\forall\exists$ edges.

%% file: cbcftlsyntax.tex
\begin{figure}[b]\small%\setstretch{1.5}
 % \begin{subfigure}[]{\textwidth}
  % CBCFTL - syntax
  \begin{mathpar}
    %\text{logic variable} \quad \lvar{x} \in \ldomain
    %\text{specification}\quad \SpecSet \bnfdef \text{true} \bnfalt \ltrule \land \SpecSet
    \text{\newls{} specification}\quad \SpecSet \bnfdef \text{true} \bnfalt \SpecSet_1 \land \SpecSet_2 \bnfalt \ltrule

    \text{history implication}\quad \ltrule \bnfdef \specOnly{\absmsg}{\directive}%\quad\text{(with $\fv{\directive} \subseteq \fv{\absmsg}$)}
    % \begin{array}{rrcll}
    % % Forall exists
    % \text{history implication}  & \ltrule        & \bnfdef & \specOnly{\absmsg}{\directive}
    %   & \text{ with } \fv{\directive} \subseteq \fv{\absmsg} \text{ and } \wellformed{\directive}\\
    % \end{array}
    % %\text{specification} \quad s \bnfdef  \forall\ \lvars{x_1}.\ (\specOnly{\absmsg}{\exists\ \lvars{x_2}.\directive})\\

    \begin{array}[t]{@{}rr@{\;}r@{\;}l@{}}
    \text{temporal formula} & \directive & \bnfdef &
      \iDir{\ltmsg} \bnfalt \notiDir{\ltmsg}
      \bnfalt \niDir{\ltmsg_2}{\ltmsg_1}
      \bnfalt \exists \symval.\directive
      \bnfalt \forall \symval.\directive
      \bnfalt \directive_1 \land \directive_2
      \bnfalt \directive_1 \lor \directive_2
      \bnfalt \symval_1 = \symval_2
      \bnfalt \symval_1 \neq \symval_2
      % \bnfalt \text{true}
      % \bnfalt \text{false}
    \end{array} 

    \text{symbolic messages}\quad\ltmsg \bnfdef \absmsg \bnfalt \exists\symval.\ltmsg
    % \begin{array}{rrcll}
    % \text{weak abstract messages} & \ltmsg &\bnfdef & \exists \symvalseq. \absmsg\\
    % \end{array}
    \\

   % 
  % \end{mathpar}
  % \caption{
  % \newls syntax. 
  % A \newls formula $\SpecSet$ is a conjunction of specifications, and each specification is the \emph{History Implication} temporal operator ($\boxright$). Universal and existential quantifiers binds the logic variables appearing in $\absmsg$ and the consequent $\directive$. The consequent $\directive$ contains past time temporal operators ($\iDir{}$ - \emph{Once}, $\notiDir{}$ - \emph{Historically Not}, and $\niDir{}{}$ - \emph{Not Since}). 
  % Weak abstract messages, \ltmsg, allow for arguments to remain unspecified.  
  % For example, $\enkwCi~\codej{s:=_.subscribe(f)}$ is short-hand for $\exists \codej{t}. \enkwCi~\codej{s:=t.subscribe(f)}$.
  % The well formed constraint, $\wellformed{\directive}$, prevents quantifier alternation to maintain decidability.
  % }
  % \label{fig:cbcftlsyntax}
  % \end{subfigure}
  %
  % CBCFTL - semantic
  %\begin{subfigure}[]{\textwidth}
  % \begin{mathpar}
    % \text{message at $i$-th} \quad \hist[i]
%
    % \text{message length} \quad \histlen{\histemp} := 0 \quad \histlen{\histcons{\hist'}{\msg}} := \histlen{\hist'} + 1
%
    %\text{assignment} \quad \assignment \bnfdef \{\} \bnfalt \assignment[x \mapsto v]
%
    % Note: \msg.\valua |= \absmsg is defined by conc and does the same thing as \valua(\absmsg)
    %\text{substitution} \quad \assignment(\absmsg) \bnfdef \absmsg[x \mapsto \assignment[x]]
  % \end{mathpar}
  %
  % Semantic - \hist \models \SpecSet
  %\begin{mathpar}
    \boxed{\hist \models \SpecSet} \quad
    \hist \models \specOnly{\absmsg}{\directive} \quad \text{iff} \quad
      \text{$\semincluded{\withvalua{\msg}}{\absmsg}$ such that $\hist[i] = \msg$
      implies $\modeltriple{\hist}{\valua}{i-1} \models \directive$}\\
    \boxed{\modeltriple{\hist}{\assignment}{i} \models \directive}\quad

    \modeltriple{\hist}{\assignment}{i} \models \niDir{\ltmsg_2}{\ltmsg_1}  \quad \text{iff} \quad
      \exists j \in [0,i].\;\modeltriple{\hist}{\assignment}{j} \models \ltmsg_1 \text{ and }
      \forall k \in (j,i].\;\modeltriple{\hist}{\assignment}{k} \not \models \ltmsg_2
    \\
    \modeltriple{\hist}{\assignment}{i} \models \iDir{\ltmsg}  \quad \text{iff} \quad  
      \exists j \in [0,i].\;\modeltriple{\hist}{\assignment}{j} \models \ltmsg

    % HN - Historically not
    \modeltriple{\hist}{\assignment}{i} \models \notiDir{\ltmsg}  \quad \text{iff} \quad 
      \forall k \in [0,i].\;\modeltriple{\hist}{\assignment}{k} \not \models \ltmsg

    \\
    \boxed{\modeltriple{\hist}{\assignment}{i} \models \ltmsg}
    \\

    % weak abstract messages
    % \modeltriple{\hist}{\assignment}{i} \models \exists \symvalseq. \absmsg \quad \textbf{iff} \quad \exists \valseq.  \semincluded{\withvalua[ \valuaext{\valua}{\symvalseq}{\valseq} ]{\hist[i]}}{\absmsg} %\valua[\symvalseq \mapsto \valseq].\hist[i] \models \absmsg
    \modeltriple{\hist}{\assignment}{i} \models \absmsg \quad \text{iff} \quad
    \text{$\semincluded{\withvalua{\msg}}{\absmsg}$ and $\hist[i] = \msg$}
  \qquad
    \modeltriple{\hist}{\assignment}{i} \models \exists \symval. \ltmsg \quad \text{iff} \quad
      \exists\val.\;\semincluded{ \modeltriple{\hist}{\valuaext{\valua}{\symval}{\val}}{i} }{ \ltmsg }

  \end{mathpar}

\caption{\label{fig:cbcftlsyntax}\label{fig:cbcfsemantics}\label{fig:cftl}%
Syntax and semantics of callback control-flow temporal logic (\newls). \newls is a subset of first-order linear temporal logic (FO-LTL)
%interpreted over finite traces
that describes a set of realizable message histories.
A \newls specification $\SpecSet$ is a conjunction of \emph{history implications} $\specOnly{\absmsg}{\directive}$.
% that each state a constraint on the \emph{past} message history with a temporal formula $\directive$ from an occurrence of a message satisfying $\absmsg$.
Temporal formulas $\directive$ include restricted versions of standard past-time temporal operators that apply only to individual symbolic messages $\ltmsg$
 with limited negation
% Yes, but not in the figure
% and universal quantification over time:
$\mathbf{O}$ (\textbf{O}nce), $\mathbf{HN}$ (\textbf{H}istorically \textbf{N}ot), and $\mathbf{NS}$ (\textbf{N}ot \textbf{S}ince).
While not shown in the syntax here, the subformula $\directive$ of universal quantification $\forall\symval.\directive$ is further restricted to $\notiDir{\ltmsg}$ or the propositional forms to limit quantifier alternation.
%The judgment $\modeltriple{\hist}{\assignment}{i} \models \ltmsg$ says that a message matching $\ltmsg$ exists at index $i$.
}
\end{figure}

\iffalse %moved syntactic restriction to previous figure TODO: Double check all this content has been merged elsewhere
\begin{figure}\small{\setstretch{1.2}
% CBCFTL - syntactic restrictions
%
\begin{mathpar}
\begin{array}{rrcll}
% Forall exists
\text{specification}  & s        & \bnfdef & \forall\ \lvars{x_1}.\ (\specOnly{\absmsg}{\exists\ \lvars{x_2}.\phi})
& \text{ with } \fv{\exists\ \lvars{x_2}.\phi} \subseteq \fv{\absmsg}  \\
% Only forall
                      &          & \bnfalt & \forall\ \lvars{x_1}.\ (\specOnly{\absmsg}{\phineg})
&  \\
%
\end{array}\\
\end{mathpar}
}
\caption{\TODO{move this information elsewhere as the spec definitions are being moved earlier}Restriction on quantifier alternation for the specification $s$.
%
We allow quantifier alternation in $s$ only if all the free variables of $\exists\ \lvars{x_2}.\phi$ (these variable must be universally quantified in $\forall \lvars{x_1}$ because $s$ is a closed formula) are also in the abstract message $\absmsg$.
%
Otherwise, we do not allow any existentially quantified variables and we prohibit the consequent $\phineg$ to contain the temporal operators \emph{Once} and \emph{Not Since}.
%
%
$\fv{\psi}$ is the set of free variables of the formula $\psi$.
%
}
%
\label{fig:cftlrestrictions}
\end{figure}
\fi

%% file: mhpl-cbcftl-together.tex
\section{Combining Abstract Message Histories with Callback Control-Flow}
%\label{sec:mhpl-cbcftl-together}
\label{sec:decision} %% caution: confusing stale label name

MHPL from \autoref{sec:applog} depends on two judgments, excludes initial $\notinit{\abstrace}$ and entailment $\entail{\abstrace}{\abstrace'}$.
Excludes initial says that abstract message history, $\abstrace$, excludes the initial, empty message history.
Message history $\abstrace$ entailing a second message history $\abstrace'$ says that all concrete traces abstracted by $\abstrace$ are also abstracted by $\abstrace'$.
Both of these judgments depend on the \newls specification from \autoref{sec:newls} for a definition of realizable message histories.
For each abstract message history, we \emph{combine} with the \newls specification in order to avoid reasoning about the specification separately.
In this section, we first show how to combine an abstract message history, $\abstrace$, with a specification, $\SpecSet$, resulting in a single temporal formula, $\directive$ (as we describe in \autoref{sec:overview-combined}).
Second, we show how to compute excludes initial and entailment for temporal formula.
Finally, we prove that defining these judgments in this way is sound.

% In \autoref{sec:applog}, we define a program logic with abstract message histories (MHPL) for reasoning about app code parametrized by some specification of realizable message histories $\SpecSet$.

% And in \autoref{sec:newls}, we define a temporal specification logic (CBCFTL) for specifying realizable message histories fully independent from a program or a program analysis.
% Here, we show how to combine them to answer queries about abstract message histories $\abstrace$ under a particular, given specification $\SpecSet$, namely to judge excludes-initial \fbox{$\notinit{\abstrace}$} and entailment \fbox{$\entail{\abstrace}{\abstrace'}$}.

The high-level intuition is that given an abstract message history $\abstrace$, we instantiate the specification of realizable message histories $\SpecSet$ with $\abstrace$ into a single temporal formula $\directive$.
Then, with this temporal formula $\directive$, we can implement these judgments on abstract message histories via queries to an off-the-shelf SMT solver (using the encoding described at the end of \secref{newls}).
In \autoref{fig:atencode}, we describe the judgment form $\encodeRel{\abstrace}{\SpecSet}{\directive}$ that captures this combining of $\abstrace$ and $\SpecSet$ into a single temporal formula $\directive$.

As we see in \autoref{fig:atencode}, the combining (or equivalent-to-a-temporal-formula) judgment form $\encodeRel{\abstrace}{\SpecSet}{\directive}$ is syntax-directed on the abstract message history $\abstrace$.
%The encode judgment is inductively defined on the structure of the abstract message history, $\abstrace$.
Under the assumption of $\SpecSet$, the abstract message history $\abstracebase$ is equivalent to the temporal formula $\temporaltrue$ (rule \TirName{temporal-okhist}).
For the ordered-implication abstract message history $\abstracecons{\abstrace}{\absmsg}$, intuitively, we want to hypothesize $\absmsg_1$ to derive any constraints from instantiating from $\SpecSet$ and to derive any constraints from ``quotienting'' the constraints from $\abstrace_2$ to ``remove $\absmsg_1$ from the end''.
Instantiating and quotienting are captured by two helper judgments.
The instantiate judgment form $\instRel{\absmsg}{\SpecSet}{\directive}$ says, ``In specification $\SpecSet$, hypothesizing abstract message $\absmsg$, temporal formula $\directive$ describe realizable message histories.''
And the quotient judgment form $\updRel{\directive}{\directive'}{\absmsg}$ says, ``Temporal formula $\directive$ is equivalent to temporal formula $\directive'$ with abstract message $\absmsg$ appended.''
We can now read the key \TirName{temporal-hypmsg} rule: if hypothesizing $\absmsg_1$ in $\SpecSet$ yields temporal formula $\directive_1'$, abstract message history $\abstrace_2$ is equivalent to temporal formula $\directive_2$, and $\directive_2$ is equivalent to temporal formula $\directive_2'$ with $\absmsg_1$ appended, then the ordered-implication abstract message history $\abstracecons{\abstrace}{\absmsg}$ is equivalent to $\directive_1' \wedge \directive_2'$.

\begin{figure}\small
  \begin{mathpar}
    \fbox{$\encodeRel{\abstrace}{\SpecSet}{\directive}$}

    %\infer[encode-okhist]{ }{
    \infer[temporal-okhist]{ }{
      \encodeRel{\abstracebase}{\SpecSet}{\temporaltrue}
    }

    %\infer[encode-next]{
    \infer[temporal-hypmsg]{
      \instRel{\absmsg_1}{\SpecSet}{\directive_1'} \\
      \encodeRel{\abstrace_2}{\SpecSet}{\directive_2} \\
      \updRel{\directive_2}{\directive_2'}{\absmsg_1}
    }{
      \encodeRel{\abstracecons{\abstrace_2}{\absmsg_1}}{\SpecSet}{ \directive_1' \wedge \directive_2'}
    }\\

    \fbox{$\instRel{\absmsg}{\SpecSet}{\directive}$

      $\updRel{\directive}{\directive'}{\absmsg}$}

    %\infer[instantiate-message-equal]{
    \infer[instantiate-yes]{
      %\SpecSet = \specOnly{\absmsg_2}{\directive} \\
      \ideq{\absmsg_1}{\absmsg_2} %\\ \instRel{\absmsg_1}{\SpecSet'}{\directive'}
    }{
      \instRel{\absmsg_1}{(\specOnly{\absmsg_2}{\directive})}{
        %\directive[\fv{\absmsg_1} / \fv{\absmsg_2}]
        [\unifysubst]\directive
      }
    }

    %\infer[instantiate-message-not-equal]{
    \infer[instantiate-no]{
      %\SpecSet = \specOnly{\absmsg_2}{\directive} \\
      \idneq{\absmsg_1}{\absmsg_2} %\\ \instRel{\absmsg_1}{\SpecSet'}{\directive}
    }{
      \instRel{\absmsg_1}{(\specOnly{\absmsg_2}{\directive})}{\text{true}}
    }\\
    \infer[quotient-once]{ }{
      \updRel{\iDir{\ltmsg}}{
	    \msgequiv{\ltmsg}{\absmsg} \vee
        (\msgnotequiv{\ltmsg}{\absmsg} \wedge \iDir{\ltmsg})
      }{\absmsg}
    }

    \infer[quotient-historically-not]{ }{
      \updRel{\notiDir{\ltmsg}}{
	    \notiDir{\ltmsg} \wedge \msgnotequiv{\ltmsg}{\absmsg}
      }{\absmsg}
    }
    \\
    \infer[quotient-not-since]{ }{
      \updRel{\niDir{\ltmsg_2}{\ltmsg_1}}{
        \msgequiv{\ltmsg_1}{\absmsg} \vee (\msgnotequiv{\ltmsg_1}{\absmsg} \wedge \niDir{\ltmsg_2}{\ltmsg_1} \wedge \msgnotequiv{\ltmsg_2}{\absmsg})
      }{\absmsg}
    }
%
    % \infer[quotient-and]{\updRel{\directive_1}{\directive_1'}{\absmsg} \\ \updRel{\directive_2}{\directive_2'}{\absmsg} }{
    %   \updRel{\directive_1 \wedge \directive_2}{\directive_1' \wedge \directive_2'}{\absmsg}
    % }
%
    % \infer[quotient-or]{\directive_1' = \updRel{\directive_1}{\directive_1'}{\absmsg} \\ \directive_2' = \updRel{\directive_2}{\directive_2'}{\absmsg} }{
    %   \updRel{\directive_1 \vee \directive_2}{\directive_1' \vee \directive_2'}{\absmsg}
    % }
%
    % \infer[quotient-eq]{ }{
    %   \updRel{\absof{x} = \absof{y}}{\absof{x} = \absof{y}}{\absmsg}
    % }
%
    % \infer[quotient-true]{ }{
    %   \updRel{\text{true}}{\text{true}}{\absmsg}
    % }
%
    % \infer[quotient-false]{ }{
    %   \updRel{\text{false}}{\text{false}}{\absmsg}
    % }
%    
    % \infer[quotient-forall]{\updRel{\directive}{\directive'}{\absmsg} }{
    %   \updRel{\forall\ \absof{x}.~\directive}{\forall\ \absof{x}.~\directive'}{\absmsg}
    % }
%
    % \infer[quotient-exists]{\updRel{\directive}{\directive'}{\absmsg} }{
    %   \updRel{\exists\ \absof{x}.~\directive}{\exists\ \absof{x}.~\directive'}{\absmsg}
    % }
  \end{mathpar}
  \caption{\label{fig:atencode}%
    Instantiating \newls{} specifications $\SpecSet$ with abstract message histories $\abstrace$ from MHPL.
    The judgment form $\encodeRel{\abstrace}{\SpecSet}{\directive}$ says, ``Under CBFTL specification $\SpecSet$, an abstract message history $\abstrace$ is equivalent to a temporal formula $\directive$.''
    We can view this judgment as giving us an encoding into a temporal formula $\directive$, the instantiation of a specification of realizable message histories $\SpecSet$ with a particular abstract message history $\abstrace$ to derive a description of the realizable message histories up to a program location.
  }
\end{figure}

% Knowing the intent of the instantiate and quotient judgments, their definitions are relatively straightforward.
Instantiation is the process of combining the hypothetical next message of an abstract message history with a history implication (e.g., \autoref{eq:inst} in the running example from \secref{overview}).
For the instantiate judgment $\instRel{\absmsg}{\SpecSet}{\directive}$, we show only the cases for single history implications $\specOnly{\absmsg_2}{\directive}$ where the hypothesized message $\absmsg_1$ either matches (\TirName{instantiate-yes}) or doesn't match (\TirName{instantiate-no}).
The other cases for $\spectrue$ and $\SpecSet_1 \land \SpecSet_2$ just yield $\temporaltrue$ and the conjunction of the instantiations in $\SpecSet_1$ and $\SpecSet_2$, respectively.
As $\specOnly{\absmsg_2}{\directive}$ implicitly binds the variables of $\absmsg_2$, we write $\ideq{\absmsg_1}{\absmsg_2}$ for a matching up to a substitution $\unifysubst$ from the variables of $\absmsg_2$ to the variables of $\absmsg_1$ and write $[\unifysubst]\directive$ for the capture-avoiding substitution with $\unifysubst$ in $\directive$. And we write $\idneq{\absmsg_1}{\absmsg_2}$ for the case where $\absmsg_1$ cannot match $\absmsg_2$.

The quotient judgment form, shown in \autoref{fig:atencode}, $\updRel{\directive}{\directive'}{\absmsg}$ is syntax-directed on $\directive$ to yield $\directive'$. 
We show the quotienting judgments for the three temporal operators
$\mathbf{O}$, $\mathbf{HN}$, and $\mathbf{NS}$.
%$\iDir{}$, $\notiDir{}$, and $\niDir{}{}$.
Quotienting the other temporal formula $\directive$ productions is straightforward.
Quotienting once, $\iDir{\ltmsg}$, with the abstract message $\absmsg$ has two possibilities:
\begin{inparaenum}
  \item $\msgequiv{\ltmsg}{\absmsg}$ --- the abstract message is equivalent to the message in the ``once'' making the temporal formula equivalent to ``true'' and
  \item $\msgnotequiv{\ltmsg}{\absmsg}$ --- the abstract message is not equivalent to the message in the ``once'' leaving the temporal formula unchanged.
\end{inparaenum}
Mirroring the quotienting of once, quotienting historically not, $\notiDir{\ltmsg}$, with the abstract message $\absmsg$ has two possibilities:
\begin{inparaenum}
  \item $\msgequiv{\ltmsg}{\absmsg}$ --- the abstract message is equivalent to the message in the ``has never'' making the temporal formula equivalent to ``false'' and
  \item $\msgnotequiv{\ltmsg}{\absmsg}$ --- the abstract message is not equivalent to the message in the ``has never'' leaving the temporal formula unchanged.
\end{inparaenum}
The operator $\niDir{\ltmsg_2}{\ltmsg_1}$ is a combination of the previous two: either the quotiented message matches the right-hand side becoming true, or it must not match the left-hand side.

% We only show the case for the temporal operator $\niDir{}{}$ in \autoref{fig:atencode} since it shows similar properties to $\iDir{}$ and $\notiDir{}$ (A full listing of quotient operators for each production in the temporal formula may be found in \refappendix{extsemantics}).
We see that for quotienting with the temporal operators, we need an analogous encoding of match or doesn't match: the meta-level functions $\msgequiv{\ltmsg}{\absmsg}$ and $\msgnotequiv{\ltmsg}{\absmsg}$ encode into propositional formula of a symbolic message $\ltmsg$ matching or not matching an abstract message $\absmsg$, respectively.
%% Evan 9/10: The previous paragraph says we're eliding the NS one.
% Quotienting the temporal formula $\niDir{\ltmsg_1}{\ltmsg_2}$ with an abstract message $\absmsg$ has 3 possible meanings:
% \begin{inparaenum}
%   \item $\msgequiv{\ltmsg}{\absmsg_1}$ -- the message abstracted by $\ltmsg$ satisfies this temporal formula making the quotient equivalent to ``true'',
%   \item $\msgequiv{\ltmsg}{\absmsg_2}$ -- the message abstracted by $\ltmsg$ is the message that should not have happened since $\ltmsg_1$ making the quotient equivalent to ``false'', and finally, 
%   \item $\msgnotequiv{\ltmsg}{\absmsg_1}$ and $\msgnotequiv{\ltmsg}{\absmsg_2}$ -- $\ltmsg$ matches neither message in the original temporal formula leaving the abstracted message histories unchanged.
% \end{inparaenum}
Since the message names, $\msgid$, and kinds are known, $\msgequiv{\ltmsg}{\absmsg}$ and $\msgnotequiv{\ltmsg}{\absmsg}$ always result in equalities and disequalities of logic variables (e.g., \autoref{eq:quot} in the running example from \secref{overview}) or ``false''.

% We show only the cases for the temporal operators (all other cases just quotient subformulas and reconstruct with same formula constructor).
% The \TirName{quotient-once} rule for $\iDir{\ltmsg}$ says either $\ltmsg$ matches $\absmsg$, which satisfies the once requirement or $\ltmsg$ does not match $\absmsg$ and thus the once requirement is still needed.
% The remaining rules shown here encode quotienting for $\notiDir{\ltmsg}$ and $\niDir{\ltmsg_2}{\ltmsg_1}$ similarly following their semantics (cf. \autoref{fig:cbcfsemantics}).
% Here, it is crucial that the temporal operators apply only to individual messages.

With the ability to combine an abstract message history $\abstrace$ from MHPL with a CBCFTL specification of realizable message histories $\SpecSet$ via the $\encodeRel{\abstrace}{\SpecSet}{\directive}$ judgment, algorithms for judging excludes-initial and entailment via SMT queries become clear.
Let us write $\toFOLRel{\directive}{\fol}$ for the encoding of a temporal formula $\directive$ into a closed first-order formula $\fol$ and use $\histAxiomsFol$ for the axioms encoding message histories from \autoref{sec:newls}.

% \TODO{Need theorems about soundness of excludes initial and entailment.}

% \TODO{discuss assumptions on excludes init and entailment in section 3}

We define procedures for judging excludes-initial $\notinit{\abstrace}$ as checking for the unsatisfiability of $\histAxiomsFol \land \fol \land \folLen = 0$ (where
$\encodeRel{\abstrace}{\SpecSet}{\directive}$ and $\toFOLRel{\directive}{\fol}$), and entailment $\entail{\abstrace}{\abstrace'}$ as checking for the unsatisfiability of $\histAxiomsFol \land \fol \land \neg \fol'$ (where
$\encodeRel{\abstrace}{\SpecSet}{\directive}$, $\toFOLRel{\directive}{\fol}$,
$\encodeRel{\abstrace'}{\SpecSet}{\directive'}$, and $\toFOLRel{\directive'}{\fol'}$).
The soundness of checking these judgments relies on the correctness of the combining judgment:
\begin{theorem}[Correct Combining of MHPL and \newls{}]
\begin{inparaenum}
\item \label{lem:encodefwd} If $\encodeRel{\abstrace}{\SpecSet}{\directive}$ such that $\inconc{\withvalua{\hist}}{\abstrace}$, then $\modeldouble{\hist}{\valua} \models \directive$.
\item \label{lem:encodebkg} If $\encodeRel{\abstrace}{\SpecSet}{\directive}$ such that $\modeldouble{\hist}{\valua} \models \directive$ and $\hist \models \SpecSet$, then $\inconc{\withvalua{\hist}}{\abstrace}$.
\end{inparaenum}
\label{lem:encode}
\end{theorem}
Note that we assume a well-formedness condition that no abstract message is vacuous (i.e., for any abstract message $\absmsg$ and any assignment $\valua$, there exists a (concrete) message $\msg$ such that $\inconc{\withvalua{\msg}}{\absmsg}$). The correctness of combining relies on correct instantiation and quotienting:
\begin{lemma}[Correct Instantiating of History Implications]
	\begin{inparaenum} 
		\item \label{lem:instfwd} If
		$\instRel{\absmsg}{\SpecSet}{\directive}$
		such that
		$\histcons{\hist}{\msg} \models \SpecSet$ and
		$\modDir{\withvalua{\msg}}{\absmsg}$,
		then $\modDir{\withvalua{\hist}}{\directive}$.
		\item \label{lem:instbkg} If
		$\instRel{\absmsg}{\SpecSet}{\directive}$
		such that
		$\hist \models \SpecSet$ and
		$\modDir{\withvalua{\msg}}{\absmsg}$ and
		$\withvalua{\hist}\models\directive$,
		then $\histcons{\hist}{\msg} \models \SpecSet$.
	\end{inparaenum}
	\label{lem:inst}
\end{lemma}
\begin{lemma}[Correct Quotienting of Temporal Formulas]
	\begin{inparaenum}
		\item \label{lem:quotfwd} If
		$\updRel{\directive}{\directive'}{\absmsg}$
		such that
		$\withvalua{\histcons{\hist}{\msg}}\models\directive$
		and
		$\withvalua{\msg}\models\absmsg$,
		then $\withvalua{\hist}\models\directive'$.
		\item \label{lem:quotbkg} If
		$\updRel{\directive}{\directive'}{\absmsg}$ such that
		$\withvalua{\hist}\models\directive'$ and
		$\withvalua{\msg}\models \absmsg$,
		then $\withvalua{\histcons{\hist}{\msg}}\models\directive$.
	\end{inparaenum}
\label{lem:quot}
\end{lemma}
\noindent Proofs for these statements may be found in \refappendix{atencproof}.

%% file: eval.tex
\section{Empirical Evaluation}
\label{sec:eval}

\newcommand{\cTimeout}{%
    \scalebox{1.5}{\ensuremath{\otimes}}%
}
\newsavebox{\cSafeBox}
\savebox{\cSafeBox}{
\hspace{-.15cm}\scalebox{0.7}{
\tikz[baseline=-0.7ex]{
\node (a) [circle,draw=black] {
};
\node (b) [draw=none] {\scalebox{1}{\checkmark}};
}}\hspace{-.1cm}
}
\newcommand{\cSafe}{
\usebox{\cSafeBox}
}

\newsavebox{\cSafeToFiveBox}
\savebox{\cSafeToFiveBox}{
\hspace{-.15cm}\scalebox{0.7}{
\tikz[baseline=-0.7ex]{
\node (a) [circle,draw=black] {
};
\node (b) [draw=none] {\scalebox{1}{5}};
}}\hspace{-.1cm}
}
\newcommand{\cSafeToFive}{
\usebox{\cSafeToFiveBox}
}

\newsavebox{\cAlarmBox}
\savebox{\cAlarmBox}{
\hspace{-.1cm}\scalebox{0.7}{
\tikz[baseline=-0.7ex]{
\node (a) [circle,draw=black] {
};
\node (b) [draw=none] {\scalebox{1}{!}};
}}\hspace{-.0cm}
}
\newcommand{\cAlarm}{\usebox{\cAlarmBox}}
As we discuss in \autoref{sec:overview}, the challenge for a program verifier is to prove the safety of assertions that depend on the callback order, while avoiding unsound models of the framework. We hypothesize that:
\begin{inparaenum}
\item thanks to the targeted callback control flow specification, \toolname can prove safe assertions while avoiding unsound results when an assertion does not hold. And
\item \toolname can be applied to real event-driven programs.
\end{inparaenum}
We validate our hypotheses with the following research questions:
\begin{enumerate}
\item[\req{1:}] \textbf{Proving Assertions:} Is it possible to write a targeted \newls specification for \toolname and prove safe assertions, while avoiding unsound framework models?
%
% [Sergio] I think the RQ about existing models is confusing here.
%
% Is it possible to write a targeted \newls specification for \toolname benchmarks representing correct usages of these patterns while avoiding unsound framework models that would cause false proof of benchmarks representing buggy usages?  Then, we ask: Assuming the best possible program analysis, could the framework modeling approaches of other state-of-the-art techniques distinguish these benchmarks?
\item[\req{2:}] \textbf{Generalizability to Real-World Applications:} Can \toolname prove assertions on real-sized, complex, and widely used Android applications?
\end{enumerate}

% First, we analyze benchmarks representing buggy and fixed versions of each pattern, and then, we evaluate \toolname on instances of these patterns in 47 widely-used open-source applications.

\paragraph{Bug Patterns.}
Checking for arbitrary assertions, such as safe \codej{null} dereference, is not interesting as most safe assertions can be proven with an intra-callback analysis. So, to find assertion locations in Android apps that require callback control flow reasoning, we identified a set of problematic API usage \textit{patterns}.
We first searched bug reports of runtime crashes for popular open source Android apps satisfying \textit{all} the following criteria:
\begin{inparaenum}[(a)]
    \item The issue had a stack trace similar the one shown in \figref{antennapod-stacktrace-issue-2855}.
    \item The issue accepted a fix that relies on the callback order. %A fix is documented in the issue tracker that depended on the possible order of callbacks.
    \item The crash involved callbacks or callins from a set of commonly used Android objects (\codej{Activity}, \codej{Fragment}, \codej{Dialog}, \codej{View} objects such as buttons and menus, \codej{AsyncTask}, and \codej{Single}/\codej{Maybe} from RxJava).
 %Components that we are familiar with played a roll in causing the crash (i.e. a callback or callin from the component appears in the stack trace).
\end{inparaenum}
Then, we classified the crashes according to 5 patterns of interaction between callbacks and callins.
%Several of our patterns come from Antennapod \cite{antennapodfix2856} as they do a superb job of documenting the exact commit that fixes a reported crash.
%We identify the benchmarks by app name and message causing the crash:, , ,  and .
\begin{inparaenum}
    \item \apGa{}~\cite{antennapodfix} --- the Android method \codej{getActivity} returns \codej{null} if called on a \codej{Fragment} that is not in the ``created'' state, and the app dereference such \codej{null} pointer (\codej{Activity} and \codej{Fragment} objects are in the ``created'' state if the \codej{onCreate} callback has been invoked, but the \codej{onDestroy} has not).
    \item \apEx{}~\cite{antennapodfix-execute} --- the app calls \codej{execute} twice on the same \codej{AsyncTask} object, ending in an exception.
    \item \ymDi{}~\cite{yambafix-dismiss-anon} --- the app calls \codej{dismiss} on a \codej{Dialog} constructed with an \codej{Activity} that is currently in the ``created'' state, ending in an exception.
    \item \cbFi{}~\cite{connectbot-finish-anon} --- the app dereference a field in an \codej{onClick} callback, the same field can be set to \codej{null} in the \codej{onPause} callback, and the app call \codej{finish} on the enclosing \codej{Activty} (we call ``nullable'' the fields that can be set to null in a callback).
    \item \synch~\cite{antennapod-synch-fix} --- the app dereferences a nullable field in a callback executed concurrently, such as \codej{Runnable} \codej{run}.
\end{inparaenum}
We name the patterns with the main message involved in the crash, followed in subscript by an ``exception property'', specifying when the bug would manifest with throwing an exception. Such exception properties may be a \newls history implication (referenced by number and listed in \refappendix{speclist}) specifying when the framework returns an exception, a \enkwNull{} value, or a nullable field dereference (indicated by \enkwNull).

\paragraph{Implementation.} \toolname implements the backward abstract interpretation with message histories of \autoref{sec:applog} for refuting callback reachability assertions in Android apps. \toolname uses Soot~\cite{vall99soot} for loading the compiled app and to
%generate code for creation of the%
implement the application only control flow graph construction (similar to \cite{DBLP:conf/ecoop/AliL13} but augmented with boundary transitions as discussed in \autoref{sec:apponly-transition-system}). \toolname implements the encoding of \autoref{sec:decision}, and uses the Z3 SMT solver~\cite{DeMoura+Bjorner/08/Z3} to check the satisfiability of temporal formulas (\autoref{sec:newls}).
\toolname further processes callbacks in parallel and pre-empts calls to Z3 when possible for performance.
We ran our experiments using Chameleon Cloud \cite{keahey2020lessons} using an AMD EPYC 7763 and 256 GB of RAM.

\newcommand{\twocol}[1]{\multicolumn{2}{c}{#1}}
\newcommand{\onecol}[1]{\multicolumn{1}{c}{#1}}
\newcommand{\callCoGa}{$3^{[\ref{spec:call},\reftr{spec:onActivityCreated}]}$}
\newcommand{\coCdEx}{$3^{[\reftr{spec:createOnce},\reftr{spec:clickDisable}]}$}
\newcommand{\shDi}{$2^{[\reftr{spec:show}]}$}
\newcommand{\clfFvOac}{$3^{[\reftr{spec:clickFinish}, \reftr{spec:findView},  \reftr{spec:createOnce}]}$}
\newcommand{\synchSpec}{$5^{[\reftr{spec:subscribeDispose}, \reftr{spec:subscribeOnSame}}_{\reftr{spec:observeOnSame}, \reftr{spec:startStop}, \reftr{spec:createUnique}]}$}
\newcommand{\alspc}{\hspace{-.05cm}}
\newcommand{\alspcc}{\hspace{-.13cm}}

%% Note that this comes from log file "
\begin{table}[t]\footnotesize\centering %p{0.2cm} %possible way to limit column size, but may be broken
\begin{footnotesize}
%                                                   | bench            | historia        res time | infer
\begin{tabular*}{\linewidth}{@{\extracolsep{\fill}}l p{1.00cm} *{2}{r} p{.40cm} r @{\hspace{0.75\tabcolsep}} r r @{\hspace{0.75\tabcolsep}} r r @{\hspace{0.75\tabcolsep}} r @{\hspace{0.75\tabcolsep}} c c c l l l @{}} 
\multicolumn{4}{c}{Pattern} & \multicolumn{10}{c}{\toolname} & \multicolumn{1}{c}{no-order} & \multicolumn{1}{c}{eager}\\ 
\cmidrule[\heavyrulewidth](r){1-4} \cmidrule[\heavyrulewidth](rl){5-14} \cmidrule[\heavyrulewidth](l){15-15} \cmidrule[\heavyrulewidth](l){16-16}
          &  & \onecol{$\enkwCb,\enkwRet$} & \onecol{$\enkwCi$}      & specs     & \twocol{$\enkwCb$}         & \twocol{$\enkwCb\enkwRet$} & \twocol{$\enkwCi$}     & time   & depth & res              & res                           & res\\
          &  &         (n)                        &   (n)                   & (n)       &    (n)     & (\%)          &    (n)     &    (\%)       & (n)     &  (\%)        & (s)    & (n)   &                  &     \\
%\toprule
\cmidrule(r){1-4} \cmidrule(rl){5-14} \cmidrule(l){15-15} \cmidrule(l){16-16}
% Bug        &      cb,ret                        &   ci                    &    specs  &\\  mcb     &  %            &   cbret    &     %         &  ci     &     %        & time  & depth &                                                                                                                                                                
Bug & \apGa  &      9                             &   15                    & \callCoGa &     3      &  33           &     1      &     11        &  2      &     13       &   9   & 3     & \cAlarm          & \cAlarm                       & false-\alspcc\cSafe \\ %Row 1
    & \apEx  &      7                             &    6                    & \coCdEx   &     2      &  29           &     0      &     0         &  2      &     33       &   12   & 3     & \cAlarm          & \cAlarm                       & \cAlarm \\ %Row 2
    & \ymDi  &      7                             &    6                    & \shDi     &     1      &  14           &     1      &    14         &  1      &     17       &   18   & 4     & \cAlarm          & \cAlarm                       & false-\alspcc\cSafe \\ %Row 5
    & \cbFi  &      7                             &    9                    & \clfFvOac &     3      &  43           &     1      &    14         &  3      &     33       &   52   & 3     & \cAlarm          & \cAlarm                       & false-\alspcc\cSafe \\ %Row 4
    & \synch &      9                             &   17                    & \synchSpec&     3      &  33           &     0      &    0          &  5      &     29       &   24   & 3     &  \cAlarm         & \cAlarm                       & false-\alspcc\cSafe\\
%\midrule
\cmidrule(r){1-4} \cmidrule(rl){5-14} \cmidrule(l){15-15} \cmidrule(l){16-16}
% Fix\\                                                                                                                                                                       
Fix & \apGa  &      9                             &   16                    & \callCoGa &     3      &  33           &     1      &     11        &  3      &     19       &   16   & 3     & \cSafe  & false-\alspc\cAlarm           & \cSafe \\ %Row 1
    & \apEx  &      7                             &   7                     & \coCdEx   &     2      &  29           &     0      &     0         &  3      &     43       &   27   & 4     & \cSafe  & false-\alspc\cAlarm           & false-\alspc\cAlarm \\ %Row 2
    & \ymDi  &      7                             &   6                     & \shDi     &     1      &  14           &     1      &    14         &  1      &     17       &   79   & 6     & \cSafe  & false-\alspc\cAlarm           & \cSafe \\ %Row 5
    & \cbFi  &      7                             &   10                    & \clfFvOac &     3      &  43           &     1      &    14         &  4      &     40       &   1800 & 5     & \cSafeToFive         & false-\alspc\cAlarm           & \cSafe \\ %Row 4
    & \synch &      9                             &   18                    & \synchSpec&     3      &  33           &     0      &     0         &  6      &     33       &   150  & 5     & \cSafe  & false-\alspc\cAlarm           & \cSafe\\
%\midrule
\cmidrule(r){1-4} \cmidrule(rl){5-14} \cmidrule(l){15-15} \cmidrule(l){16-16}
& total      &      66                            &   73                    &   10      &     22     &  \textbf{33}  &   6        & \textbf{9}    &  21     & \textbf{29}  &        &    &                  &    \\
\bottomrule
\end{tabular*}
\end{footnotesize}
\caption{
% Each row is either an existing bug (first 4 rows) or a bug-fix version (last 4 rows) of open source Android apps
% (AntennaPod [AP], Yamba [Y], ConnectBot [C]).
% A bug-fix version is a fix for the bug version with the same label (e.g., the bug-fix AP \codej{getAct} is a fix for the bug version AP \codej{getAct}), and each bug bug-fix pair correspond to specific pull requests (\apGa{}~\cite{antennapodfix}, \apEx{}~\cite{antennapodfix-execute}, \ymDi{}~\cite{yambafix-dismiss-anon}, \cbFi{}~\cite{connectbot-finish-anon}).
% %
The rows of this table are split into Bug and Fix benchmarks for each pattern. We first list the number of callbacks, callback returns, or callins that could be captured by a history implication in the framework model.  Next, we list the number of history implications (specs) written for the benchmark (listed in \refappendix{speclist}), then, we show how many of the messages are in the specification.  
The depth captures how many times \toolname needed to step backwards through a callback (e.g. \autoref{fig:refgraph} shows 4 steps back).
%In the worst case, \toolname must consider $[cb,ret]^{[depth]}$ abstract states to produce the results in each column.
\toolname alarms on all the bug versions ($\cAlarm$),
and refutes reachability of the bug assertion for 4 out of the 5 bug-fixes ($\cSafe$).
In the last case, \toolname explored up-to 5 callbacks before timing out at 30 min (\cSafeToFive).
For the comparison with ideal tools using the ``no-order'' and ``eager'' modeling approaches, some results are labeled false-\alspcc\cSafe and false-\alspc\cAlarm.
}
% The first 5 rows refer to benchmark implementations of the pattern replicating the crash, the second 5 to safe implementations.
% \refappendix{speclist} lists details for each benchmark as well as the specifications referenced in brackets (e.g. [\reftr{spec:getActivityNull}]).
% We report the callback and return count, \textit{$\enkwCb/\enkwCb\enkwRet$} and the callin call site count, \textit{$\enkwCi$}.
%
%
% The \emph{specs} column counts the history implications sufficient to prove the fixed version correct.
% The subsequent columns show the number of messages appearing both in the specification and the app.
% For the Infer column, we reduce each bug to a null dereference.  
%We reduced each bug to a null dereference for \toolinfer.
%In the case of \toolflowdroid, we manually inspected the generated main method for the callbacks in our benchmarks.
%
\label{tbl:benchmarks}
\end{table}

% Apps were selected by searching for github issues 

% select bugs found in the wild
% each bug has a fix
% bugs that cannot be differentiated by flowdroid or infer models 
% depend on event ordering
% infer cannot distinguish because it uses the top model
% flowdroid cannot distinguish because the artificial "main" method restricts the crashing callback.
%\paragraph{Reachable and Unreachable Assertions From Bugs and Bug-fixes}
\subsection{RQ1: Proving Event-Driven Patterns}
In~\autoref{tbl:benchmarks}, we evaluate the ability of \toolname and the representative state of-the art framework modeling (\emph{no-order}, \emph{eager}) to prove safe fixes of the bug patterns, while correctly alarming on instances containing the bug.
% syntetic apps
For each one of the 5 bug patterns, we distilled a \emph{Bug} and a \emph{Fix} benchmark application from the real app code mentioned in the representative bug reports  (slicing the app code to remove all the components and code non-necessary to reproduce the bug). The Bug version demonstrates the usage of the framework callbacks and callins causing the crash in the original application, while the Fix version applies the fix from the bug report. A sound analysis should always alarm on the Bug version.
%These 10 benchmark applications are used to compare \toolname and the approaches taken by \toolinfer and \toolflowdroid.

We manually wrote a \newls specification \textit{sufficient} to prove the assertion safe for each fix, and then we run \toolname with this specification (\textit{specs} column) on both the Bug and Fix version.
%We record both the number of history implications written for each benchmark and summarize the number and percentage of messages captured.
We compare \toolname with the main framework modeling approaches, which either do not assume any callback ordering (\emph{no-order}), or provide an \emph{eager} modeling of the framework.
\toolinfer~\cite{DBLP:conf/nfm/CalcagnoD11} and \toolflowdroid~\cite{DBLP:conf/pldi/ArztRFBBKTOM14} are used as representatives for the first and second approach, respectively. 
% \toolinfer handles crashes (as \toolflowdroid targets dataflow for security), while the exception properties describing the first 3 bug patterns depend on the framework (i.e. whether a callin return a $\enkwNull$ value or an exception). 
Of the 5 bug patterns, only the 4th and 5th patterns are supported by \toolinfer and none are supported by \toolflowdroid.
We note that \toolflowdroid is the only open source tool we could run in the \emph{eager} category but does not natively support these properties.
Therefore, in order to compare with the no-order model, the first three exception properties were reduced to a nullable field and checked with \toolinfer (i.e., we manually wrote code that would throw a null pointer exception just before the actual exception was thrown). For the remaining two, we added nullability annotations on the affected fields (because \toolinfer will not alarm on a null value from a field without this annotation).
For the \emph{eager} model, we manually examine the artificial main method generated by \toolflowdroid.
This is a main method that should behave as the original app composed with the framework. 
We evaluate whether any sound and precise whole-program static analysis could prove the fix while alarming on the bug with this main method.

%% Note that for the first 3 patterns, the safety property itself is caused by a message that the framework can control (i.e. whether a callin may return a $\enkwNull$ value or an exception).
%% Of the 5 patterns, only the nullable fields are supported by \toolinfer natively and none are natively supported by \toolflowdroid.

% We first ask, can \toolname prove fixed benchmarks by adding handwritten control flow specifications as needed while avoiding unsound proofs on the buggy benchmarks? We then ask, could other approaches to framework modeling handle these bug patterns?
% These benchmarks are described in more detail in \refappendix{speclist} and will be included with the artifact.

\paragraph{Discussion of the Results} 
% Writing such specifications required understanding the framework messages involved in the bug and the fix version to sufficiently refine the message history.
\toolname always (and correctly) alarms ($\cAlarm$) on all the Bug versions, while either refutes ($\cSafe$) or does not terminate before exhausting a run-time budget of 30 minutes (\cSafeToFive result in the \cbFi benchmark).
For the Fix version of the \cbFi benchmark, \toolname still does not alarm, but provides the partial result proving that no program execution containing less than 5 callback invocations can reach the assertion.
Interestingly, such a partial proof rules out the (abstract) execution \toolname found when failing to refute the assertion in the Bug version for \cbFi, which visits $4$ callbacks (see the \textit{depth} column).
Targeted refinement of the framework specific control-flow specifications was required in each case for \toolname to avoid false alarms.

% For each fixed benchmark, we wrote \newls specifications until we reached a proof or timeout with \toolname.
% The bottom line result, as shown in \autoref{tbl:benchmarks} is that we were able to fully prove 4 out of 5 of the fixed benchmarks and proved the 5th to a depth of 5 callbacks (i.e. no sequence of 5 or fewer callbacks can reach the crash).
% Additionally, we checked that \toolname still alarms on the buggy versions of the benchmarks and that the expected sequence of messages reaching the assertion failure during runtime was not excluded by the framework model.

% For \toolinfer to raise an alarm on a null value from a field, the field must be annotated as nullable.
% Once a field is annotated as nullable, \toolinfer alarms on any dereference where a local null check has not occurred.
Unsurprisingly, the no-order model results in false alarms on each fixed benchmark.
%In RQ2, we will see that the no-order model is a reasonable starting point and proves many instances of these patterns before needing control-flow refinement, but the remaining alarms are still significant.
%
Additionally, for the eager model, we found that in all but one case the artificial main method generated by \toolflowdroid rules out the sequence of callbacks reaching the real bug.
In three of these cases, a callback that has to be executed to reach the bug was missing from the call graph. In one case, the main method over-constrained the callback order.
For the remaining case, the eager model did not generate code that changed state when \codej{setEnabled(false)} was invoked, disabling a button.  Therefore, no program analysis could distinguish the state where \codej{onClick} could not occur on that button.

\subsection{RQ2: Generalizability to Real-World Applications}
Next, we evaluate the generalizability of \toolname by analyzing a set of 47 widely used applications containing over 2 million lines of code.
These apps were found and retrieved from the F-Droid repository \cite{fdroid} by filtering for apps updated in the last 2 years and that are more than 8 years old (rejecting obfuscated or otherwise difficult to inspect apps).
We answer this question by searching for the five bug patterns and attempting to verify the 1090 locations found.
First, we run \toolname on each location with only the exception property, and then, we sample 8 locations that could not be proven for targeted specification refinement including timeouts and alarms.
% We measure the tool's effectiveness by how many locations can be proven and by the difficulty of writing the specifications needed for such a proof.
% This search covered 47 open-source applications from the FDroid repository \cite{fdroid} containing over 2 million lines of code and finding a total of 1090 locations. % 2058418 counted using cloc adding together Java, Kotlin, and Scala, jupyter notebook "RunSpecs.ipynb"
% We chose such an extremely large sample set to avoid sampling bias.

%and why we believe that the application-only control flow graph \TODO{app only call graph is the only reasonable way to do this.}

% does not track downloads for privacy reasons, we used other metrics to find well-used apps for our evaluation.
% Our main criteria was that these are actively maintained apps that have a long history of use.
% Therefore, we chose apps that have had new releases pushed to FDroid within the last 2 years and were first uploaded more than 8 years ago. We additionally filtered apps that we can reasonably analyze and inspect (e.g. no code obfuscation, up-to-date documentation in English, uses a recent version of Android, and can be loaded by Soot).

% We chose apps that have been updated in the last 2 years, were added more than 8 years ago, use SDK 29 or later, have an active web page, have an English language version, maintain code and issue trackers on Github, and where Soot could reasonably load the compiled app.

We searched for the five patterns described in RQ1 using the application-only control flow graph.
For the \codej{execute} and \codej{dismiss} patterns, we searched for the callins in the call graph.
The remaining three patterns use an intraprocedural data flow analysis to find nullable values that were dereferenced.
This value comes from either \codej{getActivity} for the first pattern or a nullable field for the remaining patterns. The \codej{finish} pattern looks for such dereference commands in the \codej{onClick} callback when the \codej{finish} method is used, and the \codej{subs} pattern looks for dereferences in common concurrency callbacks.

% \TODO{Rm below}
% Specifically, we looked for:
% \begin{inparaenum}
%     \item A call to \codej{getActivity} where the return value was dereferenced.
%     \item Invocations of \codej{execute}.
%     \item Invocations of \codej{dismiss}.
%     \item \codej{onClick} callback dereferences a nullable field in apps that also use the \codej{finish} callin.
%     \item Concurrency related callback such as \codej{Runnable} \codej{run} that dereferenced a nullable field.
% \end{inparaenum}

\begin{table}[t]\footnotesize\centering %p{0.2cm} %possible way to limit column size, but may be broken
\begin{footnotesize}
%                                                   | bench            | historia        res time | infer
\begin{tabular*}{\linewidth}{@{\extracolsep{\fill}}l r r r r r r r r r r r}
                    % & \multicolumn{4}{c}{No Specification}                      \\
                    % \cmidrule(r){2-6}  %\cmidrule(r){8-11} 
        \multicolumn{4}{c}{Pattern}     &                                                 \multicolumn{7}{c}{\toolname} & \multicolumn{1}{c}{\toolflowdroid}\\
        \cmidrule[\heavyrulewidth](r){1-4}    \cmidrule[\heavyrulewidth](rl){5-11} \cmidrule[\heavyrulewidth](l){12-12}
            & locations    & apps & \multicolumn{1}{c}{KLOC} &     \multicolumn{2}{c}{alarm} & \multicolumn{2}{c}{timeout} & \multicolumn{2}{c}{safe} & \multicolumn{1}{c}{K methods} & \multicolumn{1}{c}{K methods} \\
            &  (n)         & (n)  & (n*1000) &     (n) & (\%)              &   (n) & (\%)                &  (n) & (\%) & (n*1000) & (n*1000) \\
    \midrule
    \apGa   &  558         & 24   & 1,655 &   261   &   47              &  97   &    17               & 200  & 36 & 105 & 22 \\
    \apEx   &  155         & 31   & 1,669 &     0   &    0              &  2    &     1               & 153  & 99 & 92  & 18\\
    \ymDi   &  291         & 38   & 1,853 &    43   &   15              &  208  &    71               &  40  & 14 & 102 & 21\\
    \cbFi   &   31         &  8   & 323   &     3   &   10              &  3    &    10               &  25  & 81 & 29  & 6\\
    \synch  &   55         & 10   & 1,108 &     1   &    2              &  7    &    13               &  47  & 85 & 16  & 3\\
    \midrule
    total   & 1090         & 47   & 2,058 &   308   &   28              &  317  &   29                & 465  & 43 & 121   & 28 \\
    \bottomrule 
\end{tabular*}
% note: app methods counted using code in bounder branch 'count_app_cg' and a branch of the same name on git@github.com:cuplv/FlowDroid_ftc.git
% Package filter was manually added to this code for each app
\end{footnotesize}
\caption{Verifying usages of the multi-callback patterns among 47 open-source Android apps with only the exception property. We list the results by alarms where \toolname finished but could not prove the property, timeouts where \toolname took over half an hour, and safe where no further specification was needed. We list the number of app methods that are contained in the call graphs of both \toolname and \toolflowdroid. There were 9 apps that timed out with \toolflowdroid. Among the 38 apps that \toolflowdroid could finish on, it found 28k application methods as compared to 70k application methods found by \toolname.}
\label{tbl:bigbenchmarks}
\end{table}

% This search identified 1090 locations implementing these patterns among the 47 open source Android applications.
Results are reported in \autoref{tbl:bigbenchmarks}.  The first column lists the individual patterns while the second column lists the locations found for each.  
For scale, we list the thousands of lines of code contained by the apps the patterns were found in KLOC.
% Additionally, we list the number of apps those patterns were found in and how many lines of code the apps contain (KLOC, thousands of lines of code). 
We then report the number and percentage of the locations that \toolname alarms on, timeouts on, and is able to prove safe.
%is able to prove usages of these patterns as well as alarms and timeouts.
As the most common unsoundness in RQ1 was missing methods from the call graph, we compare the number of application methods found in the call graph of \toolname and \toolflowdroid. A higher number of methods indicates more code is being analyzed.
% Since these are real apps containing tens of thousands of lines of code each, the manual inspection and reductions used for comparison to other modeling techniques was not possible.
% Instead, we compare the number of methods in the application package found by the application only control flow graph used by \toolname and the call graph generated by \toolflowdroid.

% In order to capture a representative sample of the locations for refinement, we randomly sampled locations that could not be proven from distinct applications (alarms and timeouts).
\autoref{tbl:bigsamples} lists 8 randomly sampled locations from distinct apps that \toolname could not prove without refinement.
For each sample, we recorded the time required to write the \newls specification in the "spec time" column.
The time to understand the callbacks being specified is not included in the recorded time, as this would be required for any modeling approach.
% The column "spec time" records the time it took us to write the \newls specifications (excluding the time to read the documentation as this is required for any framework modeling).
If it took more than an hour to run \toolname or if we took more than an hour to write the specification time, we record a timeout $\cTimeout$ in the result (res) column. % it took us more than an hour to write the \newls specification, we considered it a timeout (indicated by \cTimeout).
For perspective on the modeling difficulty, we list the number of messages that could be captured by the specification (the cb,ret and ci columns under Sample), as well as the total number of specifications as history implications we wrote (under the specs column) and the number and percentage of app messages that could be matched (the cb, cbret, and ci columns under \toolname).

% \TODO{=====}

% ( listed by the app name in the first column of \autoref{tbl:bigsamples})

% For each sample, we list the number of callbacks considered by \toolname (in one case, we manually prune the call graph to reduce timeouts due to the call graph imprecision indicated by * on the $\enkwCb$ and $\enkwCi$ count).
% When we were able to refine the specification sufficiently for a proof, we recorded the number of specifications along with the percentage of the callbacks and callins from the app that are captured in the specifications themselves.

\begin{table}[t]%\footnotesize\centering %p{0.2cm} %possible way to limit column size, but may be broken
\begin{footnotesize}% \hspace{-0.3cm}% \scalebox{1}[1]{
\begin{tabular*}{\linewidth}{@{\extracolsep{\fill}}l l r r r r r r r r r r l @{\hspace{0.75\tabcolsep}} r @{}}
\multicolumn{4}{c}{Sample} & \multicolumn{10}{c}{Historia}\\
\cmidrule[\heavyrulewidth](r){1-4} \cmidrule[\heavyrulewidth](l){5-14}
App           & Pattern & \multicolumn{1}{c}{\enkwCb,\enkwRet} & \multicolumn{1}{c}{ci}& \multicolumn{1}{c}{specs} & \multicolumn{2}{c}{cb} & \multicolumn{2}{c}{cbret} & \multicolumn{2}{c}{ci} & time & res                & spec time\\%             & pattern & cb/ret & ci      & specs &  cb         & cbret           & ci         & time & res                & sp time
              &         & (n)    & (n)     & (n)   & (n)  & (\%) & (n)   & (\%)    & (n) & (\%) & (s)  &                    & (m)\\
\midrule
Vanilla Music & \apGa   & 473    & 5,440   &       &      &      &       &         &     &      &   & \cTimeout          & 60\\
OpenVPN       & \apGa   & 938    & 9,628   & 2     & 49   & 5    & 5     & 1       & 167 & 2    & 0    & \cSafe             & 5\\
Seafile       & \apGa   & 2,464  & 19,562  & 2     & 54   & 2    & 9     & 0       & 68  & 0    & 1    & \cSafe             & 5\\
Syncthing     & \apGa   & 709    & 5,573   &       &      &      &       &         &     &      & 3,600& \cTimeout          & 9\\
Navit         & \cbFi   & 279    & 2,424   &       &      &      &       &         &     &      & 3,600& \cTimeout          & 54\\
Connectbot    & \ymDi   & 19$^\ast$ & 248$^\ast$   & 1     & 0    & 0    & 0     & 0       & 2   & 1    & 754  & true-\alspc\cAlarm & 4\\
BatteryBot    & \apGa   & 171    & 2,470   & 1     & 7    & 4    & 4     & 2       & 55  & 2    & 1    & \cSafe             & 3\\
Antennapod    & \apGa   & 3,906  & 2,3056  &       &      &      &       &         &     &      & 3,600& \cTimeout          & 35\\
\bottomrule
\end{tabular*}% }
\end{footnotesize}
\caption{We sampled 8 locations that could not be proven from \autoref{tbl:bigbenchmarks} and attempted to add \newls specifications to prove them safe. These are listed by app name as all samples were chosen so the apps are unique; specific locations app versions and links may be found in \refappendix{speclist}. $^\ast$The Connectbot benchmark here was a timeout in \autoref{tbl:bigbenchmarks}; we manually removed callbacks to help find the alarm and understand the bug, while the benchmarks were unmodified.}
\label{tbl:bigsamples}
\end{table}

\paragraph{Discussion}
 
Before manual refinement of the specification of the framework model, our tool was able to prove 43\% of the locations safe, raise alarms on 28\% and times out on 29\%.
We note that for \apEx, \cbFi, and \synch, we get few alarms as developers appear to use these patterns defensively.
Of the 8 samples, we found that we were able to correctly classify 4 locations within the hour budget of specification writing time (and always in within 5 minutes) and the hour budget of \toolname run time (and from a few seconds to a few minutes).
Of the 4 timeouts, one was from the specification writing time, and the rest were \toolname taking more than an hour.

In the 3 cases where we were able to prove locations, the specifications ignored 95\% or more of the boundary transitions in the app (i.e., no abstract message can match the majority of transitions in the applications).
This highlights a performance benefit to targeted-refinement --- although our analysis is unbounded in the worst case, in practice the majority of the messages that boundary transitions in the app can produce do not affect the encoded meaning of the message history, and most abstract states are immediately merged via entailment. 
Calling back to \autoref{sec:decision}, the specification ignoring most messages means that most of the time the quotient judgments result in an equivalent message history.
%(the encoded $\msgnotequiv{}{}$ case holds and not $\msgequiv{}{}$) and, additionally, no specification is instantiated.
Ignoring most messages allows most abstract states to be merged.
%improving the performance. % merging of abstract states becomes extremely important as our approach is exponential in the worst case.
With benchmarks containing hundreds to thousands of callbacks among thousands to tens of thousands of app methods and SMT calls for each new abstract pre-state, this means that we are avoiding the exponential explosion in the typical case.

%less than 2\% of the possible callback and callin names were captured by the specification.

It is also noteworthy that the application-only control flow graph used by \toolname captures significantly more applications methods than \toolflowdroid.
Among the applications that we could use \toolflowdroid to build call graphs for, it found 28K app methods.
In these same apps, \toolname found 70K app methods. 
This seems to reflect our observation from RQ1 that it is very challenging to capture all possible callbacks while eagerly modeling the framework and thus an argument for the targeted modeling approach of \toolname.

\subsection{Threats to Validity}

The main threat to validity of our experiments is the shifting behavior and authors understanding of the Android framework for which we used as a case study.  As noted in the Introduction, manual modeling of the Android framework is extremely difficult.
Even though the application-only control flow graph appears more sound from these experiments, we found that it is possible to miss callbacks without a complete list of objects the framework can instantiate with reflection.
To reduce the risk of an unsound call graph, we ensured that each location in RQ1 and a sampling of locations from RQ2 cannot be proven unreachable for any state (i.e., is reachable under \emph{some} app state) unless the location appears to actually be unreachable through manual inspection.

%% file: relatedwork.tex
\section{Related Work}
% DBLP:conf/pepm/PayetS14 an operational semantics for android activities
% DBLP:conf/eurosp/CalzavaraGM16 - horn droid - includes fwk semantics in design of analysis
% DBLP:journals/ase/YangWZWSYR18 - "static window transition graphs for android" - example of baking model into analysis - high precision, difficult to maintain
% - 
%\JEDI{Our approach of goal directed modeling and message history abstraction is unique among program analysis approaches.}
%Our approach to analyzing and modeling event driven applications is unique in three key ways: 
%\begin{inparaenum}
%  \item We may analyze the application independently of the framework while abstracting callback control flow.
%  \item Modeling of the framework is goal directed and compositional only necessitating rules that constrain behavior relevant to the property of interest.
%  \item Our abstraction of message histories is unique and extremely precise compared to other approaches.
%\end{inparaenum}
%Static analysis of event-driven systems use varying levels of precision in models of the framework.  
\JEDI{The related work focuses on modeling and abstraction of event order, i.e. message histories.}
%%% [Evan: This paragraph is nice but maybe not so needed.]
% There has been much interest in analysis of event-driven systems for finding defects, such as security vulnerabilities, crashes, race conditions, and more.
% For each analysis targeting a specific domain, the framework is modeled and then some abstraction is applied during analysis to capture the possible run-time order of callbacks.
% In this section, we use the term \emph{modeling} to refer to the developer of an analysis capturing the possible order of callbacks for a framework and the term \emph{abstraction} to refer to how the analysis itself reasons about the possible callback order within a specific application under that framework.
% In many cases, the choice of how the framework is modeled directly affects how it is abstracted.

\JEDI{Approaches use varying levels of callback control flow precision from none to highly precise models.}
Depending on the analysis domain, precise models are often included for some components but elided for others. 
As described in \autoref{sec:intro}, a common approach, particularly in industrial Android analysis tools, is to use no model at all (i.e., the most over-approximate model).
Verifiers with no callback order modeling have the advantage of performance and are the easiest to maintain but have a high false alarm rate requiring heuristic filtering~\cite{DBLP:conf/nfm/CalcagnoDDGHLOP15}.  
Precision is added to the callback control flow models for a range of different domains of static analysis. 
Awareness of the Activity lifecycle and other user interface callbacks can improve taint analysis for security \cite{DBLP:conf/pldi/ArztRFBBKTOM14,DBLP:conf/ndss/GordonKPGNR,DBLP:conf/eurosp/CalzavaraGM16}. 
Other program analysis tools will use the \codej{Activity} lifecycle in addition to precise models of other user interface components for verifying user interface properties \cite{DBLP:conf/icse/YangYWWR15,DBLP:journals/ase/YangWZWSYR18,DBLP:journals/tse/PerezL21}. 
Framework precision with respect to objects used for concurrency such as \codej{AsyncTask} and thread pools is often captured for race detection \cite{DBLP:conf/icse/YangYWWR15, DBLP:journals/ase/YangWZWSYR18, DBLP:conf/asplos/HuN18,DBLP:conf/icst/WuLSCX19} and other tools that detect concurrency issues \cite{DBLP:conf/sigsoft/PanCLYWYZ20}.  
For our experiments, we added precision for some callbacks from the \codej{Activity} lifecycle, other user interface components, and objects for concurrency such as \codej{AsyncTask}.  
The benefit of our compositional modeling approach is that components may be added on an as-needed basis as opposed to eagerly modeling a large portion of the framework.

% \JEDI{Different ways of writing down models: 1. hard code into analysis (main generator, integrating into analysis), 2. automata (meier 19)}
% Existing techniques for eagerly modeling the callback control flow of an event-driven framework fall into three general techniques: 
% \begin{inparaenum}
%   \item models built into the program semantics,
%   \item code generators that compile a main method to invoke callbacks, and
%   \item automata or graph-based representations of callback control flow.
% \end{inparaenum}
%
%\JEDI{Hard coded models give the most expressiveness and control but are the hardest to maintain.}
Building the model of the framework directly into the program semantics used for the analysis has the advantage that the subsequent abstraction may be precisely chosen based on the modeled behavior.  
Many of the tools that build the model into the analysis capture UI elements, inter-component communication, and the stack like behavior of windows \cite{DBLP:journals/ase/YangWZWSYR18,DBLP:conf/eurosp/CalzavaraGM16,DBLP:conf/pepm/PayetS14,DBLP:conf/cgo/RountevY14}.  
The drawback to building the model directly into the analysis is that adding or updating behaviors~\cite{DBLP:conf/kbse/HuangWLC18} requires modifying the analysis itself.
%
%\JEDI{Artifician main method is the most common but presents challenges with interleaving.}
The most common approach to model callback orders is by generating an artificial \codej{main} method \cite{DBLP:conf/pldi/ArztRFBBKTOM14,DBLP:conf/issta/PanCYMYZ19,DBLP:conf/asplos/HuN18,DBLP:conf/icse/ArztB16,DBLP:conf/ndss/GordonKPGNR}.
%, which is also related to patching the available non-native and non-statically loaded framework code to create a main method~\cite{DBLP:conf/pldi/BlackshearGC15}.
An artificial main method has the advantage that modeling can be decoupled from the program analysis by generating code that enforces a callback order to link with the application.
When analyzing with such a main method, the normal abstraction used by the program analysis captures the callback control flow (e.g., through context sensitivity).
The generation of main methods that can be abstracted precisely is a challenge. Capturing behavior such as arbitrary interleaving between callbacks (e.g., multiple simultaneous activities) can be difficult while avoiding language features that cause imprecision in analysis such as dynamic dispatch.
We note that race detectors often combine some aspects of hard coding the callback control flow into the program semantics with utilizing an artificial main method (often for call graph construction).
%
%\JEDI{automata methods}
Automata and graph based approaches to modeling and abstraction \cite{DBLP:conf/oopsla/BlackshearCS15, DBLP:journals/tse/PerezL21} are compositional and only rely on knowledge of relative order between callbacks.
A difficulty with any modeling approach that eagerly models components is that the more components are modeled, the more likely the model is unsound. Unsoundness is common among any approach we listed that captures some callback order~\cite{DBLP:conf/ecoop/MeierMC19,DBLP:conf/pldi/WangZR16,DBLP:conf/ndss/CaoFBEKVC15}.
Our approach can lessen the risk here by enabling a targeted approach to callback control-flow modeling to avoid modeling more than necessary.
%Our \newls language is related to Lifestate \cite{DBLP:conf/ecoop/MeierMC19} opening the possibility of validating our model against a large set of recorded traces.  Such validation could improve the measurable soundness.

%% file: appendix.tex
\input{appendix-mhpl}

\section{Proofs for Message History Program Logic}
\label{app:mhplproof}
\newcommand{\proofhead}[1]{\textbf{#1}}

% \TODO{define all propositions as things that have not been proven yet or we don't prove because its boring. Could also use "property".}

% \TODO{make theorem like environment for assumptions}

% \TODO{What to call the assumptions that we could prove but is too boring? Call these "propositions" for now.  They can be formalized and proven but not worth the time.}

In this section, we prove the theorems for the message history program logic (MHPL) that we describe in \autoref{sec:applog}.
We make the following assumptions:

% \TODO{rm once in premises}
% \begin{assumption}[Sound Framework Model]
%   If $\hist \in \wfhist$ then $\hist \models \SpecSet$.
%   \label{asm:sound}
% \end{assumption}

\begin{assumption}[Application Step Soundness]

If $\jtransok{\invpstate}{\trans}$ and $\jappeval{\pstate'}{\trans}{\pstate}$ and $\pstate \models_{\SpecSet} \maplookup{\invpstate}{\postof{\trans}}$ then $\pstate' \models_{\SpecSet} \maplookup{\invpstate}{\preof{\trans}}$.
  % If $\jhoare{\absappstate'}{\trans}{\absappstate}$ and $\jeval{\appstate'}{\trans}{\appstate}$ and $\appstate \models \absappstate$ then $\appstate' \models \absappstate'$.
% If $\jtransok[]{\invpstate}{\trans}$ and $\jstep[\prog]{\pstate'}{\pstate}$ and $\pstate \models_{\SpecSet} \maplookup{\invpstate}{\postof{\btrans}}$ then $\pstate' \models_{\SpecSet} \maplookup{\invpstate}{\preof{\btrans}}$.
  \label{asm:apptriplesound}
\end{assumption}

Providing a sound specification is the responsibility of the user of \toolname.

\begin{property}[All Prefixes are Realizable]
  If $\histcons{\hist}{\msg} \in \wfhist$ then $\hist \in \wfhist$.
\end{property}
\begin{property}[Sound Excludes Init]
  If $\notinit{ \abspstate }$ then $\initpstate \not \models_{\SpecSet} \abspstate$.
  \label{prop:soundexcludesinit}
\end{property}

\begin{property}[Entailment Soundness]
 If $\entail{\abspstate}{\abspstate'}$ and $\pstate \models_{\SpecSet} \abspstate'$ then $\pstate \models_{\SpecSet} \abspstate$.
 \label{prop:entailsound}
\end{property}

The sound excludes init \reftxt{Property}{prop:soundexcludesinit} is a consequence of the sound encoding of message history program logics \secref{decision}, a sound first order logic encoding, and a sound SMT solver.

\begin{figure}[H]
  \begin{mathpar}
    \fbox{
      $\pstate \models_{\SpecSet} \abspstate$
    }

    \mklocstate{\mem}{\loc_1} \models_{\SpecSet} \mklocstate{\absmem}{\loc_2} \quad\text{iff}\quad  \loc_1 = \loc_2 \text{ and } \mem \models_{\SpecSet} \absmem

    \\
    \fbox{
      $\mem \models_{\SpecSet} \absmem$
    }
    \\

  %\absmem\bnfdef\mkabsmem{\abstrace}{\absappstore}{\absmsgstack}{\puredom}\bnfalt\absmemFalse\bnfalt\absmemOr{\absmem_1}{\absmem_2}

    \mem \models_{\SpecSet} \absmemOr{\absmem_1}{\absmem_2} \quad\text{iff}\quad \mem \models_{\SpecSet} \absmem_1 \text{ or } \mem \models_{\SpecSet} \absmem_2

    \mkmem{\hist}{\appstore}{\msgstack} \models_{\SpecSet} \mkabsmem{\abstrace}{\absappstore}{\absmsgstack}{\puredom} \quad\text{iff}\quad \text{ there exists $\valua$ such that }\withvalua{\hist} \models_{\SpecSet} \abstrace \text{ and } \withvalua{\appstore} \models \absappstore \text{ and } \withvalua{\msgstack} \models \absmsgstack

    \\
    \fbox{
      $\withvalua{\appstore} \models \absappstore$
    }
    \\

% \text{abstract app stores}\quad\absappstore\bnfdef\absappstoreTrue\bnfalt\absappstoreAnd{\absappstore_1}{\absappstore_2}
  % \bnfalt\absappstorePtsto{\var}{\symval}\bnfalt\cdots
  % \bnfalt\absmemFalse
  % \bnfalt\absmemOr{\absappstore_1}{\absappstore_2}

    \withvalua{\appstore} \models \absappstoreTrue

    \withvalua{\appstore} \models \absappstoreAnd{\absappstore_1}{\absappstore_2} \quad\text{iff}\quad \domof{\absappstore_1} \cap \domof{\absappstore_2} = \emptyset \text{ and } \withvalua{\appstore} \models \absappstore_1 \text{ and } \withvalua{\appstore} \models \absappstore_2

    \withvalua{\appstore} \models \absappstorePtsto{\var}{\symval} \quad\text{iff}\quad \maplookup{\appstore}{\var} = \val \text{ and } \maplookup{\valua}{\symval} = \val

    \withvalua{\appstore} \models \absmemOr{\absappstore_1}{\absappstore_2} \quad\text{iff}\quad \withvalua{\appstore} \models \absappstore_1 \text{ or } \withvalua{\appstore} \models \absappstore_2
    \\

  % \text{abstract boundary stacks}\quad\absmsgstack\bnfdef\absmsgstackemp
  % \bnfalt\absmsgstackcons{\absmsgstack_1}{\absmsgstack_2}
  %\msgcb{\msgid}{\symvalseq}
    \fbox{
      $\withvalua{\msgstack} \models \absmsgstack$
    }
    \\
    \withvalua{\msgstack} \models \absmsgstackemp

    % \withvalua{\msgstack} \models \absmsgstackcons{\absmsgstack_1}{\absmsgstack_2} \quad\text{iff}\quad ???
    \withvalua[\valuaextseq{\valua}{\symval}{\val}]{\msgstackcons{\msgstack}{\msgcb{\msgid}{\valseq}}} \models \absmsgstackcons{\absmsgstack}{\msgcb{\msgid}{\symvalseq}}

    \\
    \fbox{
      $\jstepstar[p]{\initpstate}{\pstate}$
    }

    \infer[c-transitive-step]{
      % \jstep[p]{\pstate}{\pstate'}
      \pstate = \pstate'
    }{
      \jstepstar[p]{\pstate}{\pstate'}
    }

    \infer[c-trnasitive-step-inductive]{
      \jstepstar[p]{\pstate}{\pstate''}
      \\
      \jstep[p]{\pstate''}{\pstate'}
    }{
      \jstepstar[p]{\pstate}{\pstate'}
    }
  \end{mathpar}
  \caption{Extended concretization and transitive step for application only transition system.}
\end{figure}
% \begin{inparaenum}
  % \item \label{lem:excludesinitfol} The first order logic encoding of message history excludes init is sound, i.e. if $\toFOLRel{\directive}{\fol}$ and $\unsat{\histAxiomsFol \land \fol \land \folLen = 0}$ then $\withvalua{\histemp} \not\models \directive$.
  % \item \label{lem:excludesinitfolrest} Similar sound first order logic encodings for excludes init on the app store and boundary stacks exist.
  % \item The first order logic encoding of entailment is sound \TODO{}
  % \item Similar sound first order logic encodings exist for entalment on the app store and boundary stacks exist.
% \end{inparaenum}
% The first order logic encoding is straightforward from the definitions of temporal formula.

\textbf{Lemma 3.1} (\TirName{hoare triple soundness}) If~$\jhoare{\abspstate_{\text{pre}}}{\btrans}{\abspstate_{\text{post}}}$ and $\jeval{\pstate_{\text{pre}}}{\btrans}{\pstate_{\text{post}}}$ such that $\pstate_{\text{post}} \models_{\SpecSet} \abspstate_{\text{post}}$ and $\wfhist \subseteq \wfhistspec$, then $\pstate_{\text{pre}} \models_{\SpecSet} \abspstate_{\text{pre}}$.

\begin{proof}[\proofhead{Proof of \reftxt{Lemma}{lem:stepsound}}]
  By cases on $\mathcal{D} = \jeval{\pstate}{\btrans_{\text{pre}}}{\pstate_{\text{post}}}$

Case $\mathcal{D} =     \infer[c-callback-invoke]{
      %\msgci{\val'}{\msgid'}{...\appval...} \in \hist\;\text{for all $\appval \in \valseq$} \\
      \wfhistjudge{\histcons{\hist}{\msgcb{\msgid}{\valseq}}}
      % \jrealhist{ \histcons{\hist}{\msgcb{\msgid}{\valseq}} }
    }{
      \jeval
        { \mkpstate{\hist}{\appstore}{ \msgstack }{\quies} }
        { \cfedge{\quies}{ \msgcb{\msgid}{\varseq} }{\apploc} }
        { \mkpstate{ \histcons{\hist}{\msgcb{\msgid}{\valseq}} }{ \appstore\fmtseq{\extmap{\var}{\val}} }{ \msgstackcons{\msgstack}{\msgcb{\msgid}{\valseq}} }{\apploc}  }
    }$ 
    \vspace{0.5em}

    The only judgment that applies is: 
    
    $\infer[a-callback-invoke]{
  }{
    \jhoare
      { \mkabspstate{ \abstracecons{\abstrace}{\msgcb{\msgid}{\symvalseq}} }{ \absappstore }{ \absmsgstack }{\quies}{ \pureandseq{\puredom}{ \pureseen{\symval} }} }
      { \cfedge{\quies}{ \msgcb{\msgid}{\varseq} }{\apploc} }
      { \mkabspstate
          { \abstrace }
    { \absappstoreAndSeq{\absappstore}{ \absappstorePtsto{\var}{\symval} } }
    { \absmsgstackcons{\absmsgstack}{\msgcb{\msgid}{\symvalseq}} }
    {\apploc}{\puredom}
      }
  }$

  Let $\abspstate_{\text{post}} = { \mkabspstate
          { \abstrace }
    { \absappstoreAndSeq{\absappstore}{ \absappstorePtsto{\var}{\symval} } }
    { \absmsgstackcons{\absmsgstack}{\msgcb{\msgid}{\symvalseq}} }
    {\apploc}{\puredom}
      }$.

  Let $\abspstate_{\text{pre}} = { \mkabspstate
          { \abstracecons{\abstrace}{\msgcb{\msgid}{\symvalseq}} }
    { \absappstore }
    { \absmsgstack }
    {\quies}{\puredom}
      }$.

  Let $\pstate_{\text{post}} = \mkpstate{ \histcons{\hist}{\msgcb{\msgid}{\valseq}} }{ \appstore\fmtseq{\extmap{\var}{\val}} }{ \msgstackcons{\msgstack}{\msgcb{\msgid}{\valseq}} }{\apploc}  $

  Let $\pstate_{\text{pre}} = \mkpstate{ \hist }{ \appstore }{ \msgstack }{\quies}  $

  We have that $\withvalua[\valua]{\pstate_{\text{post}}} \models_{\SpecSet} \abspstate_{\text{post}}$ holds for $\valua = \valuaextseq{\valua'}{\symval}{\val}$ by the definition of concretization.

  We prove that same assignment, $\valua$ ensures that $\withvalua[\valua]{\pstate_{\text{pre}}} \models_{\SpecSet} \abspstate_{\text{pre}}$.  We break this down into sub-cases for each part of the abstract state:

  \quad Sub Case: If $\withvalua{\histcons{\hist}{\msgcb{\msgid}{\valseq}}} \models_{\SpecSet} \abstrace$ and $\wfhistjudge{\histcons{\hist}{\msgcb{\msgid}{\valseq}}}$ then $\withvalua{\hist} \models_{\SpecSet} \abstracecons{\abstrace}{\msgcb{\msgid}{\symvalseq}}$. %

  \quad\quad $\withvalua{\msgcb{\msgid}{\valseq}} \models \msgcb{\msgid}{\symvalseq}$ since $\valua = \valuaextseq{\valua'}{\symval}{\val}$.  
  
  \quad\quad Since the premise of the lemma states $\wfhist \subseteq \wfhistspec$ then for all $\hist$, $\hist \in \wfhist$ implies $\hist \models \SpecSet$.

  \quad\quad Then from $\wfhistjudge{\histcons{\hist}{\msgcb{\msgid}{\valseq}}}$ we have that $\histcons{\hist}{\msgcb{\msgid}{\valseq}} \models \SpecSet$.
 
  \quad\quad Therefore, $\withvalua{\hist} \models_{\SpecSet} \abstracecons{\abstrace}{\msgcb{\msgid}{\symvalseq}}$ holds by the definition of concretization.

  \quad Sub Case: If $\withvalua{\msgstackcons{\msgstack}{\msgcb{\msgid}{\valseq}}} \models \absmsgstackcons{\absmsgstack}{\msgcb{\msgid}{\symvalseq}}$ then $\withvalua{\msgstack} \models \absmsgstack$.

  \quad\quad Since $\valua = \valuaextseq{\valua'}{\symval}{\val}$ this case holds by the definition of concretization.

  \quad Sub Case: If $\withvalua{\appstore\fmtseq{\extmap{\var}{\val}}} \models \absappstoreAndSeq{\absappstore}{ \absappstorePtsto{\var}{\symval} }$ then $\withvalua{\appstore} \models \absappstore$.

  \quad\quad Since $\valua = \valuaextseq{\valua'}{\symval}{\val}$ this case holds by the definition of concretization.

Case $\mathcal{D} =     \infer[c-callback-return]{
      \msg = \msgcbret{ \maplookup{\appstore}{\var'} }{\msgid}{\valseq} \\
      \valseq = \fmtseq{ \maplookup{\appstore}{\var} } \\
      \wfhistjudge{\histcons{\hist}{\msg}}
      % \jrealhist{ \histcons{\hist}{\msg} }
    }{
      \jeval
        { \mkpstate{\hist}{\appstore}{ \msgstackcons{\msgstack}{\msgcb{\msgid}{\valseq}} }{ \apploc } }
        { \cfedge{\apploc}{ \msgcbret{\var'}{\msgid}{\varseq} }{\quies} }
        { \mkpstate{ \histcons{\hist}{\msg} }{\appstore}{ \msgstack }{\quies}  }
    }$ 
    \vspace{0.5em}

    The only judgment that applies is:

  \infer[a-callback-return]{
    %\symval',\symvalseq \notin \mkabsmem{ \abstrace }{\absappstore}{ \absmsgstack }{\puredom} 
    \absappstore = \absappstoreAndSeq{\absappstoreAnd{\absappstore'}{\absappstorePtsto{\var'}{\symval'}}}{ \absappstorePtsto{\var}{\symval} }
  }{
    \jhoare
      { \mkabspstate
          { \abstracecons{\abstrace}{ \msgcbret{\symval'}{\msgid}{\symvalseq} } }
    %{ \absappstoreAndSeq{ \absappstoreAnd{\absappstore}{ \absappstorePtsto{\var'}{\symval'} } }{ \absappstorePtsto{\var}{-}} }
    { \absappstore }
          { \absmsgstackcons{\absmsgstack}{\msgcb{\msgid}{\symvalseq}} }
          { \apploc }{\puredom}
      }
      { \cfedge{\apploc}{ \msgcbret{\var'}{\msgid}{\varseq} }{\quies} }
      { \mkabspstate{ \abstrace }{\absappstore}{ \absmsgstack }{\quies}{\puredom}  }
  }

  Let $\abspstate_{\text{post}} = { \mkabspstate{ \abstrace }{\absappstore}{ \absmsgstack }{\quies}{\puredom}  }$.

  Let $\abspstate_{\text{pre}} = { \mkabspstate
          { \abstracecons{\abstrace}{ \msgcbret{\symval'}{\msgid}{\symvalseq} } }
    %{ \absappstoreAndSeq{ \absappstoreAnd{\absappstore}{ \absappstorePtsto{\var'}{\symval'} } }{ \absappstorePtsto{\var}{-}} }
    { \absappstoreAnd{\absappstore}{ \absappstorePtsto{\var'}{\symval'} } }
          { \absmsgstackcons{\absmsgstack}{\msgcb{\msgid}{\symvalseq}} }
          { \apploc }{\puredom}
      }$.

  Let $\pstate_{\text{post}} = { \mkpstate{ \histcons{\hist}{\msgcbret{\val'}{\msgid}{\valseq}} }{\appstore}{ \msgstack }{\quies}  }$

  Let $\pstate_{\text{pre}} = { \mkpstate{\hist}{\appstore}{ \msgstackcons{\msgstack}{\msgcb{\msgid}{\valseq}} }{ \apploc } }$

  We have that $\withvalua[\valua]{\pstate_{\text{post}}} \models_{\SpecSet} \abspstate_{\text{post}}$ holds for $\valua = \valuaextseq{\valuaext{\valua'}{\symval'}{\val'}}{\symval}{\val}$ by the definition of concretization.

  We prove that $\withvalua[\valua]{\pstate_{\text{pre}}} \models_{\SpecSet} \abspstate_{\text{pre}}$.

  We break this down into sub-cases for each part of the abstract state:

  \quad Sub Case: If $\withvalua[\valua]{\histcons{\hist}{\msgcbret{\val'}{\msgid}{\valseq}}} \models_{\SpecSet} \abstrace$ then $\withvalua{\hist} \models_{\SpecSet}  \abstracecons{\abstrace}{ \msgcbret{\symval'}{\msgid}{\symvalseq} }$ 

  \quad\quad $\withvalua[\valua]{\msgcbret{\val'}{\msgid}{\valseq}} \models \msgcbret{\symval'}{\msgid}{\symvalseq}$ since $\valua = \valuaextseq{\valuaext{\valua'}{\symval'}{\val'}}{\symval}{\val}$. 
  
  \quad\quad Since the premise of the lemma states $\wfhist \subseteq \wfhistspec$ then for all $\hist$, $\hist \in \wfhist$ implies $\hist \models \SpecSet$.

  \quad\quad Then from $\wfhistjudge{\histcons{\hist}{\msgcbret{\val'}{\msgid}{\valseq}}}$ we have that $\histcons{\hist}{\msgcbret{\val'}{\msgid}{\valseq}} \models \SpecSet$.

  \quad\quad Therefore, $\withvalua{\hist} \models_{\SpecSet} \abstracecons{\abstrace}{\msgcbret{\symval'}{\msgid}{\symvalseq}}$ holds by the definition of concretization.

  \quad Sub Case: If $\withvalua{\msgstack} \models \absmsgstack$ then $\withvalua{\msgstackcons{\msgstack}{\msgcb{\msgid}{\valseq}}} \models \msgstackcons{\msgstack}{\msgcb{\msgid}{\symvalseq}}$.
  
  \quad \quad This case holds by the definition of concretization of the stack.

  \quad Sub Case: If $\withvalua{\valuaextseq{\appstore\extmap{\var'}{\val'}}{\var}{\val}} \models \absappstoreAndSeq{\absappstoreAnd{\absappstore'}{\absappstorePtsto{\var'}{\symval'}}}{ \absappstorePtsto{\var}{\symval}}$ 

  \quad \quad then $\withvalua{\valuaextseq{\appstore\extmap{\var'}{\val'}}{\var}{\val}} \models \absappstoreAndSeq{\absappstoreAnd{\absappstore'}{\absappstorePtsto{\var'}{\symval'}}}{ \absappstorePtsto{\var}{\symval}}$.

  \quad \quad This case holds trivially.

Case $\mathcal{D} =     \infer[c-callin-invoke]{
      \valseq = \fmtseq{ \maplookup{\appstore}{\var} } \\
      \msg = \msgci{ \val' }{\msgid}{\valseq} \\
      %\msgci{\val}{\msgid'}{...\appval...} \in \hist\;\text{if $\val' = \appval$} \\
      %\jseen{\hist}{\val'} \\
      \wfhistjudge{ \histcons{\hist}{\msg} }
      % \jrealhist{ \histcons{\hist}{\msg} }
    }{
      \jeval
        { \mkpstate{\hist}{\appstore}{ \msgstack }{ \apploc } }
        { \cfedge{\apploc}{ \msgci{\var'}{\msgid}{\varseq} }{\apploc'} }
        { \mkpstate{ \histcons{\hist}{\msg} }{\appstore\extmap{\var'}{\val'} }{ \msgstack }{\apploc'}  }
    }$ 
    \vspace{0.5em}

    The only judgment that applies is:

  \infer[a-callin-invoke]{
    \absappstore = \absappstoreAndSeq{ \absappstore' }{ \absappstorePtsto{\var}{\symval} }
  }{
    \jhoare
      { \mkabspstate
          { \abstracecons{\abstrace}{ \msgci{ \symval' }{\msgid}{\symvalseq} } }
    %{ \absappstoreAnd{\absappstore}{ \absappstorePtsto{\var'}{-} } }
    { \absappstore }
    { \absmsgstack }{ \apploc }{\pureand{\puredom}{\pureseen{\symval'}}}
      }
      { \cfedge{\apploc}{ \msgci{\var'}{\msgid}{\varseq} }{\apploc'} }
      { \mkabspstate
          {\abstrace}
          { \absappstoreAnd{\absappstore}{ \absappstorePtsto{\var'}{\symval'} } }
          {\absmsgstack}{\apploc'}{\puredom}
      }
  }

  Let $\abspstate_{\text{post}} = { \mkabspstate
          {\abstrace}
          { \absappstoreAnd{\absappstore}{ \absappstorePtsto{\var'}{\symval'} } }
          {\absmsgstack}{\apploc'}{\puredom}
      }$.

  Let $\abspstate_{\text{pre}} = { \mkabspstate
          { \abstracecons{\abstrace}{ \msgci{ \symval' }{\msgid}{\symvalseq} } }
    %{ \absappstoreAnd{\absappstore}{ \absappstorePtsto{\var'}{-} } }
    { \absappstore }
    { \absmsgstack }{ \apploc }{\pureand{\puredom}{\pureseen{\symval'}}}
      }$.

  Let $\pstate_{\text{post}} = { \mkpstate{ \histcons{\hist}{\msgci{ \val' }{\msgid}{\valseq}} }{\appstore\extmap{\var'}{\val'} }{ \msgstack }{\apploc'}  }$

  Let $\pstate_{\text{pre}} = { \mkpstate{\hist}{\appstore}{ \msgstack }{ \apploc } }$

  We have that $\withvalua[\valua]{\pstate_{\text{post}}} \models_{\SpecSet} \abspstate_{\text{post}}$ holds for any $\valua$ such that $\valua = \valuaext{\valuaextseq{\valua'}{\symval}{\val}}{\symval'}{\val'}$ by the definition of concretization.

  We prove that $\withvalua[\valua]{\pstate_{\text{pre}}} \models_{\SpecSet} \abspstate_{\text{pre}}$.  

  We break this down into sub-cases for each part of the abstract state:

  \quad Sub Case: If $\withvalua[\valua]{\histcons{\hist}{\msgci{ \val' }{\msgid}{\valseq}}} \models_{\SpecSet} \abstrace$ then $\withvalua{\hist} \models_{\SpecSet} \abstracecons{\abstrace}{ \msgci{ \symval' }{\msgid}{\symvalseq}}$

  \quad \quad $\withvalua[\valua]{\msgci{ \val' }{\msgid}{\valseq}} \models_{\SpecSet} \msgci{ \symval' }{\msgid}{\symvalseq}$ since $\valua = \valuaext{\valuaextseq{\valua'}{\symval}{\val}}{\symval'}{\val'}$

  \quad\quad Since the premise of the lemma states $\wfhist \subseteq \wfhistspec$ then for all $\hist$, $\hist \in \wfhist$ implies $\hist \models \SpecSet$.

  \quad\quad Then from $\wfhistjudge{\histcons{\hist}{\msgci{ \val' }{\msgid}{\valseq}}}$ we have that $\histcons{\hist}{\msgci{ \val' }{\msgid}{\valseq}} \models \SpecSet$.

  \quad\quad Therefore, $\withvalua{\hist} \models_{\SpecSet} \abstracecons{\abstrace}{\msgci{ \symval' }{\msgid}{\symvalseq}}$ holds by the definition of concretization.

  \quad Sub Case: If $\withvalua{\msgstack} \models \absmsgstack$ then $\withvalua{\msgstack} \models \absmsgstack$ holds trivially.

  \quad Sub Case: If $\withvalua{\appstore\extmap{\var'}{\val'} } \models \absappstoreAnd{\absappstore}{ \absappstorePtsto{\var'}{\symval'} }$ then $\withvalua{\appstore} \models \absappstore$.

  \quad \quad This case holds by the definition of concretization of app stores.

\end{proof}

\textbf{Lemma 3.2} (\TirName{boundary-step soundness}) If $\jtransok[\SpecSet]{\invpstate}{\btrans}$ and $\jeval{\pstate'}{\btrans}{\pstate}$ such that $\pstate \models_{\SpecSet} \maplookup{\invpstate}{\postof{\btrans}}$ and $\wfhist \subseteq \wfhistspec$, then $\pstate' \models_{\SpecSet} \maplookup{\invpstate}{\preof{\btrans}}$.

\begin{proof}[Proof of \reftxt{Lemma}{lem:stepsound2}]

 By inversion on $\jtransok[\SpecSet]{\invpstate}{\btrans}$ we have $\entail{ \maplookup{\invpstate}{\postof{\btrans}} }{ \abspstate }$ and $\jhoare{\abspstate'}{\btrans}{\abspstate}$ and $\entail{ \abspstate' }{ \maplookup{\invpstate}{\preof{\btrans}} }$.

 By applying \reftxt{Property}{prop:entailsound}, we have $\pstate \models_{\SpecSet} \abspstate$.

 By applying \reftxt{Lemma}{lem:stepsound}, we have $\pstate' \models_{\SpecSet} \abspstate'$.

 By applying \reftxt{Property}{prop:entailsound}, we have $\pstate' \models_{\SpecSet} \maplookup{\invpstate}{\preof{\btrans}}$.

\end{proof}

\textbf{Lemma 3.3} (\TirName{inductive soundness}) If~$\jmaywitness[\prog]{\invpstate}{\loc}{\abspstate}$
and $\jstepstar[p]{\pstate'}{\pstate}$
such that $\pstate \models_{\SpecSet} \abspstate$
and $\wfhist \subseteq \wfhistspec$,
then $\pstate' \models_{\SpecSet} \maplookup{\invpstate}{\locof{\pstate'}}$.

\begin{proof}[\proofhead{Proof of \reftxt{Lemma}{lem:inductivesound}}]

Proof by induction on the derivation $\mathcal{D} =~\jstepstar[p]{\initpstate}{\pstate}$.

Case $\mathcal{D} = \infer[c-transitive-step]{
      \pstate = \pstate'
    }{
      \jstepstar[p]{\pstate'}{\pstate}
    }$

\quad By inversion on \TirName{c-transitive-step}, we have $\pstate' =\pstate$.  

\quad From $\jmaywitness[\prog]{\invpstate}{\loc}{\abspstate}$ and by inversion on \TirName{A-Inductive}, we have  $\entail{ \abspstate }{ \maplookup{\invpstate}{\locof{\abspstate}} }$. %and
    %$\jtransok[\SpecSet]{\invpstate}{\btrans}\;\;\text{for all $\btrans \in \prog$}$ and
    %$\jtransok{\invpstate}{\trans}\;\;\text{for all $\trans \in \prog$}$.
% \quad Since $\pstate = \pstate'$ we have that $ \pstate' = \maplookup{\invpstate}{\locof{\abspstate}}$.
\quad From $\pstate \models_{\SpecSet} \abspstate$ and $\pstate' =\pstate$ and \reftxt{Property}{prop:entailsound} we have $\pstate' \models_{\SpecSet} \maplookup{\invpstate}{\locof{\pstate'}}$ satisfying the consequent of Lemma 3.3.

\vspace{1em}

%     \TODO{Apply property 2, choosing a state in sigma hat means the state is in invariant map}
% \quad By inversion on $\jstep[p]{\pstate'}{\pstate}$ there are two sub cases for application and boundary steps:

Case $\mathcal{D} =     \infer[c-trnasitive-step-inductive]{
      \jstepstar[p]{\pstate'}{\pstate''}
      \\
      \jstep[p]{\pstate''}{\pstate}
    }{
      \jstepstar[p]{\pstate'}{\pstate}
    }$

\quad The inductive hypothesis states: If~$\jmaywitness[\prog]{\invpstate''}{\loc}{\abspstate''}$ and $\pstate'' \models_{\SpecSet} \abspstate''$ and $\jstepstar[p]{\pstate'}{\pstate''}$ then  $\pstate' \models_{\SpecSet} \maplookup{\invpstate''}{\locof{\pstate'}}$. 

\quad From the premise of Lemma 3.3, we assume $\jmaywitness[\prog]{\invpstate}{\loc}{\abspstate}$ and $\pstate \models_{\SpecSet} \abspstate$ and $\jstepstar[p]{\pstate'}{\pstate}$.

\quad By inversion on \TirName{c-transitive-step-inductive} we have $\jstepstar[p]{\pstate'}{\pstate''}$ and $\jstep[p]{\pstate''}{\pstate}$.

By inversion on $\jstep[p]{\pstate''}{\pstate}$ we have two cases for an application and boundary step.

\quad\quad Sub Case \TirName{c-app-stepp}:
\quad\quad\quad By inversion on \TirName{c-app-step} we have $\jeval{\pstate''}{\trans}{\pstate}$ and $\trans \in \prog$.

\quad\quad\quad By inversion from $\jmaywitness[\prog]{\invpstate}{\loc}{\abspstate}$ we have 
    $\entail{ \abspstate }{ \maplookup{\invpstate}{\locof{\abspstate}} }$ and
    $\jtransok{\invpstate}{\trans}\;\;\text{for all $\trans \in \prog$}$.

\quad\quad\quad Applying \reftxt{Assumption}{asm:apptriplesound} to $\jappeval{\pstate''}{\trans}{\pstate}$ and $\jtransok{\invpstate}{\trans}$ and $\trans \in \prog$ we have $\pstate'' \models_{\SpecSet} \maplookup{\invpstate}{\preof{\trans}}$ and $\preof{\trans} = \locof{\pstate''}$.

\quad\quad\quad By the inductive hypothesis, we have $\pstate' \models_{\SpecSet} \maplookup{\invpstate}{\locof{\pstate'}}$.

\quad\quad Sub Case \TirName{c-boundary-step} we have $\jeval{\pstate''}{\btrans}{\pstate}$ and $\btrans \in \prog$.

\quad\quad\quad By inversion from $\jmaywitness[\prog]{\invpstate}{\loc}{\abspstate}$ we have
    $\entail{ \abspstate }{ \maplookup{\invpstate}{\locof{\abspstate}} }$ and
    $\jtransok[\SpecSet]{\invpstate}{\btrans}\;\;\text{for all $\btrans \in \prog$}$.

\quad\quad\quad Applying \reftxt{Lemma}{lem:stepsound2} to $\jeval{\pstate''}{\btrans}{\pstate}$ and $\jtransok{\invpstate}{\btrans}$ and $\btrans \in \prog$ we have $\pstate'' \models_{\SpecSet} \maplookup{\invpstate}{\preof{\btrans}}$.

\quad\quad\quad From the abstract boundary step rules $\preof{\btrans} = \locof{\pstate''}$.

\quad\quad\quad By the inductive hypothesis, we have $\pstate' \models_{\SpecSet} \maplookup{\invpstate}{\locof{\pstate'}}$.

\end{proof}

% \begin{lemma}[excludes init hist FOL encoding sound]
%   "If $\toFOLRel{\directive}{\fol}$ and $\unsat{\histAxiomsFol \land \fol \land \folLen = 0}$ then $\withvalua{\histemp} \not\models \directive$."
%   \label{lem:folsound}
% \end{lemma}
% \TODO{Can we just assume this?}

% \TODO{Inferrence rule is individual thing like tirname a-refute}

% \TODO{"By inversion" gives you top two things from bottom.}

% \TODO{Precise definitions of excludes init and entailment.}

\textbf{Theorem 3.4} (\TirName{refute soundness}) If~$\jrefute[\prog]{\loc}{\abspstate}$ and $\jstepstar[p]{\pstate'}{\pstate}$ and $\pstate \models_{\SpecSet} \abspstate$ and $\wfhist \subseteq \wfhistspec$ then $\pstate' \not = \usebox{\InitStateBox}$.

\begin{proof}[\proofhead{Proof of \autoref{thm:refutesound}}]
% If~$\jrefute[\prog]{\loc}{\abspstate}$ and $\jstepstar[p]{\pstate'}{\pstate}$ and $\pstate \models_{\SpecSet} \abspstate$ then $\pstate' \not = \usebox{\InitStateBox}$.

Proof by contradiction: if we assume that $\jrefute[\prog]{\loc}{\abspstate}$ and $\jstepstar[p]{\initpstate}{\pstate}$ and $\pstate \models_{\SpecSet} \abspstate$.

By inversion on $\jrefute[\prog]{\loc}{\abspstate}$ we know \jmaywitness[\prog]{\invpstate}{\loc}{\abspstate} and \notinit{ \maplookup{\invpstate}{\quies} } then we can derive a contradiction.

From \jmaywitness[\prog]{\invpstate}{\loc}{\abspstate} and $\jstepstar[p]{\initpstate}{\pstate}$ and \reftxt{Lemma}{lem:inductivesound} we have that $\initpstate \models_{\SpecSet} \maplookup{\invpstate}{\quies}$ (note that $\locof{\initpstate} = \quies$).

By \reftxt{Property}{prop:soundexcludesinit} it must be the case that $\initpstate \not \models \maplookup{\invpstate}{\quies}$.

% \text{\TODO{replace with property}}
% From \notinit{ \maplookup{\invpstate}{\quies} }  we have
% $\encodeRel{\maplookup{\invpstate}{\quies}}{\SpecSet}{\directive}$ and $\toFOLRel{\directive}{\fol}$ and 
% $\unsat{\histAxiomsFol \land \fol \land \folLen = 0}$.

% From $\encodeRel{\maplookup{\invpstate}{\quies}}{\SpecSet}{\directive}$ and $\toFOLRel{\directive}{\fol}$ and 
% $\unsat{\histAxiomsFol \land \fol \land \folLen = 0}$ and \reftxt{Theorem}{lem:encode} and our assumptions about the first order logic encodings (\reftxt{Assumption}{lem:excludesinitfol} and \reftxt{Assumption}{lem:excludesinitfolrest}) then $\initpstate \not\models_{\SpecSet} \maplookup{\invpstate}{\quies}$.

From $\initpstate \not\models_{\SpecSet} \maplookup{\invpstate}{\quies}$ and $\initpstate \models_{\SpecSet} \maplookup{\invpstate}{\quies}$ we derive a contradiction. Therefore, any $\pstate \models_{\SpecSet} \abspstate$ cannot be reached from the initial state.

\end{proof}

\section{Proofs for Combining Abstract Message Histories with Callback Control-Flow}
\label{app:atencproof}
% \TODO{typeset this proof}
% \includegraphics[width=\textwidth]{{figures/proof5.1}.png}

% \begin{lemma}[correct encode]
% \hspace{\linewidth}
% \begin{enumerate}
% \item If $\encodeRel{\abstrace}{\SpecSet}{\directive}$ and $\inconc{\withvalua{\hist}}{\abstrace}$ then $\modeldouble{\hist}{\valua} \models \directive$.
% \item If $\encodeRel{\abstrace}{\SpecSet}{\directive}$ and $\modeldouble{\hist}{\valua} \models \directive$ and $\hist \models \SpecSet$ then $\inconc{\withvalua{\hist}}{\abstrace}$.
% \end{enumerate}
% \label{lem:encodeeee}
% \end{lemma}

% \TODO{maybe remove name D1}
% \TODO{maybe rephrase so sentences aren't so repetitive}

\begin{proof}[\proofhead{Proof of Theorem~\ref{lem:encode}}(\ref{lem:encodefwd})]
In the forward direction (\ref{lem:encodefwd}) we proceed by induction on the derivation $\mathcal{D}=~\encodeRel{\abstrace}{\SpecSet}{\directive}$

Case $\mathcal{D} = \infer[temporal-okhist]{ }{
      \encodeRel{\abstracebase}{\SpecSet}{\text{true}}
    }$

\quad If $\encodeRel{\abstracebase}{\SpecSet}{\text{true}}$ and $\inconc{\withvalua{\hist}}{\abstracebase}$ then $\modDir{\withvalua{\hist}}{\text{true}}$. This case holds trivially.

Case $\mathcal{D} = \infer[temporal-hypmsg]{
      \instRel{\absmsg_1}{\SpecSet}{\directive_1'} \\
      \encodeRel{\abstrace_2}{\SpecSet}{\directive_2} \\
      \updRel{\directive_2}{\directive_2'}{\absmsg_1}
    }{
      \encodeRel{\abstracecons{\abstrace_2}{\absmsg_1}}{\SpecSet}{ \directive_1' \wedge \directive_2'}
    }\\$

\quad Let $\mathcal{D}_1$ be the premise $\encodeRel{\abstrace_2}{\SpecSet}{\directive_2}$.

\quad Assume $\inconc{\withvalua{\hist}}{\abstracecons{\absmsg_1}{\abstrace_2}}$.

\quad Pick a concrete message $\msg_1$ such that $\modDir{\withvalua{\msg_1}}{\absmsg_1}$ as concrete message always exists for any $\absmsg$.

\quad By the definition of concretization, $\inconc{}{}$, and since  $\inconc{\withvalua{\hist}}{\abstracecons{\absmsg_1}{\abstrace_2}}$ if $\modDir{\withvalua{\msg_1}}{\absmsg_1}$ then $\inconc{\withvalua{\histcons{\hist}{\msg_1}}}{\abstrace_2}$.

\quad By the inductive hypothesis on $\mathcal{D}_1$ with $\inconc{\withvalua{\histcons{\hist}{\msg_1}}}{\abstrace_2}$, we have $\modDir{\withvalua{\histcons{\hist}{\msg_1}}}{\directive_2}$.

\quad From $\inconc{\withvalua{\histcons{\hist}{\msg_1}}}{\abstrace_2}$, the base case of concretization, and the fact that realizable message histories are closed under prefix, we have that $\histcons{\hist}{\msg_1}\models \SpecSet$.

\quad By Lemma \ref{lem:inst}(\ref{lem:instfwd}), and the premise $\instRel{\absmsg_1}{\SpecSet}{\directive_1'}$ of $\mathcal{D}$, and $\histcons{\hist}{\msg_1}\models \SpecSet$ we have that $\modDir{\withvalua{\hist}}{\directive_1'}$.

\quad By Lemma \ref{lem:quot}(\ref{lem:quotfwd}), and the premise $\updRel{\directive_2}{\directive_2'}{\absmsg_1}$ of $\mathcal{D}$, and $\modDir{\withvalua{\histcons{\hist}{\msg_1}}}{\directive_2}$ we have that $\modDir{\withvalua{\hist}}{\directive_2'}$.

\quad Therefore $\modDir{\withvalua{\hist}}{\directive_1'\wedge\directive_2'}$.

\end{proof}

\begin{proof}[\proofhead{Proof of Theorem~\ref{lem:encode}}(\ref{lem:encodebkg})]
In the backward direction (\ref{lem:encodebkg}) we use Proof by induction on the derivation $\mathcal{D}=\encodeRel{\abstrace}{\SpecSet}{\directive}$

Case $\mathcal{D} =  \infer[temporal-okhist]{ }{
      \encodeRel{\abstracebase}{\SpecSet}{\text{true}}
    }$

\quad If $\encodeRel{\abstracebase}{\SpecSet}{\text{true}}$, $\modDir{\withvalua{\hist}}{\text{true}}$ and $\hist \models \SpecSet$ then $\inconc{\withvalua{\hist}}{\abstracebase}$. This case holds trivially by the definition of $\inconc{}{}$. 

Case $\mathcal{D} = \infer[temporal-hypmsg]{
      \instRel{\absmsg_1}{\SpecSet}{\directive_1'} \\
      \encodeRel{\abstrace_2}{\SpecSet}{\directive_2} \\
      \updRel{\directive_2}{\directive_2'}{\absmsg_1}
    }{
      \encodeRel{\abstracecons{\abstrace_2}{\absmsg_1}}{\SpecSet}{ \directive_1' \wedge \directive_2'}
    }\\$

\quad Let $\mathcal{D}_1$ be the premise $\encodeRel{\abstrace_2}{\SpecSet}{\directive_2}$.

\quad Assume $\modDir{\withvalua{\hist}}{\directive_1' \wedge \directive_2'}$.

\quad Pick $\msg_1$ such that $\modDir{\withvalua{\msg_1}}{\absmsg_1}$ as a message always exists for any $\absmsg$.

\quad From $\modDir{\withvalua{\hist}}{\directive_1' \wedge \directive_2'}$ we have $\withvalua{\hist}\models\directive_1'$.

\quad From Lemma \ref{lem:inst} (\ref{lem:instbkg}), $\hist\models\SpecSet$, $\modDir{\withvalua{\msg_1}}{\absmsg_1}$, the premise $\instRel{\absmsg_1}{\SpecSet}{\directive_1'}$, and $\withvalua{\hist}\models\directive_1'$, we have that $\histcons{\hist}{\msg_1}\models \SpecSet$.

\quad From $\modDir{\withvalua{\hist}}{\directive_1' \wedge \directive_2'}$ we have $\withvalua{\hist}\models\directive_2'$.

\quad From Lemma \ref{lem:quot} (\ref{lem:quotbkg}), $\modDir{\withvalua{\msg_1}}{\absmsg_1}$, and the last premise of $\mathcal{D}$,  $\updRel{\directive_2}{\directive_2'}{\absmsg_1}$, we have that $\modDir{\histcons{\hist}{\msg_1}}{\directive_2}$.

\quad By the induction hypothesis on $\mathcal{D}_1$, $\histcons{\hist}{\msg_1}\models \SpecSet$, and $\modDir{\withvalua{\histcons{\hist}{\msg_1}}}{\directive_2}$, we have that $\inconc{\withvalua{\histcons{\hist}{\msg_1}}}{\abstrace_2}$.

\quad From the right to left direction of the definition of concretization $\inconc{}{}$, message $\msg_1$ can be appended and model the abstract trace $\inconc{\withvalua{\histcons{\hist}{\msg_1}}}{\abstrace_2}$ and $\msg_1$ is in the abstract message $\modDir{\withvalua{\msg_1}}{\absmsg_1}$, we have that $\inconc{\withvalua{\hist}}{\abstracecons{\abstrace_2}{\absmsg_1}}$.

\end{proof}

Note that the following two proofs are inductive due to the ``and'' case of the specification, $\SpecSet$.

\begin{proof}[\proofhead{Proof of Lemma \ref{lem:inst}} in the forward direction \ref{lem:instfwd}]
  Proof by induction over the derivation $\mathcal{D}=\instRel{\absmsg}{\SpecSet}{\directive}$.

  Case $\mathcal{D}=\infer[instantiate-yes]{
      %\SpecSet = \specOnly{\absmsg_2}{\directive} \\
      \ideq{\absmsg_1}{\absmsg_2} %\\ \instRel{\absmsg_1}{\SpecSet'}{\directive'}
    }{
      \instRel{\absmsg_1}{(\specOnly{\absmsg_2}{\directive})}{[\unifysubst]\directive}
    }$

  \quad From the lemma we have a history with a message appended models the specification $\histcons{\hist}{\msg_1}\models(\specOnly{\absmsg_2}{\directive})$, there is a message in the abstract message $\withvalua[\valua_1]{\msg_1}\models\absmsg_1$. 

  \quad From the premise $\ideq{\absmsg_1}{\absmsg_2}$ we have the capture avoiding substitution $\unifysubst$ such that $\absmsg_1$ is equal to $[\unifysubst]\absmsg_2$.

  \quad From the semantics of \newls in \autoref{fig:cbcfsemantics}, we have that $\withvalua[\valua_2]{\msg_2} \models \absmsg_2$ such that $\hist[i] = \msg_2$ implies $\modeltriple{\hist}{\valua_2}{i-1} \models \directive$.

  % \quad Pick $\msg_2$ such that $\modDir{\withvalua{\msg_2}}{\absmsg_1}$.

  \quad Since the free variables of $\directive$ are a subset or equal to the free variables of $\absmsg_2$, the substitution, $[\unifysubst]\directive$, captures all free variables of $\directive$. Since this substitution is capture avoiding, all the relevant bindings in $\valua_2$ are swapped for the corresponding $\absmsg_1$ values and we can apply the implication from the \newls semantics getting $\modeltriple{\histcons{\hist}{\msg_1}}{\valua_1}{\histlen{\hist}-1} \models [\unifysubst]\directive$ which is equivalent to our theorem goal, $\modeltriple{\hist}{\theta_1}{\histlen{\hist}} [\unifysubst]\models\directive$.

  % \TODO{prove} $\modDir{\withvalua[\valua_1]{\hist}}{[\unifysubst]\directive}$

  Case $\mathcal{D}=\infer[instantiate-no]{
      %\SpecSet = \specOnly{\absmsg_2}{\directive} \\
      \idneq{\absmsg_1}{\absmsg_2} %\\ \instRel{\absmsg_1}{\SpecSet'}{\directive}
    }{
      \instRel{\absmsg_1}{(\specOnly{\absmsg_2}{\directive})}{\text{true}}
    }\\$

    \quad Trivially, $\modeltriple{\hist}{\valua}{\histlen{\hist}}\models \text{true}$.

  Case $\mathcal{D}=\infer[instantiate-and]{
      \instRel{\absmsg_1}{(\SpecSet_1)}{\directive_1} \\ 
      \instRel{\absmsg_1}{(\SpecSet_2)}{\directive_2} \\ 
    }{
      \instRel{\absmsg_1}{(\SpecSet_1 \wedge \SpecSet_2)}{\directive_1 \wedge \directive_2}
    }$

    \quad Since the semantics of intended for $\wedge$ (elided from figure \ref{fig:cbcfsemantics} for brevity) are that it is a conjunction of each instantiated specification, this case holds trivially.

  Case $\mathcal{D}=\infer[instantiate-true]{ }{
      \instRel{\absmsg_1}{\text{true}}{\text{true}}
  }$

   \quad Since any history, $\hist$, is feasible under the ``true'' specification, this case holds trivially.
\end{proof}

\begin{proof}[\proofhead{Proof of Lemma \ref{lem:inst}} in the backward direction \ref{lem:instbkg}]
  Proof by induction over the derivation $\mathcal{D}=\instRel{\absmsg}{\SpecSet}{\directive}$.

  Case $\mathcal{D}=\infer[instantiate-yes]{
      %\SpecSet = \specOnly{\absmsg_2}{\directive} \\
      \ideq{\absmsg_1}{\absmsg_2} %\\ \instRel{\absmsg_1}{\SpecSet'}{\directive'}
    }{
      \instRel{\absmsg_1}{(\specOnly{\absmsg_2}{\directive})}{[\unifysubst]\directive}
    }$
  
  \quad From the lemma, we have that $\withvalua[\valua_1]{\hist}\models(\specOnly{\absmsg_2}{\directive})$, $\withvalua[\valua_1]{\msg_1}\models\absmsg_1$, and $\withvalua[\valua_1]{\hist}\models[\unifysubst]\directive$.

  \quad From the premise $\ideq{\absmsg_1}{\absmsg_2}$ we have the capture avoiding substitution $\unifysubst$ such that $\absmsg_1$ is equal to $[\unifysubst]\absmsg_2$.

  \quad From the semantics of \newls in \autoref{fig:cbcfsemantics}, we have that $\withvalua[\valua_2]{\msg_2} \models \absmsg_2$ such that $\hist[i] = \msg_2$ implies $\modeltriple{\hist}{\valua_2}{i-1} \models \directive$.

  \quad By similar reasoning as the proof for Lemma \ref{lem:inst} (\ref{lem:instfwd}) we know that any history and assignment such that $\withvalua[\valua_1]{\hist}\models[\unifysubst]\directive$ captures the variables shared between $[\unifysubst]\absmsg_2$ and $[\unifysubst]\directive$ so any message such that $\withvalua[\valua_1]{\msg}\models[\unifysubst]\absmsg_2$ may be appended to a message history, $\hist$ such that $\hist\models(\specOnly{\absmsg_2}{\directive})$ and we know that $\histcons{\hist}{\msg}\models(\specOnly{\absmsg_2}{\directive})$.

  The other cases are similar to the forward proof.

\end{proof}

For the following proofs, we give the definition of $\msgequiv{\ltmsg}{\absmsg}$ as the following:
\begin{enumerate}
\item \label{def:meq} If the names and message types of $\ltmsg$ and $\absmsg$ are the same, then $\msgequiv{\ltmsg}{\absmsg}$ results in a conjunction of comparisons between each free variable of $\ltmsg$ and the corresponding $\absmsg$ variable.
\item \label{def:meqno} If the names or message types of $\ltmsg$ and $\absmsg$ are different, then $\msgequiv{\ltmsg}{\absmsg}$ is false.
\end{enumerate}
We define $\msgnotequiv{\ltmsg}{\absmsg}$ as the negation of $\msgequiv{\ltmsg}{\absmsg}$.

\begin{proof}[\proofhead{Proof of Lemma \ref{lem:quot}} in the forward direction \ref{lem:instfwd}]
  Proof by induction on the derivation $\mathcal{D}=\updRel{\directive}{\directive'}{\absmsg}$. We reference the \newls semantics in \autoref{fig:cbcfsemantics} and the judgments in \autoref{fig:atencode}.

  \quad In the following, let $\fv{\msg} = \fv{\ltmsg}$ be the conjunction comparing each variable not quantified in the expansion of $\ltmsg$ into its definition $\exists \symvalseq.\absmsg'$ to the corresponding value of $\msg$.

  Case $\mathcal{D}=\infer[quotient-once]{
      %\directive = (\msgnotequiv{\ltmsg}{\absmsg} \wedge \iDir{\ltmsg}) \vee \msgequiv{\ltmsg}{\absmsg}
    }{
      \updRel{\iDir{\ltmsg}}{
	\msgequiv{\ltmsg}{\absmsg} \vee (\msgnotequiv{\ltmsg}{\absmsg} \wedge \iDir{\ltmsg})
      }{\absmsg}
    }$

  \quad From the lemma we have that, $\modDir{\withvalua{\histcons{\hist}{\msg}}}{\iDir{\ltmsg}}$ and $\withvalua{\msg}\models\absmsg$. There are two subcases to consider:

  \quad Subcase 1: the message names and types are equal, $\ideq{\msg}{\ltmsg}$.

  \quad\quad From the semantics of Once, $\fv{\msg}{\ltmsg}$ implies $\withvalua{\hist} \models \iDir{\ltmsg}$.  By the definition of message equivalence (\ref{def:meq}), this is equivalent to $\msgequiv{\ltmsg}{\absmsg} \vee (\msgnotequiv{\ltmsg}{\absmsg} \wedge \iDir{\ltmsg})$

  \quad Subcase 2: the names or message types are not equal, $\idneq{\msg}{\ltmsg}$ are not equal.  

  \quad\quad From the semantics of Once, $\withvalua{\hist} \models \iDir{\ltmsg}$ must hold since $\msg$ cannot match $\ltmsg$.

  \quad\quad Under the assumption that the name or message type does not match $\msgequiv{\ltmsg}{\absmsg} \vee (\msgnotequiv{\ltmsg}{\absmsg} \wedge \iDir{\ltmsg}) \equiv \iDir{\ltmsg}$ by the definition of message equivalence (\ref{def:meqno}).
  
  Case $\mathcal{D}=\infer[quotient-historically-not]{ }{
      \updRel{\notiDir{\ltmsg}}{
	\notiDir{\ltmsg} \wedge \msgnotequiv{\ltmsg}{\absmsg}
      }{\absmsg}
    }$

  \quad From the lemma we have that, $\modDir{\withvalua{\histcons{\hist}{\msg}}}{\notiDir{\ltmsg}}$ and $\withvalua{\msg}\models\absmsg$.

  \quad\quad From the semantics of Has Not, $\withvalua{\hist}\models\notiDir{\ltmsg}$. This holds for the formula  $\notiDir{\ltmsg} \wedge \msgnotequiv{\ltmsg}{\absmsg}$.

  Case $\mathcal{D}= \infer[quotient-not-since]{
      %\directive = (\msgequiv{\ltmsg_2}{\absmsg} \vee (\msgnotequiv{\ltmsg_2}{\absmsg} \wedge \niDir{\ltmsg_1}{\ltmsg_2})) \wedge \msgnotequiv{\ltmsg_1}{\absmsg}
    }{
      \updRel{\niDir{\ltmsg_2}{\ltmsg_1}}{
        \msgequiv{\ltmsg_1}{\absmsg} \vee (\msgnotequiv{\ltmsg_1}{\absmsg} \wedge \niDir{\ltmsg_2}{\ltmsg_1} \wedge \msgnotequiv{\ltmsg_2}{\absmsg})
      }{\absmsg}
    }$

  \quad From the lemma we have that, $\modDir{\withvalua{\histcons{\hist}{\msg}}}{\niDir{\ltmsg_1}{\ltmsg_2}}$ and $\withvalua{\msg}\models\absmsg$. There are three subcases to consider:

  \quad Subcase 1: the message names and types are equal between the positive message, $\ideq{\msg}{\ltmsg_2}$. 

  \quad\quad From the semantics of Not Since: not $\fv{\msg}=\fv{\ltmsg_2}$ implies $\withvalua{\hist}\models\niDir{\ltmsg_1}{\ltmsg_2}$. This is implied by the formula $\msgequiv{\ltmsg_1}{\absmsg} \vee (\msgnotequiv{\ltmsg_1}{\absmsg} \wedge \niDir{\ltmsg_2}{\ltmsg_1} \wedge \msgnotequiv{\ltmsg_2}{\absmsg})$ since the first disjunction is false due to the definition of message equivalence.

  \quad Subcase 2: the message names and types are equal between the negative message, $\ideq{\msg}{\ltmsg_1}$. 

  \quad\quad From the semantics of Not Since: not $\fv{\msg}=\fv{\ltmsg_1}$ implies $\withvalua{\hist}\models\niDir{\ltmsg_1}{\ltmsg_2}$.  This is implied by the formula $\msgequiv{\ltmsg_1}{\absmsg} \vee (\msgnotequiv{\ltmsg_1}{\absmsg} \wedge \niDir{\ltmsg_2}{\ltmsg_1} \wedge \msgnotequiv{\ltmsg_2}{\absmsg})$ since the first disjunction is false due to the definition of message equivalence.

  \quad Subcase 3: the message names and types are not equal to either message, $\idneq{\msg}{\ltmsg_1}$ and $\idneq{\msg}{\ltmsg_2}$.

  \quad\quad From the semantics of Not Since: $\withvalua{\hist}\models\niDir{\ltmsg_1}{\ltmsg_2}$.  This is implied by the formula $\msgequiv{\ltmsg_1}{\absmsg} \vee (\msgnotequiv{\ltmsg_1}{\absmsg} \wedge \niDir{\ltmsg_2}{\ltmsg_1} \wedge \msgnotequiv{\ltmsg_2}{\absmsg})$ since $\msgequiv{\ltmsg_1}{\absmsg}$ and $\msgequiv{\ltmsg_2}{\absmsg}$ are both false by the definition of message equivalence.

  Case $\mathcal{D}= \infer[quotient-and]{
    \updRel{\directive_1}{\directive_1'}{\absmsg} \\ \updRel{\directive_2}{\directive_2'}{\absmsg}
  }{
    \updRel{\directive_1 \wedge \directive_2}{
      \directive_1' \wedge \directive_2'
    }{\absmsg}
  }$  

  \quad From the lemma and the judgment, quotient-and, we have $\withvalua{\msg}\models\absmsg$, $\withvalua{\histcons{\hist}{\msg}}\models \directive_1 \wedge \directive_2$, $\withvalua{\histcons{\hist}{\msg}}\models \directive_1$ implies $\withvalua{\hist}\models\directive_1'$, and $\withvalua{\histcons{\hist}{\msg}}\models \directive_2$ implies $\withvalua{\hist}\models\directive_2'$.

  \quad By the semantics of and (elided from \autoref{fig:cbcfsemantics}), $\withvalua{\histcons{\hist}{\msg}}\models \directive_1 \wedge \directive_2$, and the two implications, we have that $\withvalua{\hist}\models{\directive_1' \wedge \directive_2'}$.

  The cases for $\vee$, $\forall \symval.\directive$, and $\exists \symval.\directive$ are similar to $\wedge$.

  For all the following productions, the message history is not affected: $\symval_1 = \symval_2$, $\symval_1 \neq \symval_2$, and ``\text{true}''.  Therefore, they model any history $\withvalua{\hist}$.

  For the case ``false'', it cannot model $\withvalua{\hist}$ therefore holds vacuously.

\end{proof}

\begin{proof}[\proofhead{Proof of Lemma \ref{lem:quot}} in the backward direction \ref{lem:instbkg}]
  Proof by induction on the derivation $\mathcal{D}=\updRel{\directive}{\directive'}{\absmsg}$. We reference the \newls semantics in \autoref{fig:cbcfsemantics} and the judgments in \autoref{fig:atencode}.

  \quad Let $\fv{\msg} = \fv{\ltmsg}$ be the conjunction of comparing each variable not quantified in the expansion of $\ltmsg$ into its definition $\exists \symvalseq.\absmsg'$ to the corresponding value of $\msg$.

  Case $\mathcal{D}=\infer[quotient-once]{
      %\directive = (\msgnotequiv{\ltmsg}{\absmsg} \wedge \iDir{\ltmsg}) \vee \msgequiv{\ltmsg}{\absmsg}
    }{
      \updRel{\iDir{\ltmsg}}{
	\msgequiv{\ltmsg}{\absmsg} \vee (\msgnotequiv{\ltmsg}{\absmsg} \wedge \iDir{\ltmsg})
      }{\absmsg}
    }$

  \quad From the lemma we have that $\withvalua{\hist}\models\msgequiv{\ltmsg}{\absmsg} \vee (\msgnotequiv{\ltmsg}{\absmsg} \wedge \iDir{\ltmsg})$ and $\withvalua{\msg}\models{\absmsg}$. We consider two different subcases, in the case the message names are the same or not.

  \quad Assume the message names matches, $\ideq{\msg}{\ltmsg}$. It is either the case that the arguments match, $\fv{\msg} = \fv{\ltmsg}$, and $\valua(\ltmsg)=\msg$ making $i$ for Once equals $\histlen{\histcons{\hist}{\msg}}$ or the arguments don't match, not $\fv{\msg} = \fv{\ltmsg}$, and the $i$ for once is in $\hist$.  In both cases $\withvalua{\histcons{\hist}{\msg}}\models \iDir{\ltmsg}$.

  \quad Assume the message names do not match, $\idneq{\msg}{\ltmsg}$. In this case $\msgequiv{\ltmsg}{\absmsg}$ is false so the $i$ for once must be in $\hist$, therefore $\withvalua{\histcons{\hist}{\msg}}\models \iDir{\ltmsg}$.

  Case $\mathcal{D}=\infer[quotient-historically-not]{ }{
      \updRel{\notiDir{\ltmsg}}{
	\notiDir{\ltmsg} \wedge \msgnotequiv{\ltmsg}{\absmsg}
      }{\absmsg}
    }$

    \quad From the lemma we have that $\withvalua{\hist}\models\notiDir{\ltmsg} \wedge \msgnotequiv{\ltmsg}{\absmsg}$ and $\withvalua{\msg}\models{\absmsg}$.

    \quad From the definition of message not equals, we know either $\idneq{\msg}{\ltmsg}$ or not $\fv{\msg} = \fv{\ltmsg}$ so $\valua(\ltmsg) \neq \msg$.

    \quad From the definition of Historically Not we have that $\withvalua{\hist} \models \notiDir{\ltmsg}$.

    \quad Since we have $\valua(\ltmsg) \neq \msg$ and $\withvalua{\hist} \models \notiDir{\ltmsg}$, by the definition of Historically Not, we have that $\withvalua{\histcons{\hist}{\msg}}\models \notiDir{\ltmsg}$.

    Case $\mathcal{D}= \infer[quotient-not-since]{
      %\directive = (\msgequiv{\ltmsg_2}{\absmsg} \vee (\msgnotequiv{\ltmsg_2}{\absmsg} \wedge \niDir{\ltmsg_1}{\ltmsg_2})) \wedge \msgnotequiv{\ltmsg_1}{\absmsg}
    }{
      \updRel{\niDir{\ltmsg_2}{\ltmsg_1}}{
        \msgequiv{\ltmsg_1}{\absmsg} \vee (\msgnotequiv{\ltmsg_1}{\absmsg} \wedge \niDir{\ltmsg_2}{\ltmsg_1} \wedge \msgnotequiv{\ltmsg_2}{\absmsg})
      }{\absmsg}
    }$  

    \quad From the lemma we have that $\withvalua{\hist}\models\msgequiv{\ltmsg_1}{\absmsg} \vee (\msgnotequiv{\ltmsg_1}{\absmsg} \wedge \niDir{\ltmsg_2}{\ltmsg_1} \wedge \msgnotequiv{\ltmsg_2}{\absmsg})$ and $\withvalua{\msg}\models{\absmsg}$. We have two subcases:

    \quad Subcase $\ideq{\msg}{\ltmsg_2}$ and $\fv{\msg}=\fv{\ltmsg_2}$: In this case the $i$ from the definition of Not Since is at the last message, $\msg$, and $\withvalua{\histcons{\hist}{\msg}}\models\niDir{\ltmsg_1}{\ltmsg_2}$.

    \quad Subcase $\idneq{\msg}{\ltmsg_2}$ or $\fv{\msg}\neq\fv{\ltmsg_2}$: In this case, we have $\msgnotequiv{\ltmsg_1}{\absmsg}$ so the $i$ in Not Since must be in $\hist$. 

    \quad by the definition of message not equals, we know that $\idneq{\ltmsg_1}{\absmsg}$ or $\fv{\absmsg} \neq \fv{\ltmsg_1}$ so $\msg$ must not be equal to $\ltmsg$ under assignment $\valua$ and therefore $\withvalua{\histcons{\hist}{\msg}} \models \niDir{\ltmsg_1}{\ltmsg_2}$.

    Case $\mathcal{D}= \infer[quotient-and]{
      \updRel{\directive_1}{\directive_1'}{\absmsg} \\ \updRel{\directive_2}{\directive_2'}{\absmsg}
    }{
      \updRel{\directive_1 \wedge \directive_2}{
        \directive_1' \wedge \directive_2'
      }{\absmsg}
    }$  

    \quad From the lemma and the quotient-and judgment, we have that $\withvalua{\msg}\models\absmsg$, $\withvalua{\hist}\models \directive_1' \wedge \directive_2'$, $\withvalua{\hist}\models{\directive_1'}$ implies $\withvalua{\histcons{\hist}{\msg}}\models{\directive_1}$, and $\withvalua{\hist}\models{\directive_2'}$ implies $\withvalua{\histcons{\hist}{\msg}}\models{\directive_2}$.

    \quad By the straightforward semantics of and (elided from \autoref{fig:cbcfsemantics}), $\withvalua{\hist}\models \directive_1 \wedge \directive_2$, and the two implications, we have that $\withvalua{\histcons{\hist}{\msg}}\models{\directive_1 \wedge \directive_2}$.

   The cases for $\vee$, $\forall \symval.\directive$, and $\exists \symval.\directive$ are similar to $\wedge$.

   For all the following productions, the message history is not affected: $\symval_1 = \symval_2$, $\symval_1 \neq \symval_2$, and ``\text{true}''.  Therefore, they model any history $\withvalua{\histcons{\hist}}$.

   For the case ``false'', it cannot model $\withvalua{\histcons{\hist}{\msg}}$ therefore holds vacuously.

\end{proof}

\section{Extended Explanations of the Benchmark Applications and Specifications}
\label{app:speclist}

Here, we give a listing of all the specifications we wrote for \secref{eval} RQ1, 
details on the crashes, and reasons the \toolflowdroid model would not be able to classify the benchmarks correctly.

The first benchmark, \apGa, is a slightly more complex version of the motivating example in \secref{overview}.  
The main difference is that instead of an \codej{act} field, the app invokes the callin \codej{getActivity}.
Calling \codej{getActivity} before \codej{onCreate} or after \codej{onDestroy} results in a null value as captured by History Implication \ref{spec:getActivityNull}.
As a minor difference, \apGa uses the \codej{onActivityCreated} callback instead of \codej{onDestroy} which has slightly more complicated behavior (History Implication \ref{spec:onActivityCreated}).
The behavior of the \codej{call} callback was explained in \secref{overview} with History Implication \ref{spec:call}.

\begin{specDef}\label{spec:createOnce}
For all \codej{f}, if the framework invokes \codej{f.onCreate()}, the same message $\enkwCb~\codej{f.onCreate()}$ is \textbf{H}istorially \textbf{N}ot possible (or, \textbf{H}as \textbf{N}ever been invoked in the past).
\specbox{\createOnceSpec}
\end{specDef}

\begin{specDef} The method \codej{getActivity} returns null when invoked on an \codej{Activity} in the paused state.

  \begin{small}
    \begin{align*}
	\getActivityNullSpec
    \end{align*}
  \end{small}
\label{spec:getActivityNull}
\end{specDef}

\begin{specDef} For a given \codej{Fragment} instance, either \codej{onActivityCreated} has not been invoked or \codej{onDestroy} has not been invoked since \codej{onActivityCreated}.

\begin{small}
  \begin{align*}
    \ensuremath{\enkwCb~\codej{f.onActivityCreated()}} \boxright & \ensuremath{\notiDir{\enkwCb\enkwRet~\codej{f.onDestroy()}} \wedge \\
    & \notiDir{\enkwCb~\codej{f.onActivityCreated()}} \wedge \notiDir{\enkwCb~\codej{f.onActivityCreated()}}}
  \end{align*}
\end{small}

	\label{spec:onActivityCreated}
\end{specDef}

\toolflowdroid would not alarm on the buggy version of \apGa because \codej{call} was not in the call graph making the assertion unreachable.
If \codej{call} was added to the framework model generation (e.g. by \citet{DBLP:conf/ndss/CaoFBEKVC15}), then the \toolflowdroid model would not be able to capture the effect of \codej{unsubscribe} (i.e. History Implication \ref{spec:call}) resulting in a false positive for the fix.

The second benchmark, \apEx, has a button that starts an \codej{AsyncTask} to perform an action in the background using the \codej{execute} callin.
\codej{AsyncTask} is an abstract class that can be overridden by the app to encapsulate long running tasks (similar to \codej{Single}).
In order to prevent concurrency issues in the state of the overridden class, the framework enforces that \codej{execute} crashes if called twice on the same instance of the overridden class.
We capture this exceptional return with History Implication \ref{spec:execute}.
If the button was clicked twice quickly, the task could be executed twice crashing the application.
The fix was to disable the button using \codej{b.setEnabled(false)} (History Implication \ref{spec:clickDisable}).
Additionally, if the listener could be registered to two different buttons via two calls to \codej{onCreate} which calls \codej{setOnClickListener}, then the second button could be pressed crashing the app.
To rule out this case, we needed History Implication \ref{spec:createOnce} (also needed by \apGa).

\begin{specDef} For any given instance of \codej{AsyncTask}, the \codej{execute} method throws an exception if \codej{execute} has been invoked in the past.

\begin{small}
  \begin{align*}
	\enkwErr~\executeSpec
  \end{align*}
\end{small}
	\label{spec:execute}
\end{specDef}

\begin{specDef}
Every time the \codej{onClick} callback occur, the associated button has not been disabled.
\begin{small}
  \begin{align*}
    \enkwCb~\codej{l.onClick()} \boxright
      \exists\ \codej{v}.\ &
        \left( \iDir{\enkwCi~\codej{v.setOnClickListener(l)}} \land \right. \\
        & \left. (\notiDir{\enkwCi~\codej{v.setEnabled(} \enkwFalse \codej{)}} \lor
                  \niDir{\enkwCi~\codej{v.setEnabled(}\enkwFalse\codej{)}} {\enkwCi~\codej{v.setEnabled(}\enkwTrue \codej{)}} ) \right)
  \end{align*}
  \label{spec:clickDisable}
\end{small}
\end{specDef}

The fix for \apEx would be misclassified by \toolflowdroid's model because the effect of the callin \codej{b.setEnable(false)} on the button is not captured.

The third benchmark, \ymDi, uses a background task (via \codej{AsyncTask}) and when the task finishes, the \codej{onPostExecute} callback dismisses a \codej{ProgressDialog}. If the \codej{ProgressDialog} is dismissed after the parent has been paused, it throws an exception (History implication \ref{spec:dismiss}).
The fix was to check if the parent UI was visible before dismissing the dialog using the application state.
Additionally, we needed to restrict the method used to create the dialog, \codej{show}, to return a fresh dialog instance each time (History Implication \ref{spec:show}).

\begin{specDef} The \codej{dismiss} callin throws an exception if invoked while the attached \codej{Activity} is paused.

	\begin{small}
    \begin{align*}
		\enkwErr~\dismissSpec		
    \end{align*}
	\end{small}
	\label{spec:dismiss}
\end{specDef}

\begin{specDef} When the \codej{show} callin returns an object, it must have been the case that the object was not returned by a different \codej{show} in the past.
 
  \begin{small}
    \begin{align*}
	\showSpec
    \end{align*}
  \end{small}

\label{spec:show}
\end{specDef}

The bug for \ymDi would not be detected by \toolflowdroid because the \codej{onPostExecute} callback was not in the call graph.
Adding \codej{onPostExecute} to the \toolflowdroid would result in a false alarm on the fix because the interaction between \codej{show}, \codej{dismiss}, \codej{onPause}, and \codej{onResume} captured by History Implication \ref{spec:dismiss} could not be captured due to the callins.

The fourth benchmark, \cbFi, has a button that dereferences a field in a \codej{onClick} callback.  This field is set to \codej{null} when the \codej{Activity} is paused (via \codej{onPause}) to save memory.  However, via what may be considered a bug in some versions of the Android framework itself, calling \codej{finish} on an \codej{Activity} can result in the \codej{onClick} callback occurring after \codej{onPause} (History Implication \ref{spec:clickFinish}).
Similar to other benchmarks, there is a \codej{onCreate} that registers the button and can only happen once (History Implication \ref{spec:createOnce}).
Since an \codej{Activity} may have multiple simultaneous instances in an Android app, we also need a spec that says a button may only come from one \codej{Activity} instance (History Implication \ref{spec:findView}).

\begin{specDef}
Every time the \codej{onClick} callback happens on the listener object \codej{l}, the listener was registered on a \codej{Button} object \codej{v}, and the \codej{Activity} object \codej{a} was either in the ``resumed state'' (i.e., after a \codej{onResume} but before than a \codej{onPause}) or the message \codej{finish} happened.

  \begin{small}
    \begin{align*}
      \enkwCb~\codej{l.onClick()} \boxright
        \exists\ \codej{a},\codej{v}.\  &
          \left(\niDir{\enkwCi~\codej{v.setOnClickList(}\enkwNull\codej{)}}{\enkwCi~\codej{v.setOnClickList(l)}} \land \right. \\
          & \left. \iDir{\specciwithret{\codej{v}}{\codej{a}}{findViewById}{\codej{_}}} \land  \right. \\
          & \left. \left( \niDir{\enkwCb\enkwRet~\codej{a.onPause()}}{\enkwCb~\codej{a.onResume()}} \vee \iDir{\enkwCi~\codej{a.finish()}} \right)  \right)
    \end{align*}
  \end{small}
  \label{spec:clickFinish}
\end{specDef}

\begin{specDef} 
Every time a \codej{findViewById} on an \codej{Activity} object \codej{a} returns a \codej{View} object \codej{v},
any previous invocation of \codej{findViewById} that returned \codej{v} was invoked on the same \codej{Activity} \code{a}.
  \begin{small}
  \begin{align*}
    \specciwithret{\codej{v}}{\codej{a}}{findViewById}{\codej{id}} \boxright \forall\ \codej{a2} .\ 
    \left( \notiDir{\specciwithret{\codej{v}}{\codej{a2}}{findViewById}{\codej{_}}}  \vee \codej{a} = \codej{a2} \right)
  \end{align*}
  \end{small}
  \label{spec:findView}
\end{specDef}

The bug for \cbFi would be missed by \toolflowdroid's model because this model assumes a \codej{onClick} may only occur after an \codej{onCreate} and before an \codej{onPause}. If \codej{onClick} were allowed to occur after \codej{onPause}, then the \toolflowdroid model would still need to reason about the callin \codej{v.setOnClickListener(null)} which prevents \codej{onClick}.

The fifth benchmark \synch is caused by a dereference of a nullable field callback that is used by RXJava for synchronizing the results of a background task with the UI thread.
Similar to the first benchmark, this bug would not be detected by a whole-program analysis using the \toolflowdroid main method because the synchronization callback is missing from the call graph.

\begin{specDef}
  The \codej{dispose} callin prevents calls to the subscribe callback.
  \begin{small}
  \begin{align*}
      \enkwCb~\codej{l.}  {\colordullx{0} \codex{language=Java,breaklines=true}{subscribe}} \codej{()}  \boxright
        \exists\ \codej{m},\codej{s}.\  & \iDir{\enkwCi~\codej{m} = \codej{create(l)}} \land \\
        & \niDir{\enkwCi~\codej{s.dispose()}}{\enkwCi~\codej{s} = \codej{m.subscribe()}}
  \end{align*}
  \end{small}
  \label{spec:subscribeDispose}
\end{specDef}

\begin{specDef}
  The \codej{subscribeOn} callin always returns its recieiver (Java builder pattern).
  \begin{small}
    \begin{align*}
      \enkwCi~\codej{t} = \codej{v.subscribeOn()} \boxright \codej{t} = \codej{v}
    \end{align*}
  \end{small}
  \label{spec:subscribeOnSame}
\end{specDef}

\begin{specDef}
  The \codej{observeOn} callin always returns its recieiver (Java builder pattern).
  \begin{small}
    \begin{align*}
      \enkwCi~\codej{t} = \codej{v.observeOn()} \boxright \codej{t} = \codej{v} 
    \end{align*}
  \end{small}
  \label{spec:observeOnSame}
\end{specDef}

\begin{specDef}
  The \codej{onStart} callback may only occur first or after an \codej{onStop}.
  \begin{small}
    \begin{align*}
      \enkwCb~\codej{f.onStart()} \boxright 
        & \left( \notiDir{\enkwCb~\codej{f.onStart()}} \land \notiDir{\enkwCb~\codej{f.onStop()} \right) \vee \\ 
        & }{ \niDir{\enkwCb~\codej{f.onStart()}}{\enkwCb~\codej{f.onStop()}}}
  \end{align*}
\end{small}
  \label{spec:startStop}
\end{specDef}

\begin{specDef}
A value may only be returned from the \codej{create} callin once.
  \begin{small}
    \begin{align*}
      \enkwCi~\codej{m} = \codej{create()} \boxright 
        \notiDir{\enkwCi~\codej{m} = \codej{create()}} 
    \end{align*}
  \end{small}
\label{spec:createUnique}
\end{specDef}
% \begin{specDef}\label{specapdx:call}
% For all objects \codej{l} and \code{m}, if the framework invokes \codej{l.call(m)}, then for some (subscription) object \code{s}, the message $\enkwCi~\codej{s.unsubscribe()}$ has \textbf{N}ot happened \textbf{S}ince $\enkwCi~\codej{s = _.subscribe(l)}$.
% \begin{small}
%   \begin{align*}
% \callSpec
%   \end{align*}
% \end{small}
% \end{specDef}

% \begin{specDef}\label{specapdx:createOnce}
% For all \codej{f}, if the framework invokes \codej{f.onCreate()}, the same message $\enkwCb~\codej{f.onCreate()}$ is \textbf{H}istorially \textbf{N}ot possible (or, \textbf{H}as \textbf{N}ever been invoked in the past).
% \begin{small}
%   \begin{align*}
% \createOnceSpec
%   \end{align*}
% \end{small}
% \end{specDef}

%%%Note: old appendices "What do we express with CBCFTL?", "Empirical Evaluation Addtional material" and "wak points in the paper to improve" are in "appendix_appendix.txt"

%% file: appendix-mhpl.tex
\section{Assumptions of the Application-Only Transition System}
\label{app:assume}

In practice, the translation from a compiled application to the application-only transition system makes some assumptions about the execution environment that we list in this section.
% \TODO{Exceptional control flow disabled. This is try/catch stuff.}
These assumptions simplify reasoning about concurrency, reflection, and library behavior.
First, we assume that all callbacks occur on the same thread (i.e., individual statements inside two separate callbacks cannot be interleaved during execution).
In practice, some callbacks do occur concurrently.  
For example, the callback passed to \codej{Single.create(...)} in the place of the \codej{...} is typically executed on a background thread.
Specific kinds of program analysis exist to address the interleaving caused by threaded concurrency, but that is out of scope for our paper.
Additionally, we do not consider exceptional control flow involving try/catch blocks.

For the application only transition system, we assume that the framework cannot directly modify the app store except through the invocation of a callback.
In theory, the framework may directly modify fields such as \codej{this.act} generating an execution not represented by the application only transition system.
However, such direct access is rare.
Since the framework must be compiled without the application, any method or field that the framework accesses must typically extend a type declared by the framework (e.g., \codej{Fragment} which declares an abstract method \codej{onCreate}).
The only exception is reflection, which may be used to access the app store such as the \codej{this.act} field directly, but we assume such uses of reflection are rare, as it breaks encapsulation.
More commonly, the framework uses reflection to create instances of objects such as \codej{PlayerFragment}.
\citet{DBLP:conf/ecoop/AliL12} coined the phrase \emph{separate-compilation assumption} to describe how a reasonable set of assumptions on framework behavior may be used to generate an application-only call graph.  Our application-only transition system follows the same principle and is based on the same assumptions.

As mentioned in the Overview (\secref{overview}), we assume callin invocations do not \emph{synchronously} call back to the app (i.e., a callin stays in framework code until it returns). 
This assumption is frequently mirrored by other static analysis, such as \citet{DBLP:conf/icse/ArztB16}, by framework stub implementations that do not invoke callbacks.
In general, we find that not many Android methods have synchronous callbacks.
In fact, none of our examples use a method exhibiting this behavior.

This assumption is reflected by the single $\cfedge{\apploc}{ \msgci{\var'}{\msgid}{\varseq} }{\apploc'}$ boundary transition.
Our boundary stacks are degenerate here in that they will only ever have at most one activation $\msgstackframe$ (for the active callback).
The boundary stack $\msgstack$ is conceptually the subsequence of the run-time call stack corresponding to boundary transitions (i.e., the pending calls alternating between callbacks and callins), which ensures that the message history consists of matching calls and returns.

One can extend the language to support synchronous callbacks by splitting the callin invocation boundary transition in two: one for the callin invocation into the framework (i.e., from $\apploc$ to $\quies$) and one for a call return back from the framework (i.e., from $\quies$ to $\apploc'$), analogous to callback invocations and callback returns.
Note that if synchronous callbacks can be soundly modeled as asynchronous ones, then it is also unnecessary to add the complexity of synchronous callbacks (which is generally the case for Android).